\documentclass[article,english,12pt]{article}

\usepackage{amscd,amsfonts,amsmath,amssymb,amsthm,bbm,bm,latexsym,mathrsfs}
\usepackage{epsfig,graphics,graphicx,longtable}

\usepackage[pagebackref=true, hyperfootnotes=true, citecolor=black, colorlinks=true, urlcolor=black, linkcolor=black]{hyperref}

\usepackage[]{natbib}
\usepackage[english]{babel}
\usepackage[usenames]{color}
\usepackage{bookmark}
\usepackage{sgame}
\allowdisplaybreaks
\usepackage[top=1.4in, bottom=1.7in, left=0.7in, right=0.7in]{geometry}
\usepackage{adjustbox}
\usepackage{gensymb}
\usepackage{epstopdf}
\usepackage{tikz}
\usepackage[toc,page]{appendix}
\usepackage{chapterbib}
\usepackage[normalem]{ulem}
\usepackage{cancel}
\usepackage{float}
\usepackage{caption}
\usepackage{subcaption}
\usepackage{placeins}

\usepackage[colorinlistoftodos,color=orange!60]{todonotes}

\makeatother

\theoremstyle{remark}

\theoremstyle{plain}

\newcommand{\blue}[1]{\color{blue} #1} %

\graphicspath{ {figures/} }

\begin{document}

\title{Type 2 Tobit Sample Selection Models with Bayesian Additive Regression Trees}
\author{\hspace{-0pt} Eoghan O'Neill$^{a}$\thanks{Email: eoghan.oneill1@ucd.ie . Acknowledgements: The author gratefully acknowledges helpful comments from Mikhail Zhelonkin, Chen Zhou, Richard Paap, Dennis Fok, Adam Iqbal, F. Javier Rubio, and Emmanuel O. Ogundimu.}
        \\ 
        \vspace{-7pt}
        \\
        {\centering {\small{$^a$University College Dublin}}}}     
\vspace{-5pt}
\date{\today}
\maketitle

\vspace{-10pt}

\begin{abstract}
This paper introduces Type 2 Tobit Bayesian Additive Regression Trees (TOBART-2). BART can produce accurate individual-specific treatment effect estimates. However, in practice estimates are often biased by sample selection. We extend the Type 2 Tobit sample selection model to account for nonlinearities and model uncertainty by including sums of trees in both the selection and outcome equations. A Dirichlet Process Mixture distribution for the error terms allows for departure from the assumption of bivariate normally distributed errors. Soft trees and a Dirichlet prior on splitting probabilities improve modeling of smooth and sparse data generating processes. We include a simulation study and an application to the RAND Health Insurance Experiment dataset.
\end{abstract}

\newpage



\section{Introduction}

Treatment effect estimates obtained from observational data are often biased by non-random sample selection. \citet{heckman1974shadow} proposed to jointly model the outcome and the selection of variables to produce unbiased estimates.{\footnote{The Heckman selection model was named the Type 2 Tobit model by {\citet{amemiya1984tobit}} . 
}} However, this model assumes a linear functions of covariates in the selection and outcome equations, and joint normality of the error terms.   %

To address these issues, we introduce a type 2 Tobit model with separate sums of trees in the selection and outcome equations instead of linear combinations of covariates. This model is henceforth referred to as TOBART-2.  This builds on the Bayesian formulation of type 2 Tobit models introduced by \cite{omori2007efficient, chib2009estimation}, and \cite{van2011bayesian}. The Markov chain Monte Carlo implementation is most similar to that of \cite{van2011bayesian}.  However, Type 2 Tobit models that impose a normality assumption can produce biased estimates when applied to data with small deviations from normality.\footnote{Many other papers have provided alternatives to the assumption of joint normality of the errors in sample selection models. \cite{gallant1987semi} use Hermite series, \cite{marchenko2012heckman} use a t-distribution, \cite{ding2014bayesian} applies a t-distribution in a Bayesian model, and \cite{ogundimu2016sample} consider a skew-normal distribution. \cite{zhelonkin2016robust} propose an  outlier-robust estimator.} A more robust method, TOBART-2-NP, contains a Dirichlet Process Mixture (DPM) of bivariate Gaussian distributions for the error term, as described by \cite{van2011bayesian}.\footnote{A similar DPM of normal distributions is used to model the error terms in the instrumental variable BART model of \cite{mcculloch2021causal}, which is based on \cite{conley2008semi}.} 

The advantages of TOBART-2 are that it accounts for nonlinearity, variable selection, and model uncertainty, while flexibly modelling the errors to allow for non-normality. In addition to point estimates, the  method produces credible intervals for outcome predictions and treatment effect estimates. As noted by \cite{van2011bayesian}, posterior inference can be applied to the covariance of error terms to indicate the extent of sample selection. 
Variable importance measures can be obtained for the selection and outcome equations. 

Data analysis competitions and other simulation studies have demonstrated the impressive performance of BART \citep{hahn2019atlantic, dorie2019automated, wendling2018comparing} in estimating treatment effects. BART models have been formulated to account for confounding \citep{hill2011bayesian, hahn2020bayesian, kim2022bayesian}, sparsity \citep{linero2018bayesianB}, smoothness \citep{linero2018bayesianA}, and unobserved confounders with instrumental variables \citep{mcculloch2021causal}. To the best of our knowledge, the only other paper that describes a BART model that allows for a form of selection on unobservables is the multivariate missing not at random outcome model of \cite{goh2024joint}, created concurrently with TOBART-2.\footnote{Some papers describe BART-based methods for data with selection on observables, e.g. \cite{wang2023improving, gao2021treatment, elliott2023improving, bisbee2019barp}. } However, \cite{goh2024joint} model missingness as a function of covariates and latent outcomes, and do not model selection through a joint distribution of error terms as in standard Heckman selection model formulations.\footnote{TOBART-2 is arguably a more interpretable model with a more clearly specified form of selection on unobservables, and it allows for direct specification of the prior on selection and allows for inference on the level of selection on unobservables. The method of \cite{goh2024joint} is not generalized to non-normal errors, although such an extension would be straightforward. The general approach to selection modelling of TOBART-2 is more common in the econometrics literature, whereas the approach of \cite{goh2024joint} is more often applied in other disciplines.}



Bayesian Model Averaging of linear Type 2 Tobit models has been considered by \cite{jordan2012tobit} and \cite{ eicher2012robust}. Nonlinear approaches to Type 2 Tobit models include kernel regression \citep{ahn1993semiparametric, newey1990semiparametric, lee1994semiparametric, chen2010semiparametric}, splines \citep{marra2013estimation,das2003nonparametric, newey2009two}, and neural networks \citep{zhang2021deep}.\footnote{However,  \cite{zhang2021deep} do not evaluate their methods using simulations with non-zero correlation of the error terms in the selection and outcome equations. i.e. there is no selection on unobservables.} \cite{chib2009estimation} formulate a semiparametric Bayesian model with sample selection and endogeneity by using a second-order Markov Process prior. However, \cite{chib2009estimation} apply the restrictive assumption of jointly normally distributed errors. \cite{kim2019bayesian} formulate a Bayesian sample selection model with a scale mixture of bivariate normal distributions for the error term, and a Berstein polynomial regression model for the outcome. However, the selection equation contains a linear function of covariates. \cite{wiesenfarth2010bayesian} make use of penalized splines in the outcome and selection equations of a Bayesian sample selection model, although the errors are restrictively modelled as being jointly normally distributed.  \cite{wiemann2022correcting} introduce a Bayesian method with splines and a bivariate copula.

\cite{fan2007sample} apply trees and model averaging to data with sample selection. \cite{zadrozny2004learning, tran2017selection}, and many others combine machine learning methods with inverse probability weighting approaches that cannot account for selection on unobservables. \cite{alaimo20222} apply gradient boosted trees to tax revenue data with sample selection, and account for model uncertainty by using a bootstrap procedure. However, this approach is also based on the inverse propensity methods of \cite{zadrozny2004learning} and therefore does not account for selection on unobservables.

\cite{brewer2024addressing} describe control function approaches that allow for machine learning methods to be combined with a Heckman-style two-step framework. This method leads to improved predictive accuracy in simulations with moderate to high levels of selection. However, this approach relies on the assumption of joint normality and \cite{brewer2024addressing} do not discuss efficiency or methods for inference. \cite{zadrozny2001learning} similarly apply a control function approach with decision trees. \cite{zhu2017nonasymptotic} apply a LASSO penalty using a control function approach and semiparametric framework. \cite{ogundimu2022lasso, ogundimu2024lasso} applies LASSO and adaptive LASSO in both the selection and outcome equations. \cite{iqbal2023bayesian} apply spike-and-slab priors to the coefficients in the selection and outcome equations to produce a Bayesian sample selection model for high dimensional data. \cite{bia2024double} describe a double machine learning approach to estimation of the average treatment effect for high-dimensional data with sample selection.

In addition to TOBART-2, this paper introduces the following:
\begin{itemize}
    \item To allow for data-informed calibration of the prior variance of the outcome equation errors, the prior of \cite{ding2014bayesian} is altered to a bivariate normal prior with an additional data-informed hyperparameter. This is motivated by the fact that good performance of standard BART without hyperparameter tuning is partly attributable to well-calibrated hyperparameters. The sampler of \cite{ding2014bayesian} is also improved by removing unnecessary imputation of the missing outcomes.\footnote{This was noted as an area for future research by \cite{ding2014bayesian}.}
    \item We provide new insights pertaining to the \cite{van2011bayesian} and \cite{omori2007efficient} Tobit-2 priors. The CDF of the correlation between selection and outcome errors is derived for both priors. We derive the conditions in which the implied prior on the correlation specified by \cite{van2011bayesian} is unimodal, and when it is concentrated at +1 and -1. The prior CDF of the outcome variance is derived for prior calibration. The prior correlation distribution specified by \cite{omori2007efficient} can take various unimodal and bimodal forms, depending non-trivially on hyperparameters.\footnote{This contrasts with a note by \cite{van2011bayesian}, which perhaps refers to particular hyperparameter values.} These results and prior calibration are discussed in Appendix \ref{imp_details_app}.
    \item We consider a variation on the MCMC sampler of \cite{van2005bayesian, van2011bayesian}  in which the outcome equation variance and the covariance between the selection and outcome equation errors are jointly sampled. See Appendix \ref{phigamma_norm_ig_sec}.
    \item We describe an alternative TOBART-2 implementation that extends the marginalized BART method described by \cite{collins2023improved} to a sample selection model. This implementation, TOBART2marg, marginalizes out both the terminal node parameters and the covariance term determining sample selection when updating trees in the outcome equation, and then makes a joint draw of all marginalized parameters. This is motivated by the observation of \cite{van2005bayesian, van2011bayesian} in the linear model setting that there is high dependence between the outcome equation coefficient draws and the covariance parameter draws. See Appendix \ref{TOBART2Marg_app}.
\end{itemize}

An \texttt{R} package containing implementations of TOBART-2 and the Bayesian linear model described by \cite{van2011bayesian} is available at \url{https://github.com/EoghanONeill/TobitBART}. The remainder of the paper is structured as follows: In section \ref{methods_sec} we describe the TOBART-2 model and MCMC implementation, in section \ref{criteria_sec} we discuss the choice of estimand and evaluation metrics, section \ref{simulation_sec} contains a simulation study, section \ref{application_sec} contains an application to a real world data, and section \ref{conclusion_sec} concludes the paper.

\section{Methods}\label{methods_sec}

\subsection{Review of Bayesian Additive Regression Trees (BART)}

\noindent \textbf{Description of Model and Priors}

\medskip

Suppose there are $n$ observations, and the $n \times p$ matrix of explanatory variables, $X$, has $i^{th}$ row $x_i=[x_{i1},...,x_{ip}]$. Following the notation of \cite{chipman2010bart}, let $T$ be a binary tree consisting of a set of interior node decision rules and a set of terminal
nodes, and let $M = \{ \mu_1 , ..., \mu_b \}$ denote a set of parameter values associated with each of the $b$ terminal nodes of $T$. The interior node decision rules are binary splits of the predictor space into the sets $\{ x_{is} \le c \}$ and $\{ x_{is} > c \}$ for continuous $x_{s}$. Each observation's $x_i$ vector is associated with a single terminal node of $T$, and is assigned the $\mu$ value associated with this terminal node. For a given $T$ and $M$, the function $g(x_i;T,M)$ assigns a $\mu \in M$
to $x_i$. This gives the single tree model $Y \sim g(x_i;T,M) + \varepsilon \ , \ \varepsilon \sim N(0,\sigma^2)$  \citep{chipman1998bayesian}. For the standard BART model, the outcome is determined by a sum of trees,
$$Y_i = f(x_i) +\varepsilon_i = \sum_{j=1}^m g(x_i ; T_j, M_j)+\varepsilon_i \ \ , \ \ \varepsilon_i \overset{i.i.d}{\sim} N(0, \sigma^2)$$
where $g(x_i;T_j,M_j)$ is the output of a decision tree. $T_j$ refers to a decision tree indexed by $j=1,...,m$, where $m$ is the total number of trees in the model. $M_j$ is the set of terminal node parameters of $T_j$.

Prior independence is assumed across trees $T_j$ and across terminal node means $M_j = (\mu_{1j}...\mu_{b_j j})$, where $1,...,b_j$ indexes the terminal nodes of tree $j$. The prior of \cite{chipman2010bart} has the form:
$$p(M_1,...,M_m,T_1,...,T_m,\sigma) \propto \left[ \prod_j \left[ \prod_k p(\mu_{kj}|T_j) \right] p(T_j)\right]p(\sigma) $$
%
%
%
In standard BART, $\mu_{kj} | T_j \overset{i.i.d}{\sim} N(0,\sigma_0^2)$ where $\sigma_0 = \frac{0.5}{e \sqrt{m}}$ and $e$ is a user-specified hyper-parameter. 

\cite{chipman2010bart} set a regularization prior on the tree size and shape $p(T_j)$. The probability that a given node within a tree $T_j$ is split into two child nodes is $\alpha (1+d_h)^{-\beta}$, where $d_h$ is the depth of internal node $h$, and the parameters $\alpha$ and $\beta$ determine the size and shape of $T_j$ respectively. \cite{chipman2010bart} use uniform priors on available splitting variables and splitting points. 
\cite{chipman2010bart} assume that the model precision $\sigma^{-2}$ has a conjugate prior distribution $\sigma^{-2} \sim Ga(\frac{v}{2}, \frac{v \lambda}{2})$ with degrees of freedom $v$ and scale $\lambda$. 

BART predictions are averages of sum-of-tree models. Therefore model uncertainty is taken into account and there are two levels of regularization. Firstly, greater prior probability is placed on models with shallower trees with fewer splitting points. Secondly, over-fitting is further avoided through the prior on the terminal node parameters $\mu_{kj}$, as in standard Bayesian linear regression.

\bigskip

\noindent \textbf{BART Implementation}


\medskip

Samples can be taken from the posterior distribution $p((T_1, M_1),...,(T_m,M_m), \sigma | y)$ by a Bayesian backfitting MCMC algorithm. This algorithm is a Gibbs sampler, involving $m$ successive draws from 
$(T_j , M_j )| T_{(j)} , M_{(j)} , \sigma , y $ for $j=1,...,m$, where $T_{(j)} , M_{(j)} $ are the trees and parameters for all trees except the $j^{th}$ tree, followed by a draw of $\sigma $ from the full conditional $\sigma | T_1,...,T_m,M_1,...,M_m,y$.

The $q^{th}$ Gibbs iteration defines the sum of trees function $f_q^*(.)= \sum_{j=1}^m g_q(. \ ; T_{j}^*, M_{j}^*)$. After burn-in, the sequence of $f^*$ draws, $f_1^*,...,f_Q^*$ may be regarded as an approximate, dependent sample of size $Q$ from $p(f|y)$.
To estimate the unknown function $f(x)$, 
the expectation $E(f(x)|y)$ is approximated by 
$\frac{1}{Q} \sum_{q=1}^Q f_{q}^* (x)$. Posterior credible intervals can be obtained from quantiles of the draws $f_{q}^* (x)$. 

%

\subsection{Type 2 Tobit}

\subsubsection{Type 2 Tobit and TOBART Model}


In the standard (linear) Type 2 Tobit model \citep{omori2007efficient}, separate parameters and variables enter the selection and outcome equations:
$$
y_i = 
\begin{cases}
	y_i^* 			& \text{if } z_i^* \ge 0\\
	n.a.            & \text{otherwise}
\end{cases}
  , \ z_i^* = \bm{w}_i'\bm{\theta} + \xi_i    , \ y_i^* = \bm{x}_i'\bm{\beta} + \eta_i , \ \begin{pmatrix}
	\xi_i \\
	\eta_i
\end{pmatrix}
\sim  \mathcal{N}(\bm{0}, \Sigma)  \text{ where } \Sigma = 
\begin{pmatrix}
	1 & \gamma \\
	\gamma & \phi + \gamma^2 
\end{pmatrix} $$
In Type 2 TOBART, the linear combinations $\bm{w}_i'\bm{\theta}$ and $\bm{x}_i'\bm{\beta} $ are replaced by sums-of-trees denoted by $f_z(\bm{w}_i)$ and $f_y(\bm{x}_i)$ respectively.
$$z_i^* = f_z(\bm{w}_i) + \xi_i \ , \ y_i^* = f_y(\bm{x}_i) + \eta_i $$
%
%
%
For linear Type 2 Tobit, the coefficient priors are $ \bm{\theta} \sim \mathcal{N}(\bm{\theta}_0, \bm{\Theta}_0) \ , \ \bm{\beta} \sim \mathcal{N} (\bm{\beta}_0, \bm{B}_0 ) $, and for TOBART  the priors are $f_z \sim BART $ and $f_y \sim BART $. The covariance matrix parameters have the priors $ \gamma \sim \mathcal{N} (\gamma_0, G_0) $ and $ \phi \sim IG(\frac{n_0}{2}, \frac{S_0}{2})  $. An alternative prior, $\gamma  | \phi \sim \mathcal{N} (\gamma_0, \tau \phi)$, is introduced by \cite{van2011bayesian}, who notes that dependence between $\gamma$ and $\phi$ through $\tau>0$ allows various shapes of the prior on correlation between $\xi$ and $\nu$, and an induced prior correlation with less probability mass near $\pm 1 $. 
Details of the Type 2 Tobit and TOBART Gibbs sampler are provided in appendix \ref{app_mcmc_tbart2norm}.

We also consider a prior for the distribution of the error terms similar to the prior introduced by \cite{ding2014bayesian}. \cite{ding2014bayesian} assumes that the errors have a bivariate t-distribution:
$$ 
\begin{pmatrix}
	\xi_i \\
	\eta_i
\end{pmatrix}
\sim \ \mathcal{N}(\bm{0}, \alpha \Omega/q_i) \ , \ \text{where} \ q_i \sim \alpha \chi_{\nu}^2/\nu \ , \ i=1,\dots,N
$$
where $ \Omega = \begin{pmatrix}
    1 & \rho \tilde{\sigma}_2 \\
    \rho \tilde{\sigma}_2 & \tilde{\sigma}_2^2 \\
\end{pmatrix} $
and the prior is $ \alpha \sim b / \xi_{c}^2 $ and 
$ \tilde{\Sigma} = \begin{pmatrix}
    \tilde{\sigma}_1 & 0 \\
    0 & 1 \\
\end{pmatrix}  \Omega  \begin{pmatrix}
    \tilde{\sigma}_1 & 0 \\
    0 & 1 \\
\end{pmatrix} 
\sim W_2^{-1} (\nu_0, I_2) $, 
which is equivalent to   $p(\Omega) \sim (1 - \rho^2)^{-3/2} \tilde{\sigma}_2^{- (\nu_0 +3 )} \exp \Big\{ - \frac{1}{2 \tilde{\sigma}_2^2 (1- \rho^2) }  \Big\}$ and $\tilde{\sigma}_1^2|\Omega \sim \{ (1 - \rho^2) \chi_{\nu_0}^2 \}^{-1} $ , where  $\tilde{\sigma}_1 = \sqrt{\tilde{\Sigma}_{1,1}} $ , and $\tilde{\sigma}_2 = \sqrt{\tilde{\Sigma}_{2,2}} = \sqrt{\Omega_{2,2}} $.

However, instead of specifying a t-distribution, we maintain a normal specification for the error term, and allow for departures from normality through a mixture of normal distributions.
$$ 
\begin{pmatrix}
	\xi_i \\
	\eta_i
\end{pmatrix}
\sim \ \mathcal{N}(\bm{0},  \Omega)  \ , \  \tilde{\Sigma} = \begin{pmatrix}
    \tilde{\sigma}_1 & 0 \\
    0 & 1 \\
\end{pmatrix} 
\Omega  \begin{pmatrix}
    \tilde{\sigma}_1 & 0 \\
    0 & 1 \\
\end{pmatrix} 
\sim W_2^{-1} (\nu_0, I_2) 
$$
As noted by \cite{ding2014bayesian}, this prior imposes a $U([-1,1])$ marginal prior on $\rho = \text{corr} (\xi_i, \eta_i) $ when $\nu_0 = 3$. In contrast, the priors of \cite{van2011bayesian} and \cite{omori2007efficient} can be notably non-uniform.\footnote{The prior of \cite{van2011bayesian} can place most mass around zero, or closer to $\pm 1$ depending on the value of $\tau$. See Appendix \ref{imp_details_app} and the appendix of \cite{iqbal2023bayesian} for further discussion.}

\bigskip

However, this prior does not allow for adequate calibration of the (marginal) prior on $\tilde{\sigma}_2 $ relative to the data-informed prior calibration described by \cite{chipman2010bart} and \cite{mcculloch2021causal}. In particular, it can be shown that (see appendix \ref{marginal_prior_app} for derivation)
$$ p(\tilde{\sigma}_2^2 ) \propto \left( \frac{1}{\tilde{\sigma}_2^2 } \right)^{\frac{\nu_0-1}{2} + 1}
\exp \left( - \frac{1}{ 2 \tilde{\sigma}_2 ^2} \right)  $$
therefore the marginal prior is $\tilde{\sigma}_2^2  \sim \Gamma^{-1}( \frac{\nu_0-1}{2} ,  \frac{1}{2})$. This prior does not allow for both a uniform marginal prior on $\rho$ and a data-informed marginal prior on $\tilde{\sigma}_2^2 $. Therefore, we specify the following prior
$$ 
\begin{pmatrix}
	\xi_i \\
	\eta_i
\end{pmatrix}
\sim \ \mathcal{N}(\bm{0},  \Omega)  \ , \  \tilde{\Sigma} = \begin{pmatrix}
    \tilde{\sigma}_1 & 0 \\
    0 & 1 \\
\end{pmatrix} 
\Omega  \begin{pmatrix}
    \tilde{\sigma}_1 & 0 \\
    0 & 1 \\
\end{pmatrix} 
\sim W_2^{-1} (\nu_0, c I_2) 
$$
where we have introduced the hyperparameter $c$. Similarly to \cite{ding2014bayesian}, we can re-express this prior as 
$p(\Omega) \sim (1 - \rho^2)^{-3/2} \tilde{\sigma}_2^{- (\nu_0 +3 )} \exp \Big\{ - \frac{c}{2 \tilde{\sigma}_2^2 (1- \rho^2) }  \Big\}$ and $\tilde{\sigma}_1^2|\Omega \sim \{ \frac{(1 - \rho^2)}{c} \chi_{\nu_0}^2 \}^{-1} $  or $\tilde{\sigma}_1^2|\Omega \sim \Gamma^{-1} \left( \frac{\nu_0}{2}, \frac{c}{2(1-\rho^2)}  \right)  $. It can be shown (see Appendix \ref{marginal_prior_app}) that this implies the following marginal priors:
$$ p(\rho) \propto (1 - \rho^2)^{\frac{\nu - 3}{2} }  \ \text{ and } \  p(\tilde{\sigma}_2^2) \propto \left( \frac{1}{\tilde{\sigma}_2^2}   \right)^{\frac{\nu_0-1}{2} + 1} \exp \left( - \frac{ c }{ 2 \tilde{\sigma}_2^2 }  \right)  $$
This prior allows us to set $\nu_0 =3$ for a uniform marginal $\rho$ distribution and set $c$ such that the $q^{th}$ quantile of the marginal prior on $\tilde{\sigma}_2^2$, i.e. $\Gamma^{-1}( \frac{\nu_0-1}{2}, \frac{c}{2}) $, is equal to an estimated variance of $\eta_i$, denoted by $\hat{\sigma}_2^2$. This is similar to the prior calibration approach of \cite{chipman2010bart} and \cite{mcculloch2021causal}. We set $q = 0.95$. Setting $\hat{\sigma}_2^2$ equal to the sample variance of the observed outcomes might not give a well-calibrated prior if the variance of the observed outcomes is far from the variance of the full set of outcomes, therefore it is preferable to use the estimated outcome variance from a linear Tobit model estimated by maximum likelihood.\footnote{The value of $c$ is  $\hat{\sigma}_2^2$ multiplied by the $(1-q)^{th}$ quantile of the $\chi_{\nu_0}^2$ distribution. We also considered a number of data-informed prior calibration methods for the priors of \cite{omori2007efficient} and \cite{van2011bayesian}. See Appendix \ref{imp_details_app}.}

\subsection{Nonparametric Type 2 Tobit}

\subsubsection{Nonparametric Type 2 Tobit and TOBART Models}

The models presented here are based on the semiparametric Bayesian Type 2 Tobit model implementation introduced by \cite{van2011bayesian}.\footnote{The notation is more similar to that of \cite{omori2007efficient}.} In the standard (linear) Type 2 Tobit model, separate parameters and variables enter the selection and outcome equations:
$$
y_i = 
\begin{cases}
	y_i^* 			& \text{if } z_i^* \ge 0\\
	n.a.            & \text{otherwise}
\end{cases} , \ z_i^* = \bm{w}_i'\bm{\theta} + \xi_i  , \  y_i^* = \bm{x}_i'\bm{\beta} + \eta_i, \begin{pmatrix}
	\xi_i \\
	\eta_i
\end{pmatrix}
\overset{i.i.d.}{\sim}   \mathcal{N}\Bigg(
\begin{pmatrix}
	\mu_{i1} \\
	\mu_{i2}
\end{pmatrix}
, \Sigma_i = 
\begin{pmatrix}
	1 & \gamma_i \\
	\gamma_i & \phi_i + \gamma_i^2 
\end{pmatrix}\Bigg) 
$$
$$\{
\begin{pmatrix}
	\mu_{i1} &
	\mu_{i2}
\end{pmatrix}'
, \gamma_i, \phi_i \} | H \sim H  \ , \ H | \alpha, H_0  \sim \mathcal{DP} (\alpha, H_0)$$
In Type 2 TOBART, the linear combinations $\bm{w}_i'\bm{\theta}$ and $\bm{x}_i'\bm{\beta} $ are replaced by sums-of-trees denoted by $f_z(\bm{w}_i)$ and $f_y(\bm{x}_i)$ respectively.
$$z_i^* = f_z(\bm{w}_i) + \xi_i \ , \ y_i^* = f_y(\bm{x}_i) + \eta_i \ , \  
$$
%
%
%
%
%
For linear Tobit, the priors are $ \bm{\theta} \sim \mathcal{N}(\bm{\theta}_0, \bm{\Theta}_0) \ , \ \bm{\beta} \sim \mathcal{N} (\bm{\beta}_0, \bm{B}_0 ) $, and for TOBART $f_z \sim BART $ and $f_y \sim BART $. $H$ is a discrete distribution of the parameters $ \vartheta_i = \{
\begin{pmatrix}
	\mu_{i1} &
	\mu_{i2}
\end{pmatrix}'
, \gamma_i, \phi_i \} $ . The base measure $H_0$ is defined by 
$\begin{pmatrix}
	\mu_{i1} &
	\mu_{i2}
\end{pmatrix}' \sim  \mathcal{N} (\bm{0}, \Omega) \ , \  \gamma_i \sim \mathcal{N} (\gamma_0, G_0) \ , \ \phi_i \sim IG(\frac{n_0}{2}, \frac{S_0}{2})  $ and
\cite{van2011bayesian} sets $\Omega = 10 \bm{I}_2$.\footnote{An alternative would be to set it equal to $\frac{\hat{\sigma}_y^2}{0.016} \bm{I}_2$ such that the prior probability that a $\mu_{i}$ value lies in $(-10,10)$ is $0.8$. This is similar to the hyperparameter settings described by \cite{chipman1998bayesian} and \cite{conley2008semi}, albeit without a conjugate prior. The other base distribution hyperparameters are set to the same values as for standard TOBART-2. Alternatively, as suggested by \cite{george2019fully}, scale and shape of the prior variance of the outcome error can be set to values that imply a tighter base distribution, as the mixture distribution creates additional spread for the overall prior.}  However, since $ f_z(\bm{w}_i) $ is non-linear, we can set $\mu_{i1}$ to 0 and still have a flexible selection model, $\Pr(z_i^* >0) = \Phi ( f_z(\bm{w}_i))$, with a standard normal marginal distribution of $\xi_i$ for identification of $f_z(\bm{w}_i)$.
The parameter $\alpha$ determines the distribution of the weights given to each element of $H$. As $\alpha$ increases, more elements receive non-negligible weight. 
As in the parametric error model, \cite{van2011bayesian} uses the alternative prior $\gamma_i | \phi_i \sim \mathcal{N} (\gamma_0, \tau \phi_i)$. 


\cite{george2019fully} and \cite{van2011bayesian}  apply different priors to $\alpha$. \cite{george2019fully} choose the maximum and minimum number of components in $G$, denoted by $I_{min}$ and $I_{max}$, then solve for $\alpha_{min}$ such that the mode of the number of elements (denoted by $I$) of $H$ is $I_{min}$. Similarly $\alpha_{max}$ is derived from $I_{max}$. Then
 $p(\alpha) \propto (1- \frac{ \alpha - \alpha_{min}}{\alpha_{max} - \alpha_{min}} )^{\psi} $ . 
The default values are $I_{min} = 1$, $I_{max} = [(0.1)*n]$, where $[\bullet]$ denotes the floor function, and $\psi = 0.5$. 
\cite{mcculloch2021causal} also specify a DPM prior on the errors of an IV-BART model and set  $I_{min} = 2$, $I_{max} = [(0.1)*n] + 1$. 
\cite{van2011bayesian} applies the prior from \cite{escobar1994estimating}, $\alpha \sim \text{Gamma}(c_1, c_2)$, with $c_1=2$ and $c_2=2$. Details of the sampler introduced by \cite{van2011bayesian} are provided in appendix \ref{app_mcmc_tbart2np}. We apply the sampler of \cite{van2011bayesian} by default and provide an option for the prior of \cite{george2019fully}.

%
%

\subsubsection{Type 2 TOBART Gibbs Sampler}


Samples of $y_i^*$ and $z_i^*$ in the Gibbs sampler for TOBART are the same as in the standard Tobit sampler. However, in contrast to the draws of coefficient parameters for linear Tobit, each tree must be drawn separately conditional on all other trees. Detailed outlines of the full TOBART-2 and TOBART-2-NP samplers are included in appendices \ref{app_mcmc_tbart2norm} and \ref{app_mcmc_tbart2np} respectively. An alternative sampler, in which the outcome equation tree draws are made unconditional on the terminal nodes and covariance term $\gamma$, and in which the outcome equation leaf parameters and $\gamma$ are sampled jointly, is described in appendix \ref{TOBART2Marg_app}. This alternative sampler is intended to reduce the dependence between the outcome equation tree samples and $\gamma$ samples. The sampler of \cite{van2011bayesian} is similarly motivated.

\subsubsection{Inference about degree of dependence}

As noted by \cite{van2011bayesian}, the samples of the error terms allow for inference on the degree of dependence between the errors in the selection and outcome equations. 
At each iteration $t$ of the TOBART-2-NP Gibbs sampler, we may generate a pseudo-sample $\{u_{i,t} \}_{i=1}^n$ where $u_{i,t} \sim \mathcal{N}(\mu_{i,t}, \Sigma_{i,t})$ and calculate a dependence measure. The set of values of this measure across all iterations of the Gibbs sampler are an approximate sample from its posterior distribution. 

\subsection{Note on prior hyperparameters}

\cite{iqbal2023bayesian} apply the \cite{van2011bayesian} prior with $n_0 = 2$ and $S_0= 2$. It is claimed that these values give an induced prior on the outcome variance $ \phi + \gamma^2$ that is similar to an Inverse Gamma. However, the induced prior mean is $ E[ \phi + \gamma^2 ] = (1 + \tau)\frac{S_0}{n_0-2} + g_0^2  $ and therefore is undefined if $n_0=2$, which it would also be for the Inverse Gamma. We note in Appendix \ref{imp_details_app} that the prior distribution of $ \phi + \gamma^2$ is a non-trivial sum of an inverse gamma distributed variable and a (dependent) square of a location and scale shifted t-distributed variable.

Appendix \ref{imp_details_app} contains an additional discussion of some methods for calibration of $n_0$ and $S_0$, and of potential disadvantages of the recommendation of \cite{iqbal2023bayesian} to set $\tau=5$ in general. In particular, large values of $\tau$ increase the prior mean of of $\phi + \gamma^2$. Moreover, we observe that larger values of $\tau$ can result in a more peaked bimodal prior for $\rho$ with modes closer to $\pm1$.\footnote{An alternative would be to apply a hyperprior to $\tau$. We did not find this to be useful.} 









\subsection{Note on Estimands and Evaluation criteria}\label{criteria_sec}
\subsubsection{Prediction Estimands}
%
 We assume that the data generating process is as described for the Type 2 TOBART model. To predict the observed outcome, $Y_i$ for selected observations, we may use the following estimand:
$$ E[Y_i | Z_i^* \ge 0 , \bm{x}_i, \bm{w}_i ] = 
f_{y}(\bm{x}_i) + E[\eta_i |f_z(\bm{w}_i) + \xi_i \ge 0  ] = f_{y}(\bm{x}_i)  + \gamma \frac{\phi(f_z(\bm{w}_i) )}{\Phi(f_z(\bm{w}_i) )} $$
%
%
%
 We expect any nonlinear machine learning method trained on the selected observations to estimate this equally as well as TOBART-2. 
 For the nonparametric TOBART-2 DGP, we instead have
$$ E[Y_i | Z_i^* \ge 0 , \bm{x}_i, \bm{w}_i,
	\mu_{i1} ,
	\mu_{i2}
, \Sigma_i ] = $$ 
$$ f_{y}(\bm{x}_i) + E[\eta_i |f_z(\bm{w}_i) + \xi_i \ge 0 , 
	\mu_{i1} ,
	\mu_{i2}
, \Sigma_i ]  = 
f_{y}(\bm{x}_i) + \mu_{i2} + \gamma_i \frac{\phi(f_z(\bm{w}_i) + \mu_{i1})}{\Phi(f_z(\bm{w}_i) + \mu_{i1})} $$
Then we predict the outcome by averaging over MCMC draws indexed by $d=1,\dots,D$
$$ E[Y_i | Z_i^* \ge 0, \bm{x}_i, \bm{w}_i  ] \approx \frac{1}{D} \sum_{d=1}^{D} E[Y_i | Z_i^* \ge 0 ,  f_{y}^{(d)},  f_{z}^{(d)},
	\mu_{i1}^{(d)} ,
	\mu_{i2}^{(d)}
, \Sigma_i^{(d)}, \bm{x}_i, \bm{w}_i ] = $$
$$ \frac{1}{D} \sum_{d=1}^{D}  \Bigg\{ f_{y}^{(d)}(\bm{x}_i) + \mu_{i2}^{(d)} + \gamma_i^{(d)} \frac{\phi(f_z^{(d)}(\bm{w}_i) + \mu_{i1}^{(d)})}{\Phi(f_z^{(d)}(\bm{w}_i) + \mu_{i1}^{(d)})} \Bigg\} $$


If the task is to predict the latent outcome regardless of whether the test observation will be selected or not, or if there will be no selection mechanism in the test data, and instead all outcomes will be observed, then we predict $Y_i$ with the following estimand for the TOBART-2 DGP:
$$ E[Y_i^* |  \bm{x}_i, \bm{w}_i ] = f_{y}(\bm{x}_i)  $$
Therefore, for this DGP, a machine learning method trained on the selected data would target an estimand containing a bias term $ \gamma \frac{\phi(f_z(\bm{w}_i) )}{\Phi(f_z(\bm{w}_i) )} $. In contrast, a Tobit-2-based approach, such as TOBART-2, directly models $f_y$, and we can use the estimate $\frac{1}{D} \sum_{d=1}^{D} f_{y}^{(d)}(\bm{x}_i)  $ for standard TOBART-2 and $\frac{1}{D} \sum_{d=1}^{D}  \Big\{ f_{y}^{(d)}(\bm{x}_i) + \mu_{i2}^{(d)}  \Big\} $ for TOBART-2-NP.

\subsubsection{Treatment Effect Estimands}

Potential outcomes under treatment and control allocation are denoted by $Y_i(1)$ and $Y_i(0)$ respectively. Similarly, $Y_i^*(1)$ and $Y_i^*(0)$ denote potential latent outcomes, and $Z_i^*(1)$ and $Z_i^*(0)$ denote potential selection equation latent variable values. The binary treatment variable is denoted by $T_i$. 

Generally, the estimand of interest is an effect on the latent outcome $Y_i^*$, not an effect on the observed outcome $Y_i$. If treatment affects the latent outcome, but has no effect on selection, i.e. we now have $f_y(\bm{x}_i, T_i)$ and $f_z(\bm{w}_i)$, then the estimand of interest for the TOBART-2 DGP is:
$$ E[Y_i^*(1) | \bm{x}_i, \bm{w}_i ] - E[Y_i^*(0) | \bm{x}_i, \bm{w}_i ] $$
Assume that there is unconfoundedness of treatment, possibly conditional on covariates, i.e. $ Y_i^*(1), Y_i^*(0) \perp T_i | \bm{x}_i$. Then TOBART-2 directly models the above estimand as $f_y(\bm{x}_i, 1) - f_y(\bm{x}_i, 0) \approx \frac{1}{D} \sum_{d=1}^{D} f_y^{(d)}(\bm{x}_i, 1) - f_y^{(d)}(\bm{x}_i, 0) $. A nonlinear machine learning method naively trained only on selected observations targets the same estimand because \footnote{The naive method can be an ``S-learner'' with one trained model for all selected observations, with treatment included as a covariate, or a ``T-learner'' with two models separately trained on selected treatment and control observations \citep{kunzel2019metalearners}.}
$$ E[Y_i^*(1) | Z_i^* \ge 0, \bm{x}_i, \bm{w}_i ] - E[Y_i^*(0) | Z_i^* \ge 0, \bm{x}_i, \bm{w}_i ]  = f_y (\bm{x}_i, 1) - f_y (\bm{x}_i, 0)  + \gamma \Bigg(  \frac{\phi(f_z(\bm{w}_i) )}{\Phi(f_z(\bm{w}_i) )} - \frac{\phi(f_z(\bm{w}_i) )}{\Phi(f_z(\bm{w}_i) )}   \Bigg) $$
$$ =  f_y (\bm{x}_i, 1) - f_y (\bm{x}_i, 0) . $$
Now suppose that treatment enters both the outcome and selection equations, so that we have $f_y(\bm{x}_i, T_i)$ and $f_z(\bm{w}_i, T_i)$, and assume $ Y_i^*(1), Y_i^*(0), Z_i^*(1), Z_i^*(0),  \perp T_i | \bm{x}_i,\bm{w}_i$. TOBART-2 still directly models the estimand of interest $f_y (\bm{x}_i, 1) - f_y (\bm{x}_i, 0)$. However, the treatment effect estimates produced by a machine learning method naively trained only on the selected observations would target an estimand containing a selection bias term:
%
%
$$ E[Y_i^*(1) | Z_i^* (1) \ge 0, \bm{x}_i, \bm{w}_i ] - E[Y_i^*(0) | Z_i^*(0) \ge 0, \bm{x}_i, \bm{w}_i ]  = $$
$$ f_y (\bm{x}_i, 1) - f_y (\bm{x}_i, 0)  + \gamma \Bigg(  \frac{\phi(f_z(\bm{w}_i,1) )}{\Phi(f_z(\bm{w}_i,1) )} - \frac{\phi(f_z(\bm{w}_i,0) )}{\Phi(f_z(\bm{w}_i,0) )}   \Bigg) $$
Therefore TOBART-2 is expected to produce more accurate ITE estimates for DGPs in which treatment affects sample selection.

\bigskip



\section{Simulation Study}\label{simulation_sec}





\subsection{Description of Simulations}

In addition to the simulations described here, we provide a simulation study with the linear DGP of \cite{iqbal2023bayesian} in Appendix \ref{further_sims}, although this DGP does not include an excluded instrument.

\subsubsection{  \cite{brewer2024addressing} Simulations}

We simulate the data generating processes described by \cite{brewer2024addressing}, with less observations and variables for computational feasibility. We also consider non-normally distributed errors. Covariates $X_1,...,X_{10}$ are generated from a multivariate normal distribution with $\text{Cov}(X_{ki},X_{ji}) = 0.3^{|k-j|}$. The excluded instrument, $W_{exc,i}$, that appears in the selection equation, but not the outcome equation is generated by $W_{exc,i} = \sum_{j=1}^{10}  0.05 X_{ji} + e_i \ , \ e_{w,i} \sim \mathcal{N}(0,0.75) $. 
%
%

We consider the following DGPs, and two more DGPs in Appendix \ref{further_sims}:

\begin{enumerate}
    \item Outcome: $Y_i^* = \sum_{j=1}^{10} \frac{0.4}{j^2} X_{ji}  + \eta_i$ \\
    Selection: $ Z_{i}^* =  1.25 + \sum_{j=1}^{10} \frac{0.1}{(10.5 - j)^2} X_{jo} + W_{exc,i} + \xi_i  $. The errors $\eta_i$ and $\xi_i$ are bivariate standard normal with correlation $\rho \in \{ 0,0.45,0.9 \} $.
    \item Outcome: $Y_i^* = -0.25  1.25 \sin \left( \frac{\pi}{4} +  0.75 \pi \sum_{j=1}^{10} \frac{0.4}{j^2} X_{ji} \right) + \eta_i$ \\
    Selection: $ Z_{i}^* =  1.25 + \sum_{j=1}^{10} \frac{0.1}{(10.5 - j)^2} X_{jo} + W_{exc,i} + \xi_i  $. The errors $\eta_i$ and $\xi_i$ are bivariate standard normal with correlation $\rho \in \{ 0,0.45,0.9 \} $.
    
    \end{enumerate}

Appendix \ref{further_sims} includes results for the above DGPs with the following non-normal distributions for the errors. \footnote{TOBART-2-NP results will be added to an updated version of this paper at a later date.}
\begin{itemize}
    \item $\eta_i \sim \rho V_i + \rho t_{\nu = 5} $ and $\xi_i \sim t_{\nu = 5}  $, where $t_{nu = 5}$ is a standard t-distribution with $5$ degrees of freedom. The correlation is $\rho = \frac{1}{sqrt{2}} $.
    \item The errors are generated from a mixture of standard bivariate normals $ 0.3 \mathcal{MVN}_2 ( (0,-2.1)', \Sigma) + 0.7 \mathcal{MVN}_2 ( (0,0.9)', \Sigma) $  where $\Sigma_{12} = 0.85$ , and the implied correlation is $0.5$.
    \end{itemize}

The numbers of training and test observations are set to $2500$ and $500$ respective;y, and the number of repetitions for each simulation scenario is set to $10$.

\subsubsection{Heterogeneous Treatment Effect Simulations}

When designing a simulation study to evaluate heterogeneous treatment effect estimation, we consider the following:

\begin{itemize}
    \item There is unlikely to be a notable difference between TOBART-2 and a naive application of a nonparametric regression method to sample selected data unless treatment has a non-negligible effect on the inverse Mill's ratio for a substantial number of observations, assuming bivariate normally distributed errors.  Furthermore, if the bias of naive application of nonparametric regression methods is to be meaningful, it must be of non-trivial magnitude relative to the true effect on the latent outcome. 
    \item To identify the true conditional mean of the latent outcome (i.e. without the inverse Mills ratio term) and identify heterogeneous effects, we require the excluded variables to have a large impact on the inverse Mill's ratio for a range of values of both treatment and other covariates.
    \item We require sufficiently many treated, untreated, selected, and unselected observations.
\end{itemize}


We generate $p$ independent standard normally distributed covariates $X_{1},\dots, X_p \sim \mathcal{N}(0,1)$. A binary treatment variable is generated with selection on observables. The number of covariates is $p=50$, and $X_1, \dots, X_5$ are excluded from the outcome equation data, i.e. they are in $\bm{w}_i$, but not $\bm{x}_i$. There are 4000 training observations and 4000 test observations.  The propensity score function is:
$$\pi(\bm{x}_i) = 0.7 \Phi(0.5 x_{6,i}x_{7,i} + 0.3 x_{8,i}^2) +0.1 $$
The treatment variable is denoted by $T_i \sim \text{Bernoulli}(\pi(\bm{x}_i))$. The selection and outcome equations are:
$$ Y_i^* =  0.5x_{6,i}  + 0.5 x_{6,i}x_{7,i} +  0.25 (x_{3,i}-1)^2 T_i   + \eta_i $$ 
$$ Z_{i}^* =  0.5 x_{1,i}x_{6,i} +  x_{1,i}x_{6,i}T_i + \xi_i  $$
This implies that the heterogeneous effects on the inverse Mills ratio are equal to:
$$ \frac{\phi(f_z(\bm{w}_i,1) )}{\Phi(f_z(\bm{w}_i,1) )} - \frac{\phi(f_z(\bm{w}_i,0) )}{\Phi(f_z(\bm{w}_i,0) )}   = \frac{\phi (0.5 x_{1,i}x_{6,i} +  x_{1,i}x_{6,i} ) }{ \Phi(0.5 x_{1,i}x_{6,i} +  x_{1,i}x_{6,i} ) } - \frac{\phi (0.5 x_{1,i}x_{6,i}  ) }{ \Phi(0.5 x_{1,i}x_{6,i}  ) }  $$
The heterogeneous effects on the latent outcome are equal to:
$$ f_y (\bm{x}_i, 1) - f_y (\bm{x}_i, 0) =  0.25 (x_{3,i}-1)^2  $$
%
%
%
%
%
Let the potential outcomes of the selection indicator variable under treatment and control allocation be denoted by $S_i(1)$ and $S_i(0)$ respectively. We use the following measures to evaluate all methods:

\medskip

\noindent $\bullet$ MSE of estimated effect on the probability of censoring
$$ \frac{1}{n_{test}} \sum_{i=1}^{n_{test}} \Big( \big(\hat{\Pr}(S_i(1)=0|X) - \hat{\Pr}(S_i(0)=0|X) \big) - \big({\Pr}(S_i(1)=0|X) - {\Pr}(S_i(0)=0 |X) \big)\Big)^2  $$
$\bullet$ RMSE of estimated treatment effect on the latent outcome:
$$ \sqrt{\frac{1}{n_{test}} \sum_{i=1}^{n_{test}} \Big( \big(\hat{\mathbb{E}}(Y_i(1)=0|X) - \hat{\mathbb{E}}(Y_i(0)=0|X) \big) - \big({\mathbb{E}}(Y_i(1)=0|X) - {\mathbb{E}}(Y_i(0)=0 |X) \big)\Big)^2  }$$

\subsection{Simulation Results}\label{sim_results}


\subsubsection{\cite{brewer2024addressing} Simulations   }

Table \ref{Brewer_res_table} contains the results for the simulation study from \cite{brewer2024addressing}.\footnote{TOBART-2-NP results will be added to an updated version of this paper at a later date.} As anticipated, the sparse linear Bayesian model described by \cite{iqbal2023bayesian} produces better results for DGP 1, because a linear model is appropriate for a DGP with linear selection and outcome equations. BART and TOBART do not improve estimates of sample selection probabilities relative to the sparse linear model, because for both DGPs the true model for the latent selection variable is linear.

Latent outcome predictions for the nonlinear DGP2 are notably more accurate for Soft BART and TOBART models than for linear Tobit models or BART. When there is no sample selection, there is no notable difference between the SoftBART and Soft TOBART models. When there is sample selection in unobservables, TOBART models produce much more accurate nonlinear DGP predictions than the other models. Moreover, confidence interval coverage is close to 95\% and posterior means of the correlation between selection and outcome equation errors are very close to the true values. 

The best performing TOBART models are marginalized TOBART-2 and Soft TOBART-2. A combination of the two models, i.e. marginalized soft TOBART-2, produces similar estimated to unmarginalized Soft TOBART-2, although it produces more accurate estimates of $\rho$ when $\rho=0.45$.\footnote{Marginalized SoftBART surprisingly produces much less accurate esitmates of selection probabilities. Perhaps this is explainable by  differences in the implementations of unmarginalized and marginalized tree samplers.}  We do not observe large differences in results for different priors on the distribution of the error terms, and different calibrations of these priors. Perhaps the methods would be more sensitive to the prior if the sample size were much smaller than 2000. 

\begin{table}[ht]
\centering
\begin{tabular}{p{0.1cm}p{0.6cm}|p{1.08cm}p{1.08cm}p{1.08cm}p{1.0cm}p{1.0cm}p{1.08cm}p{1.08cm}p{1.08cm}p{1.08cm}p{1.08cm}p{1.08cm}p{1.08cm}}
  \hline
D G P & corr & Tobit VH  & Tobit Ding & Tobit Omori & BART & Soft BART & Sparse Tobit & TO BART 2 VH  & TO BART 2 Ding & TO BART 2 marg & Soft TO BART 2 VH & Soft TO BART 2 marg  \\ 
  \hline
  \hline
  \hline
  \multicolumn{12}{c}{$f_y(\bm{x})$ prediction RMSE, relative to BART}\\
  \hline
1 & 0.00 & 0.677 & 0.666 & 0.676 & 1.000 & 0.554 & 0.365 & 0.854 & 0.865 & 0.850 & 0.576 &   0.705  \\ 
  2 & 0.00 & 2.350 & 2.349 & 2.350 & 1.000 & 0.746 & 2.556 & 0.997 & 0.998 & 0.986 & 0.760 &  0.772  \\ 
  \hline
  1 & 0.45 & 0.396 & 0.440 & 0.395 & 1.000 & 0.792 & 0.212 & 0.496 & 0.557 & 0.487 & 0.361 &  0.396  \\ 
  2 & 0.45 & 1.543 & 1.551 & 1.543 & 1.000 & 0.832 & 1.679 & 0.645 & 0.680 & 0.643 & 0.505 &  0.505  \\ 
  \hline
  1 & 0.90 & 0.240 & 0.272 & 0.239 & 1.000 & 0.897 & 0.123 & 0.257 & 0.306 & 0.262 & 0.190 & 0.207   \\ 
  2 & 0.90 & 0.989 & 1.001 & 0.989 & 1.000 & 0.913 & 1.075 & 0.389 & 0.414 & 0.401 & 0.310 &  0.306  \\ 
   \hline
    \hline
  \hline
    \multicolumn{12}{c}{$f_y(\bm{x})$ 95\% prediction interval mean coverage}\\
   \hline  
1 & 0.00 & 0.871 & 0.868 & 0.870 & 0.985 & 0.982 & 0.956 & 0.953 & 0.949 & 0.972 & 0.978 &  0.994  \\ 
  2 & 0.00 & 0.211 & 0.208 & 0.213 & 0.973 & 0.934 & 0.157 & 0.933 & 0.930 & 0.943 & 0.934 &  0.990  \\ 
  \hline
  1 & 0.45 & 0.853 & 0.781 & 0.854 & 0.885 & 0.684 & 0.946 & 0.954 & 0.914 & 0.972 & 0.966 &  0.989  \\ 
  2 & 0.45 & 0.212 & 0.226 & 0.211 & 0.884 & 0.737 & 0.153 & 0.928 & 0.908 & 0.950 & 0.931 &  0.993  \\ 
  \hline
  1 & 0.90 & 0.809 & 0.728 & 0.811 & 0.647 & 0.511 & 0.895 & 0.947 & 0.909 & 0.985 & 0.952 &  0.993  \\ 
  2 & 0.90 & 0.197 & 0.225 & 0.196 & 0.683 & 0.550 & 0.125 & 0.897 & 0.896 & 0.932 & 0.910 &  0.991  \\ 
    \hline
     \hline
  \hline
\multicolumn{12}{c}{Selection Probability MSE, relative to BART}\\
  \hline
  1 & 0.00 & 2.154 & 2.151 & 2.153 & 1.000 & 0.727 & 0.249 & 1.409 & 1.454 & 1.808 & 0.735 &  1.173  \\ 
  2 & 0.00 & 2.143 & 2.144 & 2.155 & 1.000 & 0.727 & 0.252 & 1.387 & 1.423 & 1.775 & 0.716 &  1.162  \\ 
  \hline
  1 & 0.45 & 2.115 & 2.139 & 2.122 & 1.000 & 0.724 & 0.196 & 1.556 & 1.565 & 1.533 & 0.686 &  0.892  \\ 
  2 & 0.45 & 2.128 & 2.143 & 2.134 & 1.000 & 0.724 & 0.213 & 1.539 & 1.589 & 1.667 & 0.687 &  0.908  \\ 
  \hline
  1 & 0.90 & 2.158 & 2.258 & 2.155 & 1.000 & 0.625 & 0.232 & 1.370 & 1.422 & 1.684 & 0.566 &  1.643  \\ 
  2 & 0.90 & 2.200 & 2.274 & 2.200 & 1.000 & 0.625 & 0.248 & 1.380 & 1.464 & 1.680 & 0.603 &  0.963  \\ 
   \hline
     \hline
  \hline
         \multicolumn{12}{c}{Mean correlation estimate for selection and outcome equation errors}\\
  \hline
1 & 0.00 & -0.001 & 0.000 & -0.001 & 0.000 & 0.000 & -0.018 & -0.040 & -0.024 & -0.004 & -0.016 &  0.002  \\ 
  2 & 0.00 & -0.004 & -0.001 & -0.003 & 0.000 & 0.000 & -0.035 & -0.032 & -0.019 & -0.009 & -0.023 &  -0.005  \\ 
  \hline
  1 & 0.45 & 0.418 & 0.242 & 0.418 & 0.000 & 0.000 & 0.416 & 0.410 & 0.240 & 0.454 & 0.427 &  0.446  \\ 
  2 & 0.45 & 0.373 & 0.219 & 0.376 & 0.000 & 0.000 & 0.350 & 0.413 & 0.244 & 0.461 & 0.427 &   0.448  \\ 
  \hline
  1 & 0.90 & 0.894 & 0.703 & 0.894 & 0.000 & 0.000 & 0.891 & 0.894 & 0.692 & 0.933 & 0.911 &  0.920  \\ 
  2 & 0.90 & 0.751 & 0.542 & 0.751 & 0.000 & 0.000 & 0.709 & 0.884 & 0.682 & 0.923 & 0.900 &  0.913  \\ 
  \hline
 
\end{tabular}
\caption{Results for \cite{brewer2024addressing} simulation study. RMSE and MSE results are relative to BART.} \label{Brewer_res_table}
\end{table}

\subsubsection{CATE simulation results}

Table \ref{cate_res_table} contains the results for the treatment effect simulation study. Nonlinear methods outperform linear methods for this simple quadratic DGP with interactions. Tree-based methods with soft splitting are more accurate than their counterparts with hard splitting rules. Marginalized TOBART outperforms vanilla implementations of TOBART, suggesting that without marginalization TOBART sampler did not converge or mix well. 

Surprisingly, SoftBART with the inverse Mills ratio as a covariate, following the general approach of \cite{brewer2024addressing}, is almost as accurate as Soft TOBART-2 across all measures except PEHE of effects on selection and outcome CATE posterior interval length. Soft marginalized TOBART-2 produces the best estimates of outcome treatment effects, and surprisingly the worst predictions of selection treatment effects. It also produces the smallest posterior intervals, albeit with over-coverage. 

\begin{table}[ht]
	\centering
	\begin{tabular}{p{0.6cm}p{0.4cm}p{0.7cm}|p{0.9cm}p{1cm}p{1.3cm}p{1.3cm}p{1cm}p{1cm}p{1cm}p{1cm}p{1cm}p{1cm}}
		\hline
		n & p & corr & Tobit VH  & Sparse Bayes Tobit & BART control & Soft BART control & TO BART 2 marg & TO BART 2 VH  & TO BART 2 Ding & Soft TO BART 2 VH  & Soft TO BART 2 Ding & Soft TO BART 2 marg \\ 
		\hline
	& & &	\multicolumn{10}{l}{ { \blue 
 \textit{PEHE Selection Test}} } \\
        4000 & 50 & 0.50 & 1.000 & 0.997 & 0.703 & 0.646 & 0.544 & 0.655 & 0.626 & 0.593 & 0.409 &   1.738  \\ 
     & &	 &  \multicolumn{10}{|l}{ { \blue 
 \textit{PEHE Selection Training}} } \\
4000 & 50 & 0.50 & 1.000 & 0.997 & 0.710 & 0.646 & 0.547 & 0.656 & 0.630 & 0.594 & 0.408 &   1.739  \\
& &		&		\multicolumn{10}{|l}{{ \blue \textit{PEHE Outcome Test}}}\\
4000 & 50 & 0.50 & 1.000 & 1.013 & 0.403 & 0.295 & 0.345 & 0.451 & 0.431 & 0.295 & 0.297 &  0.270  \\ 			
& &		&		\multicolumn{10}{|l} { { \blue \textit{PEHE Outcome Training}} }\\
4000 & 50 & 0.50 & 1.000 & 1.013 & 0.402 & 0.291 & 0.341 & 0.448 & 0.428 & 0.291 & 0.293  &  0.265 \\ 			
& &		&		\multicolumn{10}{|l} { {\blue \textit{Length Outcome CATE PI Test}}}\\
4000 & 50 & 0.50 & 0.217 & 0.247 & 0.733 & 0.678 & 0.624 & 0.461 & 0.507 & 0.602 & 0.611 &  0.582  \\ 			
& &	&			\multicolumn{10}{|l}{ { \blue \textit{Coverage Outcome CATE PI Test}} }\\
4000 & 50 & 0.50 & 0.144 & 0.226 & 0.938 & 0.978 & 0.932 & 0.790 & 0.836 & 0.974 & 0.973 &  0.979  \\ 			
& &	&			\multicolumn{10}{|l}{  { \blue \textit{Length Outcome CATE PI Training}}}\\
4000 & 50 & 0.50 & 0.217 & 0.247 & 0.733 & 0.682 & 0.623 & 0.462 & 0.509 & 0.604 & 0.613 &  0.582  \\ 			
& &	&		\multicolumn{10}{|l}{   { \blue \textit{Coverage Outcome CATE PI Training}}}\\
4000 & 50 & 0.50 & 0.145 & 0.228 & 0.938 & 0.980 & 0.932 & 0.788 & 0.839 & 0.977 & 0.976 &  0.981 \\ 
\hline
	\end{tabular}
    \caption{Results for Treatment Effect simulation study. PEHE results are relative to linear Bayesian Tobit. PI = 95 \% Posterior Interval} \label{cate_res_table}
\end{table}

\FloatBarrier

\section{Data Application}\label{application_sec}











\subsubsection{Description of RAND Health Insurance Experiment}




We estimate the heterogeneous effects of allocation to a  deductible plan for health insurance. The dataset has been studied by \cite{deb2002structure}, \cite{cameron2005microeconometrics}, and others. The outcome variable is the log of medical expenses in dollars. Selected observations are those with non-zero medical expenditures. There are 15 other covariates.\footnote{The other covariates are: the log of one plus the coinsurance rate, the log of the participation incentive payment, physical limitations (binary), number of chronic diseases, good/fair/poor self-rated health (binary variables), log of family income, log of family size, education of household head in years, age in years, female (binary), child (1 if has child below 18 years), interaction of female and child variables, black (binary).} None of the covariates are excluded from the outcome equation. Further applications to female earning data \citep{mroz1987sensitivity} and a job training experiment for disadvantaged youths \citep{schochet2001national} are included in Appendix \ref{further_application_app}.

\subsubsection{Data Application Results}

The data application results are presented below.\footnote{TOBART-2-NP results will be added to an updated version of this paper at a later date.} Table \ref{mroz_dependency_table} shows that the posterior mean and 95\% posterior interval for the correlation parameter $\rho$ are comparable to the estimates produced by \cite{van2011bayesian}. Table \ref{mroz_ate_table} contains estimates of average treatment effects of individual deductible plans on probabilities of zero expenditures and the log of medical expenditures. The average treatment effect estimates are larger in magnitude for the Bayesian methods than for standard Tobit. The average treatment effect on the latent outcome estimate produced by Soft-TOBART-2 is remarkably close to the effect estimated by \cite{van2011bayesian} with a Bayesian linear Tobit model.

Appendix \ref{data_app} contains 5-fold cross validated Brier scores of selection probabilities, MSE of observed outcome predictions, and coverage of observed outcome posterior intervals. All models produce similar results. This is unsurprising because sample selection models are not expected to produce better predictions of the observed outcomes, but rather better predictions of the mean of latent outcomes, which cannot be evaluated using observed sample selected data.


\FloatBarrier

\begin{table}[ht]
\centering
\begin{tabular}{llll}
  \hline
  \hline
 & Mean & 2.5\% Quantile & 97.5\% Quantile \\ 
  \hline
 TOBART-2  & 0.679 & 0.519 & 0.763 \\ 
 \cite{van2011bayesian} & 0.712 & 0.634 & 0.784 \\ 
   \hline
\end{tabular}
\caption{Mean and posterior distributions of dependence measures for selection and outcome equation error terms of TOBART model applied to RAND Health Insurance Experiment data, and original results from \cite{van2011bayesian}.}
\label{mroz_dependency_table}
\end{table}

\begin{table}[H]
\centering
\begin{tabular}{lp{1.25cm}p{1.15cm}p{1.15cm}p{1.15cm}p{1.25cm}p{1.5cm}p{1.25cm}p{1.5cm}p{1.5cm}}
  \hline
  \hline
   & Tobit 2-step & Tobit ML & BART & Soft BART & Bayesian Splines & TO BART 2 & Soft \newline TO BART 2 & Bayes Tobit \cite{van2011bayesian} \\ 
  \hline
  Prob. Censored & -0.039 & -0.031 & -0.063 & -0.085 & -0.031 & -0.079 & -0.04 & \\ 
  Latent Outcome & -0.042 & -0.108 &  &  & -0.13 & -0.133 & -0.147 & -0.147 \\ 
  Observed Outcome & -0.025 & -0.053 & -0.13 & -0.011 &  & -0.157 & -0.084 &  \\ 
  \hline
\end{tabular}
\caption{Estimates of average treatment effects of individual deductible plans on the probability of censoring (i.e. zero expenditure), the the latent outcome, and observed log of medical expenses (conditional on expenses being positive) for RAND Health Insurance Experiment data.}
	\label{mroz_ate_table}
\end{table}

\FloatBarrier

\FloatBarrier

\section{Conclusion}\label{conclusion_sec}

The results of this paper suggest that TOBART-2 can produce more accurate outcome predictions and treatment effect estimates than linear sample selection models when the DGP is non-linear. 
However, for some DGPs, a naive application of BART or softBART to the selected data, with an estimate of the inverse Mills ratio as a covariate, also provides estimates with comparable accuracy.

This paper also describes alternative MCMC implementations and prior settings. Improvements in predictive accuracy can be obtained by the use of soft splitting rules and marginalization in MCMC draws of trees. Prior sensitivity, particularly for small sample sizes, would be an interesting area for future research. Possible extensions of the model include heteroskedastic type 2 Tobit models \citep{donald1995two, de2022generalized}, with modelling of variances by Bayesian trees \citep{pratola2020heteroscedastic}.


\bibliographystyle{agsm}
\bibliography{tobart_references}

\appendix

\section{MCMC algorithm - TOBART 2 with jointly normally distributed errors}\label{app_mcmc_tbart2norm}

\subsection{Type 2 Tobit Gibbs Sampler}

The full conditionals of $\phi, \gamma, \bm{\psi} = (\bm{\theta}', \bm{\beta}')'$ are

$$ \bm{\psi} |  \gamma, \phi, \bm{z^*}, \bm{y_c^*}, \bm{y_o}  \sim \mathcal{N}(\bm{\psi}_1 , \bm{\Psi}_1)$$
$$ \gamma | \bm{\psi} , \phi, \bm{z^*}, \bm{y_c^*}, \bm{y_o}  \sim \mathcal{N}(\gamma_1 , G_1)$$
$$ \phi | \bm{\psi} , \gamma, \bm{z^*}, \bm{y_c^*}, \bm{y_o}  \sim \mathcal{IG}(\frac{n_1}{2}, \frac{S_1}{2})$$
where $\bm{z^*} = (z_1^*, z_2^*, ..., z_n^*)' $, $n_1 = n_0 + n $, $G_1^{-1} = G_0^{-1} + \phi^{-1} \sum_{i=1}^n (z_i^* - \bm{w}_i \bm{\theta})^2 $.
$$\gamma_1 = G_1 \{ G_0^{-1} \gamma_0 + \phi^{-1} \sum_{i=1}^n (z_i^* - \bm{w}_i \bm{\theta} )(y_i^* - \bm{x}_i' \bm{\beta}) \} $$
$$ S_1 = S_0 + \gamma^2 \sum_{i=1}^n (z_i^* - \bm{w}_i' \bm{\theta} )^2 - 2 \gamma \sum_{i=1}^n (z_i^* - \bm{w}_i' \bm{\theta})(y_i^* - \bm{x}_i' \bm{\beta} ) + \sum_{i=1}^n (y_i^* - \bm{x}_i' \bm{\beta} )^2 $$
$$\bm{\Psi}_1 = \left( \bm{\Psi}_0^{-1} + \sum_{i=1}^n \tilde{X}_i' \Sigma^{-1}  \tilde{X}_i \right)^{-1} \ , \ \bm{\psi} = \bm{\Psi}_1 \left( \bm{\Psi}_0^{-1} \bm{\psi}_0 + \sum_{i=1}^n \tilde{X}_i' \Sigma^{-1}  \tilde{\bm{y}}_i^* \right)$$
$$ \tilde{\bm{y}}_i^*  = 
\begin{pmatrix}
	z_i^* \\
	y_i^*
\end{pmatrix} \ , \
\tilde{X}_i = 
\begin{pmatrix}
	\bm{w}_i' & \bm{0}' \\
 	\bm{0}'   & \bm{x}_i'\\
\end{pmatrix} \ , \ 
\bm{\psi}_0 =
\begin{pmatrix}
	\bm{\theta}_0 \\
	\bm{\beta}_0
\end{pmatrix} \ , \
\bm{\Psi}_0 = 
\begin{pmatrix}
	\bm{\Theta}_0 &  \bm{0} \\
	\bm{0} &  \bm{B}_0 \\
\end{pmatrix}
$$
If applying the prior $\gamma \sim \mathcal{N} (\gamma_0, \tau \phi)$ of \cite{van2011bayesian}, then set $G_0 =  \tau \phi$ and add $\frac{\gamma^2}{ \tau}$ to $S_1$ 
$$ S_1 = S_0 + \frac{\gamma^2}{\tau} +  \gamma^2 \sum_{i=1}^n (z_i^* - \bm{w}_i' \bm{\theta} )^2 - 2 \gamma \sum_{i=1}^n (z_i^* - \bm{w}_i' \bm{\theta})(y_i^* - \bm{x}_i' \bm{\beta} ) + \sum_{i=1}^n (y_i^* - \bm{x}_i' \bm{\beta} )^2 $$

\subsection{Type 2 TOBART Gibbs Sampler}

Most of the linear Type 2 Tobit sampler is unchanged for Type 2 Tobit BART. Conditional on the sums-of-trees, $f_z$ and $f_y$, the other full conditional distributions are the same as for linear Tobit except $  \bm{w}_i' \bm{\theta}  $ is replaced by $f_z (\bm{w}_i)$ and $\bm{x}_i' \bm{\beta} $ is replaced by $f_y (\bm{w}_i)$. 

Samples of $y_i^*$ and $z_i^*$ in the Gibbs sampler for TOBART are the same as in the standard Tobit-2 sampler. 
The sampling of trees has some similarities with Seemingly Unrelated BART \citep{chakraborty2016bayesian} because there are correlated errors for the two latent outcomes $z^*$ and $y^*$.

Trees are sampled sequentially from the full conditionals for selection equation trees and outcome equation trees:
$$( T_s^{z^*}, M_s^{z^*}) | \{(T_l^{z^*}, M_l^{z^*}) \}_{l \neq s}, \bm{z^*}, \bm{y_c^*}, \bm{y_o}, f_y (\bm{w}_i)  \ \forall s  \in \{1,...,m_z\} $$
$$( T_q^{y^*}, M_q^{y^*}) | \{(T_l^{y^*}, M_l^{y^*}) \}_{l \neq q}, \bm{z^*}, \bm{y_c^*}, \bm{y_o} , f_z (\bm{w}_i)  \ \forall q \in \{1,...,m_y\} $$
where $q= 1,...,Q$ indexes trees in $f_y$ and $s=1,...,S$ indexes trees in $f_z$.

The terminal node parameter draws for $M_q^{y^*}$ depend on $R_q^{y^*}, z^*$. The conditional is equivalent to the posterior of a regression of $R_q^{y^*}$ on $T_q^{y^*}$ terminal node indicator variables jointly with a regression of $z^*$ on $T_1^{z^*},...,T_S^{z^*}$ conditional on $f_z (\bm{w}_i)$  and the  covariance $\Sigma$.

\bigskip

\noindent \textbf{Implementation without sampling of unobserved outcomes, based on \cite{van2011bayesian}}

\medskip

The implementation below is based on \cite{van2011bayesian}. This approach does not sample missing outcomes and only augments the data with latent outcomes for the selection equation. \cite{chib2009estimation}  noted three advantages of this approach: 1. Lower computational and storage requirements, 2. improved mixing behaviour of the Markov chain, and 3. the algorithm is applicable when some or all of the covariates in the outcome model are unobserved for censored outcomes. This was also implemented by \cite{omori2007efficient} for linear Type 2 Tobit, who noted that this sampler is more efficient.

\begin{enumerate}
	\item Set initial values for all the parameters $\phi$, $\gamma$, $f_z(\bm{w}_i)  $ and $f_y(\bm{x}_i)$ .
	\item Draw the latent variable from $\bm{z}^* | \phi, \gamma, f_z(\bm{w}_i), f_y(\bm{x}_i), \bm{y}_o  $.
	\begin{enumerate}
		\item For censored observations, generate  $ z_i^* |  \phi, \gamma, f_z(\bm{w}_i), f_y(\bm{x}_i) \sim \mathcal{TN}_{(-\infty,0)} (f_z(\bm{w}_i), 1)$ where $\mathcal{TN}_{(a,b)}(\mu, \sigma^2)$ denotes a normal distribution with mean $\mu$ and variance $\sigma^2$ truncated on the interval $(a,b)$.
		\item For uncensored observations, generate $z_i^* | \bm{y}_c, \phi, \gamma, f_z(\bm{w}_i), f_y(\bm{x}_i) \sim \mathcal{TN}_{[0,\infty)} (\mu_z , \sigma_z^2)$ where $\mu_z = f_z(\bm{w}_i) + \gamma (y_i - f_y(\bm{x}_i)) / (\phi + \gamma^2)$, and $\sigma_z^2 = 1 - \gamma^2/(\phi+\gamma^2) = \frac{\phi}{\gamma^2 + \phi}$.
	\end{enumerate}
	\item Draw the sums of trees $f_z(\bm{w}_i)$. First note that the full conditional distribution $\bm{z}^* | \bm{y}_c , \phi , \gamma, f_z(\bm{w}_i), f_y(\bm{x}_i) $
 \begin{equation*}
z_i^* | \bm{y}_c , \phi , \gamma, f_z(\bm{w}_i), f_y(\bm{x}_i) = \begin{cases}
		\mathcal{N}(f_z(\bm{w}_i), 1) & \text{if $i$ censored,} \\
		\mathcal{N}(f_z(\bm{w}_i) +  \frac{\gamma}{\gamma^2 + \phi}(y_i - f_y(\bm{x}_i)),  \frac{\phi}{\gamma^2 + \phi}) & \text{if $i$ uncensored}.
	\end{cases}
\end{equation*}

	\item For $\bm{z}^*$, for each tree from $k=1$ to $k=m_z$
	\begin{enumerate}
		\item Create partial residuals
		$$ \bm{R}_{zki} = \begin{cases}
			z_i^*  - \sum_{s \neq k} g_s^z (\bm{w}_i) & \text{if $i$ censored,} \\
			z_i^* -  ( y_i^* -  \sum_{k=1}^{m_y} g_k^y (\bm{x}_i)  )\frac{\gamma}{\gamma^2 + \phi} - \sum_{s \neq k} g_s^z (\bm{w}_i) & \text{if $i$ uncensored} \\
		\end{cases}$$
		\item Draw a new tree $g_k^{z} (\bm{w})$ using the sampler described by \cite{chipman2010bart} for a regression of $\bm{R}_{zk}$ on a single tree with variance $1$ for censored observations, and $ \frac{\phi}{\gamma^2 + \phi} $ for uncensored observations.\footnote{This is implemented in the \texttt{dbarts} $\bm{R}$ package by specifying weights $1$ for censored observations and $ \frac{\gamma^2 + \phi}{\phi} $ for uncensored observations.}
	\end{enumerate}

	\item Note that for uncensored observations, conditional on ${z}_i^* , \phi , \gamma, f_z(\bm{w}_i), f_y(\bm{x}_i) $,
	$$y_i = f_y(\bm{x}_i) + \gamma (z_i^* - f_z(\bm{w}_i) ) + \eta_i  $$
	where $\eta_i |\bm{x}_i, \bm{w}_i \sim \mathcal{N}(0, \phi)$. Full conditional draws of $f_y(\bm{x}_i)$ and $\gamma$ can therefore be made as in semiparametric BART \citep{prado2021semi, zeldow2019semiparametric}.
		
	\medskip
	
	For $\bm{y}^*$, for each tree from $k=1$ to $k=m_y$.
	\begin{enumerate}
		\item Create partial residuals 
		$  \bm{R}_{yk} = 
        \bm{y}^* -  ( \bm{z}^* -  \sum_{k=1}^{m_z} g_k^z (\bm{w})  ) \gamma - \sum_{s \neq k} g_s^y (\bm{x})$			
		\item Draw a new tree $g_k^{y} (\bm{x})$ using the sampler described by \cite{chipman2010bart} for a regression of $\bm{R}_{yk}$ on a single tree with variance $ \phi $  .
	\end{enumerate}
	\item Sample $ \gamma$ from $ \mathcal{N}(\gamma_1, G_1) $, where $G_1^{-1} = G_0^{-1} + \phi^{-1} \sum_{i \text{ uncensored}} (z_i^* - f_z(\bm{w}_i))^2 $ and $\gamma_1 = G_1 \{ G_0^{-1} \gamma_0 + \phi^{-1} \sum_{i \text{ uncensored}} (z_i^* - f_z(\bm{w}_i) )(y_i^* - f_y(\bm{x}_i)) \} $, where $G_0 = \tau \phi$.
	
	\item Sample $ \phi $ from $ \mathcal{IG}(c, d)$ where $c = \frac{n_0+n_1 + 1}{2} $ and $ d = \frac{S_0}{2} + \frac{\gamma^2}{2 \tau} 
 +  \frac{1}{2}\sum_{i \text{ uncensored}}  (y_i^* - f_y(\bm{x}_i) - \gamma (z_i^* - f_z(\bm{w}_i) ) )^2$ .

    [Alternatively $\gamma$ and $\phi$ can be  sampled from a Normal-Inverse Gamma joint conditional distribution as described in Appendix \ref{phigamma_norm_ig_sec}. ]
\end{enumerate}


If instead the modified \cite{ding2014bayesian} prior is applied, steps 6 and 7 are replaced by the following:
Sample, $\sigma_1^2 | \Omega, \dots  \sim  \{ \frac{1- \rho^2}{c} \chi_{\nu_0}^2 \}^{-1} $, then calculate $ \bm{E}_i =
\begin{pmatrix}
    \sigma_1  \tilde{{z}}_i \\
    \tilde{{y}}_i \\
\end{pmatrix}$, then sample from 
$$\tilde{\Sigma } | \tilde{\bm{z}}, \tilde{\bm{y}}, \dots  \sim  \bm{W}_2^{-1} \left( n_{uncens} + \nu_0 , c \bm{I}_2  +   \sum_{i = 1}^{n_{uncens}}    \bm{E}_i  \bm{E}_i^T  \right)  $$
$$ \Omega =  \begin{pmatrix}
    1 & \omega_{12} \\
    \omega_{12} & \omega_{22} \\
\end{pmatrix} =
\begin{pmatrix}
    \frac{1}{\tilde{\sigma}_1 } & 0 \\
    0 & 1 \\
\end{pmatrix}   \tilde{\Sigma}
 \begin{pmatrix}
    \frac{1}{\tilde{\sigma}_1 } & 0 \\
    0 & 1 \\
\end{pmatrix}  = 
 \begin{pmatrix}
    1 & \frac{\tilde{\sigma}_{12}}{\tilde{\sigma}_1 }  \\
    \frac{\tilde{\sigma}_{12}}{\tilde{\sigma}_1 } & \tilde{\sigma}_2^2 \\
\end{pmatrix} $$
The values of $\phi$ and $\gamma$ required for the other steps are defined as $ \gamma = \omega_{12}$  and $\phi =  \omega_{22} - \omega_{12}^2 $.

\subsection{Sampler for binary outcomes}

If the outcome is binary, then it is necessary to also sample $y^*$ for uncensored outcomes. Note that for uncensored observations, conditional on ${z}_i^* , \phi , \gamma, f_z(\bm{w}_i), f_y(\bm{x}_i) $,
	$$y_i^* = f_y(\bm{x}_i) + \gamma (z_i^* - f_z(\bm{w}_i) ) + \eta_i  $$
	where $\eta_i |\bm{x}_i, \bm{w}_i \sim \mathcal{N}(0, \phi)$.

Therefore, conditioning on binary outcomes $\bm{y}$, the sampling distributions of $y^*$  are
$$ y_i^* | y =1 , {z}_i^* , \phi , \gamma, f_z(\bm{w}_i), f_y(\bm{x}_i) \sim  \mathcal{TN}_{[0,\infty]}( f_y(\bm{x}_i) + \gamma (z_i^* - f_z(\bm{w}_i) ) , \phi) $$ 
$$ y_i^* | y =0 , {z}_i^* , \phi , \gamma, f_z(\bm{w}_i), f_y(\bm{x}_i) \sim  \mathcal{TN}_{[-\infty,0]}( f_y(\bm{x}_i) + \gamma (z_i^* - f_z(\bm{w}_i) ) , \phi) $$

\section{MCMC Algorithm - Type 2 Tobit with Dirichlet Process Mixture for errors}\label{app_mcmc_tbart2np}

\subsection{Nonparametric Type 2 Tobit Gibbs Sampler}

%
%

Let $\bm{\psi} = (\bm{\theta}', \bm{\beta}')'$ and $\bm{z^*} = (z_1^*, z_2^*, ..., z_n^*)' $, $n_1 = n_0 + n $, $G_1^{-1} = G_0^{-1} + \phi^{-1} \sum_{i=1}^n (z_i^* - \bm{w}_i \bm{\theta})^2 $.
$$\gamma_1 = G_1 \{ G_0^{-1} \gamma_0 + \phi^{-1} \sum_{i=1}^n (z_i^* - \bm{w}_i \bm{\theta} )(y_i^* - \bm{x}_i' \bm{\beta}) \} $$
$$ S_1 = S_0 + \gamma^2 \sum_{i=1}^n (z_i^* - \bm{w}_i' \bm{\theta} )^2 - 2 \gamma \sum_{i=1}^n (z_i^* - \bm{w}_i' \bm{\theta})(y_i^* - \bm{x}_i' \bm{\beta} ) + \sum_{i=1}^n (y_i^* - \bm{x}_i' \bm{\beta} )^2 $$
$$\bm{\Psi}_1 = \left( \bm{\Psi}_0^{-1} + \sum_{i=1}^n \tilde{X}_i' \Sigma_i^{-1}  \tilde{X}_i \right)^{-1} \ , \ \bm{\psi} = \bm{\Psi}_1 \left( \bm{\Psi}_0^{-1} \bm{\psi}_0 + \sum_{i=1}^n \tilde{X}_i' \Sigma_i^{-1}  \tilde{\bm{y}}_i^* \right)$$
$$ \tilde{\bm{y}}_i^*  = 
\begin{pmatrix}
	z_i^* \\
	y_i^*
\end{pmatrix} \ , \
\tilde{X}_i = 
\begin{pmatrix}
	\bm{w}_i' & \bm{0}' \\
	\bm{0}'   & \bm{x}_i'\\
\end{pmatrix} \ , \ 
\bm{\psi}_0 =
\begin{pmatrix}
	\bm{\theta}_0 \\
	\bm{\beta}_0
\end{pmatrix} \ , \
\bm{\Psi}_0 = 
\begin{pmatrix}
	\bm{\Theta}_0 &  \bm{0} \\
	\bm{0} &  \bm{B}_0 \\
\end{pmatrix}
$$
%
%
%
The full residuals are denoted by $ u_{i1} = y_i - \bm{x}_i'\bm{\beta}  $ and $ u_{i2} = z_i^* - \bm{w}_i' \bm{\theta}  $ . The sampler introduced by \cite{van2011bayesian} proceeds as follows:
\begin{enumerate}
	\item Set initial values for all the parameters $\phi_i$, $\gamma_i$, $\bm{\psi} = (\bm{\theta}', \bm{\beta}')'$  .
	\item Draw the latent variable from $\bm{z}^* | \phi_i, \gamma_i, \bm{\theta}', \bm{\beta}', \bm{y}_o  $.
	\begin{enumerate}
		\item For censored observations, generate  $ z_i^* |  \phi_i, \gamma_i, \bm{\theta}, \bm{\beta} \sim \mathcal{TN}_{(-\infty,0)} (\bm{w}_i'\bm{\theta} + \mu_{i1}, 1)$ where $\mathcal{TN}_{(a,b)}(\mu, \sigma^2)$ denotes a normal distribution with mean $\mu$ and variance $\sigma^2$ truncated on the interval $(a,b)$.
		\item For uncensored observations, generate $z_i^* | \bm{y}_c, \phi_i, \gamma, \bm{\theta}, \bm{\beta} \sim \mathcal{TN}_{[0,\infty)} (\mu_{z,i} , \sigma_{z,i}^2)$ where $\mu_{z,i} = \bm{w}_i'\bm{\theta} +  \mu_{i1} + \gamma_i (y_i - \bm{x}_i'\bm{\beta} - \mu_{i2}  ) / (\phi_i + \gamma_i^2)$, and $\sigma_{z,i}^2 = 1 - \gamma_i^2/(\phi_i+\gamma_i^2) = \frac{\phi_i}{\gamma_i^2 + \phi_i}$.
	\end{enumerate}
	%
	\begin{equation*}
		z_i^* | \bm{y}_c , \phi_i , \gamma_i, \bm{\theta}', \bm{\beta}' = \begin{cases}
			\mathcal{N}(\bm{w}_i'\bm{\theta} +  \mu_{i1} , 1) & \text{if $i$ censored,} \\
			\mathcal{N}(\bm{w}_i'\bm{\theta} +  \mu_{i1}  +  \frac{\gamma_i}{\gamma_i^2 + \phi_i}(y_i - \bm{x}_i'\bm{\beta} - \mu_{i2}  ),  \frac{\phi_i}{\gamma_i^2 + \phi_i}) & \text{if $i$ uncensored}.
		\end{cases}
	\end{equation*}
	
	\item Sample $\bm{\beta}$ from the full conditional in Appendix \ref{app_mcmc_tbart2norm}.
	\item Sample $\bm{\theta}$ from the full conditional in Appendix \ref{app_mcmc_tbart2norm}.
	\item Sample an auxiliary variable $\kappa \sim Beta(\alpha + 1 , n)$ and sample $\alpha$ from the mixture distribution
	$$ \alpha | k \sim p_{\kappa} \text{Gamma} (c_1 +k , c_2 - \log \kappa ) + (1- p_{\kappa}) \text{Gamma}(c_1 +k -1 , c_2 - \log \kappa ) $$
	where $k$ is the current number of mixture components, i.e. unique elements of $ \{ \vartheta_i \}_{i=1}^{n}=  \{ \begin{pmatrix}		\mu_{i1} & \mu_{i2} \end{pmatrix}', \gamma_i, \phi_i \}_{i=1}^{n} $. The mixing probability is $p_{\kappa}$, which can be solved from
	$$ \frac{p_{\kappa}}{1 -p_{\kappa} } = \frac{c_1 + k - 1}{n (c_2 - \log \kappa)} \ \text{ therefore } p_{\kappa} = \frac{ c_1 + k - 1  }{ n (c_2 - \log \kappa ) + c_1 + k - 1 } $$
	If the prior on $\alpha$ is the prior applied by \cite{george2019fully, mcculloch2021causal} and \cite{conley2008semi}, then samples are obtained from $\alpha | k $ by noting that $p(\alpha | k) \propto p(k|\alpha) p(\alpha) \propto \alpha^k \frac{\Gamma(\alpha)}{\Gamma(n+\alpha)} \times (1- \frac{ \alpha - \alpha_{min}}{\alpha_{max} - \alpha_{min}} )^{\psi} $ \citep{antoniak1974mixtures}. A sample can be obtained by discretizing the support and using a multinomial draw. \cite{mcculloch2021causal} use an equally spaced grid of $100$ values from $\alpha_{min}$ to $\alpha_{max}$.
	\item This step is based on algorithm 8 of \cite{neal2000markov}. Denote the cluster to which observation $i$ belongs by $\varrho_i$.  Denote the number of observations in the same cluster as observation $i$ by $n_{\varrho_i}$. For observations $i$ with observed outcomes, $i \in N_1$,
	\begin{enumerate}
		\item If $n_{\varrho_i} > 1$ sample an auxiliary value $\tilde{\vartheta} =\{ \begin{pmatrix}		\tilde{\mu}_{1} & \tilde{\mu}_{2} \end{pmatrix}', \tilde{\gamma}, \tilde{\phi} \}  \sim H_0 $ . Sample $ \vartheta_i =  \{ \begin{pmatrix}		\mu_{i1} & \mu_{i2} \end{pmatrix}', \gamma_i, \phi_i \}  $ according to
		$$\vartheta_i = 
		\begin{cases}
			\vartheta_j  \ \text{with probability } \ \frac{C_i}{\alpha + n -1} f(z_i^*, y_i  | \bm{\theta}', \bm{\beta}' , \vartheta_j) , \ j \neq i  \\
			\tilde{\vartheta}  \ \text{with probability } \ \frac{C_i}{\alpha + n -1} \alpha f(z_i^*, y_i  | \bm{\theta}', \bm{\beta}' , \tilde{\vartheta} )   
		\end{cases}  $$
		where $C_i$ is a normalizing constant
        $$ C_i = \frac{\alpha + n - 1}{ \alpha f(z_i^*, y_i  | \bm{\theta}', \bm{\beta}' , \tilde{\vartheta} ) + \sum_{j \neq i}  f(z_i^*, y_i  | \bm{\theta}', \bm{\beta}' , \vartheta_j) } $$
        and $f(z_i^*, y_i  | \bm{\theta}', \bm{\beta}' , \vartheta_j)$ is the bivariate normal probability
		$$ 
		\begin{pmatrix}
			z_i^* \\
			y_i
		\end{pmatrix}
		\sim \ i. i. d. \ \mathcal{N}\Bigg(
		\begin{pmatrix}
			\bm{w}_i'\bm{\theta} \\
			\bm{x}_i'\bm{\beta}
		\end{pmatrix} +
	\begin{pmatrix}
			\mu_{j1} \\
			\mu_{j2}
		\end{pmatrix}
		, \Sigma_j \Bigg)
		$$
		This gives the probability
		$$f(z_i^*, y_i  | \bm{\theta}', \bm{\beta}' , \vartheta_j) = 
        \frac{1}{2 \pi \sqrt{\phi_j}} \exp \Bigg( - \frac{ \phi_j  + \gamma_j^2 }{ 2\phi_j }  \bigg[  (z_i^* - \bm{w}_i'\bm{\theta} - \mu_{j1})^2 - $$
		$$ 2 \frac{\gamma_j}{{\phi_j + \gamma_j^2}} (z_i^* - \bm{w}_i'\bm{\theta} - \mu_{j1})( y_i - \bm{x}_i'\bm{\beta} - \mu_{i2}   ) +   \frac{(y_i - \bm{x}_i'\bm{\beta} - \mu_{i2})^2}{ {\phi_j + \gamma_j^2}}   \bigg] \Bigg) $$
		and similarly
        $$f(z_i^*, y_i  | \bm{\theta}', \bm{\beta}' , \tilde{\vartheta} ) = \frac{1}{2 \pi \sqrt{\tilde{\phi}}} \exp \Bigg( - \frac{ \tilde{\phi}  + \tilde{\gamma}^2 }{ 2\tilde{\phi} }  \bigg[  (z_i^* - \bm{w}_i'\bm{\theta} - \tilde{\mu}_{1})^2 - $$
		$$ 2 \frac{\gamma_i}{{ \tilde{\phi} + \tilde{\gamma}^2}} (z_i^* - \bm{w}_i'\bm{\theta} - \tilde{\mu}_{1})( y_i - \bm{x}_i'\bm{\beta} - \tilde{\mu}_{2}   ) +   \frac{(y_i - \bm{x}_i'\bm{\beta} - \tilde{\mu}_{2})^2}{ { \tilde{\phi} + \tilde{\gamma}^2}}   \bigg] \Bigg) $$
		\item If $n_{\varrho_i} = 1$ sample $ \vartheta_i =  \{ \begin{pmatrix}		\mu_{i1} & \mu_{i2} \end{pmatrix}', \gamma_i, \phi_i \}  $ according to
		$$\vartheta_i = 
		\begin{cases}
			\vartheta_j  \ \text{with probability } \ \frac{C_i}{\alpha + n -1} f(z_i^*, y_i  | \bm{\theta}', \bm{\beta}' , \vartheta_j) , \ j \neq i  \\
			\text{unchanged}  \ \text{with probability } \ \frac{C_i}{\alpha + n -1} \alpha f(z_i^*, y_i  | \bm{\theta}', \bm{\beta}' , \vartheta_i )   
		\end{cases}  $$
	\end{enumerate}
	For observations $i$ with unobserved outcomes, $i \in N_0$, sample as in (a) and (b) with the bivariate likelihoods replaced by the univariate likelihoods of $z_i^*$.
	\begin{equation*}
		z_i^* | \bm{y}_c , \vartheta_j, \bm{\theta}', \bm{\beta}' = 
			\mathcal{N}(\bm{w}_i'\bm{\theta} +  \mu_{j1} , 1) \ \text{if $i$ censored,} 
	\end{equation*}
	$$f(z_i^*  | \bm{\theta}' , \vartheta_j)  =  \frac{1}{\sqrt{2 \pi}} \exp \bigg( - \frac{1}{2}(z_i^* - \bm{w}_i'\bm{\theta} -  \mu_{j1})^2 \bigg)  \ , \ f(z_i^*  | \bm{\theta}' , \tilde{\vartheta} )  =  \frac{1}{\sqrt{2 \pi}} \exp \bigg( - \frac{1}{2}(z_i^* - \bm{w}_i'\bm{\theta} -  \tilde{\mu}_{1})^2 \bigg) $$
	and 
	$$ C_i = \frac{\alpha + n - 1}{ \alpha f(z_i^*  | \bm{\theta}' , \tilde{\vartheta} )  + \sum_{j \neq i}  f(z_i^*  | \bm{\theta}' , \vartheta_j) } $$
	\item Let $\bm{\vartheta}* = \{ \vartheta_j^* \}_{j=1}^k$ denote the set of distinct values in $\bm{\vartheta} = \{ \vartheta_i\}_{i=1}^n$ and denote the vector of cluster variables by $\bm{\varrho} = (\varrho_1,...,\varrho_n)$. For all distinct groups in the mixture, indexed by $j=1,...,k$ either sample the entire vector $f(\vartheta_j^* | \bm{y}, \bm{z}^*, \bm{\theta}, \bm{\beta}, \bm{\varrho} )$ in one step, or blocks of 	$\vartheta_j^*$ from their conditional posterior.
	
	In this step the set of unique parameter values $\bm{\vartheta}$ are resampled or remixed. This speeds up convergence of the Markov chain.
	
	Sampling from  $f(\vartheta_j^* | \bm{y}, \bm{z}^*, \bm{\theta}, \bm{\beta}, \bm{\varrho} )$ proceeds as described by \cite{van2011bayesian}: \newline The distinct values in $\bm{\vartheta}^*$ are  drawn from $H_0$. The $\vartheta_j^*$ are conditionally independent in the posterior. Denote the set of censored observations in cluster $j$ by $N_{0j} \equiv \{ i : \varrho_i = j , z_i = 0 \}$, and denote the set of uncensored observations in cluster $j$ by $N_{1j} \equiv \{ i : \varrho_i = j , z_i = 1 \}$. For any $j$, there are three possible cases:
	
	\begin{itemize}
		\item Suppose only censored observations are in cluster $j$, i.e. $N_{1j}= \emptyset$ and $N_{0j} \neq \emptyset$. Let $ \vartheta_j^* =  \{ \begin{pmatrix}		\mu_{j1}^* & \mu_{j2}^* \end{pmatrix}', \gamma_j^* , \phi_j^*  \}  $. The new draw $\vartheta_j^*$ is sampled as follows for prior $\gamma_i | \phi_i \sim \mathcal{N} (\gamma_0, \tau \phi)$ 
		$$\phi_j^* \sim IG(\frac{n_0}{2}, \frac{S_0}{2}) \ \ , \ \ \gamma_j^* | \phi_j \sim \mathcal{N} (\gamma_0, \tau \phi_j^*)$$
		$$ \mu_{j1}^* | \bm{y}, \bm{z}^* , \bm{\theta}, \bm{\beta}, \bm{\varrho} \sim  \mathcal{N} \bigg( \frac{\sum_{i: \varrho_i = j} u_{i1} }{ \frac{1}{\omega_{11}} + n_j} \ , \ \frac{ 1 }{ \frac{1}{\omega_{11}} + n_j} \bigg)   \ , \  \mu_{j2}^* | \mu_{j1}^*  \sim \mathcal{N} \Bigg( \frac{\omega_{12}}{\omega_{11}} \mu_{j1}^* , \omega_{22} - \frac{\omega_{12}^2}{\omega_{11}}  \Bigg) $$
		where $n_j = \sum_{i=1}^n \mathbb{I} \{ \varrho_i = j \} $ and $\omega_{11}, \omega_{12}, \omega_{22}$ are the elements of the prior covariance matrix $\Omega$ for $\begin{pmatrix} \mu_{i1} & \mu_{i2} \end{pmatrix}' $.

		\item Suppose only uncensored observations are in cluster $j$, i.e. $N_{0j}= \emptyset$ and $N_{1j} \neq \emptyset$ Then exact sampling of $\vartheta_j^*$ from the posterior is not feasible. A Gibbs sampler iteration is used to sample $\phi_j^*$, $\gamma_j^*$, and $\begin{pmatrix} \mu_{i1}^* & \mu_{i2}^* \end{pmatrix}'$
		$$\begin{pmatrix} \mu_{i1}^*  & \mu_{i2}^* \end{pmatrix}' | \bm{y}, \bm{z}^*, \bm{\theta}, \bm{\beta}, \bm{\varrho} , \phi_j^* , \gamma_j^*  \sim  \mathcal{N}_2 \bigg( [\Omega^{-1} +n_j \Sigma_j^{*-1}]^{-1}  \Sigma_j^{*-1}\sum_{i : \varrho_i=j } u_i , [\Omega^{-1}+n_j \Sigma_j^{*-1}]^{-1}   \bigg) $$
		$$\phi_j^* |  \bm{y}, \bm{z}^*, \bm{\theta}, \bm{\beta}, \bm{\varrho} ,\begin{pmatrix} \mu_{i1}^* & \mu_{i2}^* \end{pmatrix}' \ , \  \gamma_j^*  \sim \mathcal{IG}\left(\frac{n_0}{2} + \frac{n_j+1}{2}, \bar{d}\right) $$
		where $\bar{d} = \frac{S_0}{2} + \frac{{\gamma_j^*}^2}{ \tau} + \frac{1}{2}\sum_{i: \varrho_i = j} \big(u_{i2} - \mu_{j2}^* - \gamma_j^* (u_{i1} - \mu_{j1}^*)\big)^2$
		$$\gamma_j^* |  \bm{y}, \bm{z}^*, \bm{\theta}, \bm{\beta}, \bm{\varrho} ,\begin{pmatrix} \mu_{i1}^* \\ \mu_{i2}^* \end{pmatrix} \ , \ \phi_j^*  \sim \mathcal{N} \left( \bar{\gamma},  \frac{\phi_j^*}{\frac{1}{\tau} + \sum_{i: \varrho_i = j} (u_{i1} - \mu_{j1}^*)^2}  \right) $$
		$$\text{ where } \ \bar{\gamma} = \frac{\sum_{i: \varrho_i = j} (u_{i1} - \mu_{j1}^*) (u_{i2} - \mu_{j2}^*) }{\frac{1}{\tau} + \sum_{i: \varrho_i = j} (u_{i1} - \mu_{j1}^*)^2}  $$
		\item Suppose $j$ contains both censored and uncensored observations.  i.e. $N_{1j} \neq \emptyset$ and $N_{0j} \neq \emptyset$. The posterior can be factored as $f(\vartheta_j^* | \bm{y}, \bm{z}^*, \bm{\theta}, \bm{\beta}, \bm{\varrho} ) = f(\mu_{j1}^* | \bm{y}, \bm{z}^*, \bm{\theta}, \bm{\beta}, \bm{\varrho} ) f(\mu_{j2}^*, \gamma_j^*, \phi_j^*  | \bm{y}, \bm{z}^*, \bm{\theta}, \bm{\beta}, \bm{\varrho}, \mu_{j1}^*  )  $.
		
		First sample directly from $f(\mu_{j1}^* | \bm{y}, \bm{z}^*, \bm{\theta}, \bm{\beta}, \bm{\varrho} )$:
		$$ \mu_{j1}^* | \bm{y}, \bm{z}^*, \bm{\theta}, \bm{\beta}, \bm{\varrho} \sim \mathcal{N} \Bigg( \frac{\sum_{i: \varrho_i = j} u_{i1} }{ \frac{1}{\omega_{11}} + n_j} , \frac{ 1 }{ \frac{1}{\omega_{11}} + n_j}  \Bigg)  $$
		It is not possible to sample directly from $ f(\mu_{j2}^*, \gamma_j^*, \phi_j^*  | \bm{y}, \bm{z}^*, \bm{\theta}, \bm{\beta}, \bm{\varrho}, \mu_{j1}^*  ) $. A Gibbs sampler is used to sample $\mu_{j2}^*, \gamma_j^*, \phi_j^*$. The conditional posteriors are:
		$$ \mu_{j2}^* | \bm{y}, \bm{z}^*, \bm{\theta}, \bm{\beta}, \bm{\varrho}, \mu_{j1}^*, \gamma_j^*, \phi_j^*  \sim \mathcal{N} \left( \bar{\mu}_2 , \bar{\Omega}_2 \right) \text{where} \ \bar{\Omega}_2 = \Bigg( \frac{\omega_{11}}{\omega_{11}\omega_{22}- \omega_{12}^2 }  + \frac{n_{j1}}{\phi_j^*}  \Bigg)^{-1} $$
		$$ \text{ and } \bar{\mu}_2 = \bar{\Omega}_2 \Bigg( \frac{\omega_{11} \mu_{j1}^*}{\omega_{11}\omega_{22}- \omega_{12}^2 }  + \frac{1}{\phi_j^*} \sum_{i: \varrho_i=j , z_i = 1} \big( u_{i2} - \gamma_j^* (u_{i1} - \mu_{j1}^* )  \big)  \Bigg) $$

		$$\gamma_{j}^* | \bm{y}, \bm{z}^*, \bm{\theta}, \bm{\beta}, \bm{\varrho}, \mu_{j1}^*. \mu_{j2}^*,  \phi_j^*  \sim \mathcal{N} \Bigg( \bar{\gamma} , \frac{\phi_j^*}{\frac{1}{\tau}  +  \sum_{i: \varrho_i=j , z_i = 1} (u_{i1} - \mu_{j1}^* )^2 } \Bigg)  $$
		 $$ \text{ where } \bar{\gamma} = \frac{ \sum_{i: \varrho_i=j , z_i = 1} (u_{i1} - \mu_{j1}^* )(u_{i2} - \mu_{j2}^* )  }{\frac{1}{\tau}  +  \sum_{i: \varrho_i=j , z_i = 1} (u_{i1} - \mu_{j1}^* )^2 }$$
		 
		$$\phi_{j}^* | \bm{y}, \bm{z}^*, \bm{\theta}, \bm{\beta}, \bm{\varrho}, \mu_{j1}^*. \mu_{j2}^*,  \gamma_j^*  \sim \mathcal{IG}(\frac{n_0}{2} + \frac{n_{j1}+1}{2}, \bar{d}) $$
		where $\bar{d} = \frac{S_0}{2} + \frac{{\gamma_j^*}^2}{ \tau} + \frac{1}{2}\sum_{i: \varrho_i = j, z_i = 1} \big(u_{i2} - \mu_{j2}^* - \gamma_j^* (u_{i1} - \mu_{j1}^*)\big)^2$

	\end{itemize}

%
%
%
%

	\item Return to step 2 and repeat.
	
	
	
\end{enumerate}

\subsection{Nonparametric Type 2 TOBART Gibbs Sampler}

\subsubsection{Sampler Details}



This implementation without sampling of unobserved outcomes is based on \cite{van2011bayesian}. Alternative implementations based on \cite{omori2007efficient} and \cite{chib2009estimation} are also feasible.

The full residuals are denoted by $ u_{i1} = y_i -  f_y(\bm{x}_i)  $ and $ u_{i2} = z_i^* - f_z (\bm{w}_i)  $ . The sampler introduced by \cite{van2011bayesian}, adjusted to include sampling of sums of trees, proceeds as follows:

\begin{enumerate}
	\item Set initial values for all the parameters $\phi_i$, $\gamma_i$, $f_z,  f_y$  .
	\item Draw the latent variable from $\bm{z}^* | \phi_i, \gamma_i, f_z,  f_y, \bm{y}_o  $.
	\begin{enumerate}
		\item For censored observations, generate  $ z_i^* |  \phi_i, \gamma_i, f_z,  f_y \sim \mathcal{TN}_{(-\infty,0)} (f_z (\bm{w}_i) + \mu_{i1}, 1)$ where $\mathcal{TN}_{(a,b)}(\mu, \sigma^2)$ denotes a normal distribution with mean $\mu$ and variance $\sigma^2$ truncated on the interval $(a,b)$.
		\item For uncensored observations, generate $z_i^* | \bm{y}_c, \phi_i, \gamma, f_z,  f_y \sim \mathcal{TN}_{[0,\infty)} (\mu_{z, i} , \sigma_{z, i}^2)$ where $\mu_{z, i} = f_z (\bm{w}_i) +  \mu_{i1} + \gamma_i (y_i - f_y(\bm{x}_i) - \mu_{i2}  ) / (\phi_i + \gamma_i^2)$, and $\sigma_{z,i}^2 = 1 - \gamma_i^2/(\phi_i+\gamma_i^2) = \frac{\phi_i}{\gamma_i^2 + \phi_i}$.
	\end{enumerate}
	%
	\begin{equation*}
		z_i^* | \bm{y}_c , \phi_i , \gamma_i, f_z,  f_y = \begin{cases}
			\mathcal{N}(f_z (\bm{w}_i) +  \mu_{i1} , 1) & \text{if $i$ censored,} \\
			\mathcal{N}\bigg(f_z (\bm{w}_i) +  \mu_{i1}  +  \frac{\gamma_i}{\gamma_i^2 + \phi_i}\big(y_i  - \mu_{i2} - f_y(\bm{x}_i)  \big),  \frac{\phi_i}{\gamma_i^2 + \phi_i}\bigg) & \text{if $i$ uncensored}.
		\end{cases}
	\end{equation*}
	\item Draw the sums of trees $f_z(\bm{w}_i)$. First note that the full conditional distribution $\bm{z}^* | \bm{y}_c , \phi , \gamma, f_z(\bm{w}_i), f_y(\bm{x}_i) $
	\begin{equation*}
		z_i^* | \bm{y}_c , \phi , \gamma, f_z(\bm{w}_i), f_y(\bm{x}_i) = \begin{cases}
			\mathcal{N}(f_z(\bm{w}_i) + \mu_{i1}, 1) & \text{if $i$ censored,} \\
			\mathcal{N}\left(f_z(\bm{w}_i) + \mu_{i1} + \frac{\gamma_i}{\gamma_i^2 + \phi_i}\big(y_i - \mu_{i2} - f_y(\bm{x}_i)\big),  \frac{\phi_i}{\gamma_i^2 + \phi_i}\right) & \text{if $i$ uncensored}.
		\end{cases}
	\end{equation*}
	For $\bm{z}^*$, for each tree from $k=1$ to $k=m_z$
	\begin{enumerate}
		\item Create partial residuals
		$$ {R}_{zki} = \begin{cases}
			z_i^* - \mu_{i1} - \sum_{s \neq k} g_s^z (\bm{w}_i) & \text{if $i$ censored,} \\
			z_i^* - \mu_{i1} - \big( y_i^* - \mu_{i2} -  \sum_{k=1}^{m_y} g_k^y (\bm{x}_i)  \big)\frac{\gamma_i}{\gamma_i^2 + \phi_i} - \sum_{s \neq k} g_s^z (\bm{w}_i) & \text{if $i$ uncensored} \\
		\end{cases}$$
		\item Draw a new tree $g_k^{z} (\bm{w})$ using the sampler described by \cite{chipman2010bart} for a regression of $\bm{R}_{zk}$ on a single tree with variance $1$ for censored observations, and $ \frac{\phi_i}{\gamma_i^2 + \phi_i} $ for uncensored observations.\footnote{This is implemented in the \texttt{dbarts} $\bm{R}$ package by specifying weights $1$ for censored observations and $ \frac{\gamma_i^2 + \phi_i}{\phi_i} $ for uncensored observations.}
	\end{enumerate}
	%
	\item Note that for uncensored observations,
	$$y_i = f_y(\bm{x}_i) + \gamma_i \big(z_i^* - \mu_{i1} - f_z(\bm{w}_i) \big) + \mu_{i2} + \eta_i  $$
	where $\eta_i |\bm{x}_i, \bm{w}_i \sim \mathcal{N}(0, \phi_i)$.
	
	Full conditional draws of $f_y(\bm{x}_i)$ and $\gamma_i$ can therefore be made as in semiparametric BART \citep{prado2021semi, zeldow2019semiparametric}. For $\bm{y}^*$, for each tree from $k=1$ to $k=m_y$ :
	\begin{enumerate}
		\item Create partial residuals
		$$  {R}_{yki} = 
        {y}_i^* - \mu_{i2} - \big( {z}_i^* - \mu_{i1} -  \sum_{k=1}^{m_z} g_k^z (\bm{w}_i)  \big) \gamma_i - \sum_{s \neq k} g_s^y (\bm{x}_i)$$			%
		\item Draw a new tree $g_k^{y} (\bm{x}_i)$ using the sampler described by \cite{chipman2010bart} for a regression of $\bm{R}_{yk}$ on a single tree with variances $ \phi_i $  .
	\end{enumerate}
	\item Sample an auxiliary variable $\kappa \sim Beta(\alpha + 1 , n)$ and sample $\alpha$ from the mixture distribution \citep{west1992hyperparameter, escobar1994estimating, escobar1995bayesian}
	$$ \alpha | k \sim p_{\kappa} \text{Gamma} (c_1 +k , c_2 - \log \kappa ) + (1- p_{\kappa}) \text{Gamma}(c_1 +k -1 , c_2 - \log \kappa ) $$
	where $k$ is the current number of mixture components, i.e. unique elements of $ \{ \vartheta_i \}_{i=1}^{n}=  \{ \begin{pmatrix}		\mu_{i1} & \mu_{i2} \end{pmatrix}', \gamma_i, \phi_i \}_{i=1}^{n} $. The mixing probability is $p_{\kappa}$, which can be solved from
	$$ \frac{p_{\kappa}}{1 -p_{\kappa} } = \frac{c_1 + k - 1}{n (c_2 - \log \kappa)} \ \text{ therefore } p_{\kappa} = \frac{ c_1 + k - 1  }{ n (c_2 - \log \kappa ) + c_1 + k - 1 } $$
	If the prior on $\alpha$ is the prior applied by \cite{george2019fully, mcculloch2021causal} and \cite{conley2008semi}, then samples are obtained from $\alpha | k $ by noting that $p(\alpha | k) \propto p(k|\alpha) p(\alpha) \propto \alpha^k \frac{\Gamma(\alpha)}{\Gamma(n+\alpha)} \times (1- \frac{ \alpha - \alpha_{min}}{\alpha_{max} - \alpha_{min}} )^{\psi} $ \citep{antoniak1974mixtures}. A sample can be obtained by discretizing the support and using a multinomial draw. \cite{mcculloch2021causal} use an equally spaced grid of $100$ values from $\alpha_{min}$ to $\alpha_{max}$.
	\item This step is based on algorithm 8 of \cite{neal2000markov}. Denote the cluster to which observation $i$ belongs by $\varrho_i$.  Denote the number of observations in the same cluster as observation $i$ by $n_{\varrho_i}$. For observations $i$ with observed outcomes, $i \in N_1$,
	\begin{enumerate}
		\item If $n_{\varrho_i} > 1$ sample an auxiliary value $\tilde{\vartheta} =\{ \begin{pmatrix}		\tilde{\mu}_{1} & \tilde{\mu}_{2} \end{pmatrix}', \tilde{\gamma}, \tilde{\phi} \}  \sim H_0 $ . Sample $ \vartheta_i =  \{ \begin{pmatrix}		\mu_{i1} & \mu_{i2} \end{pmatrix}', \gamma_i, \phi_i \}  $ according to
		$$\vartheta_i = 
		\begin{cases}
			\vartheta_j  \ \text{with probability } \ \frac{C_i}{\alpha + n -1} f(z_i^*, y_i  | f_z,  f_y , \vartheta_j) , \ j \neq i  \\
			\tilde{\vartheta}  \ \text{with probability } \ \frac{C_i}{\alpha + n -1} \alpha f(z_i^*, y_i  | f_z,  f_y , \tilde{\vartheta} )   
		\end{cases}  $$
		where $C$ is a normalizing constant
        $$ C_i = \frac{\alpha + n - 1}{ \alpha f(z_i^*, y_i  | f_z,  f_y , \tilde{\vartheta} ) + \sum_{j \neq i}  f(z_i^*, y_i  | f_z,  f_y , \vartheta_j) } $$
        and $f(z_i^*, y_i  | f_z,  f_y , \vartheta_j)$ is the bivariate normal probability
		$$ 
		\begin{pmatrix}
			z_i^* \\
			y_i
		\end{pmatrix}
		\sim \ i. i. d. \ \mathcal{N}\Bigg(
		\begin{pmatrix}
			f_z (\bm{w}_i) \\
			f_y (\bm{x}_i)
		\end{pmatrix} +
		\begin{pmatrix}
			\mu_{j1} \\
			\mu_{j2}
		\end{pmatrix}
		, \Sigma_j \Bigg)
		$$
		This gives the probability
		$$f(z_i^*, y_i  | f_z,  f_y , \vartheta_j) = 
        \frac{1}{2 \pi \sqrt{\phi_j}} \exp \bigg( - \frac{ \phi_j  + \gamma_j^2 }{ 2\phi_j }  \bigg[  (z_i^* - f_z (\bm{w}_i) - \mu_{j1})^2 - $$
		$$ 2 \frac{\gamma_j}{{\phi_j + \gamma_j^2}} (z_i^* - f_z (\bm{w}_i) - \mu_{j1}) ( y_i - f_y(\bm{x}_i) - \mu_{i2}   ) +   \frac{(y_i - f_y(\bm{x}_i) - \mu_{i2})^2}{ {\phi_j + \gamma_j^2}}   \bigg] \bigg) $$
		and similarly
		$$f(z_i^*, y_i  | f_z,  f_y , \tilde{\vartheta} ) = \frac{1}{2 \pi \sqrt{\tilde{\phi}}} \exp \bigg( - \frac{ \tilde{\phi}  + \tilde{\gamma}^2 }{ 2\tilde{\phi} }  \bigg[  (z_i^* - f_z (\bm{w}_i) - \tilde{\mu}_{1})^2 - $$
		$$ 2 \frac{\gamma_i}{{ \tilde{\phi} + \tilde{\gamma}^2}} (z_i^* - f_z (\bm{w}_i) - \tilde{\mu}_{1})( y_i - f_y(\bm{x}_i) - \tilde{\mu}_{2}   ) +   \frac{(y_i - f_y(\bm{x}_i) - \tilde{\mu}_{2})^2}{ { \tilde{\phi} + \tilde{\gamma}^2}}   \bigg] \bigg) $$
		\item If $n_{\varrho_i} = 1$ sample $ \vartheta_i =  \{ \begin{pmatrix}		\mu_{i1} &  \mu_{i2} \end{pmatrix}', \gamma_i, \phi_i \}  $ according to
		$$\vartheta_i = 
		\begin{cases}
			\vartheta_j  \ \text{with probability } \ \frac{C_i}{\alpha + n -1} f(z_i^*, y_i  | f_z,  f_y , \vartheta_j) , \ j \neq i  \\
			\text{unchanged}  \ \text{with probability } \ \frac{C_i}{\alpha + n -1} \alpha f(z_i^*, y_i  | f_z,  f_y , \vartheta_i )   
		\end{cases}  $$
	\end{enumerate}
	
	For observations $i$ with unobserved outcomes, $i \in N_0$, sample as in (a) and (b) with the bivariate likelihoods replaced by the univariate likelihoods of $z_i^*$.
	
	\begin{equation*}
		z_i^* | \bm{y}_c , \vartheta_j, f_z,  f_y = 
		\mathcal{N}(f_z (\bm{w}_i) +  \mu_{j1} , 1) \ \text{if $i$ censored,} 
	\end{equation*}
	
	$$f(z_i^*  | f_z  , \vartheta_j)  =  \frac{1}{\sqrt{2 \pi}} \exp \bigg( - \frac{1}{2}(z_i^* - f_z (\bm{w}_i) -  \mu_{j1})^2 \bigg) \ , \ f(z_i^*  | f_z  , \tilde{\vartheta} )  =  \frac{1}{\sqrt{2 \pi}} \exp \bigg( - \frac{1}{2}(z_i^* - f_z (\bm{w}_i) -  \tilde{\mu}_{1})^2 \bigg) $$
	and 
	$$ C_i = \frac{\alpha + n - 1}{ \alpha f(z_i^*  | f_z  , \tilde{\vartheta} )  + \sum_{j \neq i}  f(z_i^*  | f_z  , \vartheta_j) } $$
	
	\item Let $\bm{\vartheta}* = \{ \vartheta_j^* \}_{j=1}^k$ denote the set of distinct values in $\bm{\vartheta} = \{ \vartheta_i\}_{i=1}^n$ and denote the vector of cluster variables by $\bm{\varrho} = (\varrho_1,...,\varrho_n)$. For all distinct groups in the mixture, indexed by $j=1,...,k$ either sample the entire vector $f(\vartheta_j^* | \bm{y}, \bm{z}^*, f_z,  f_y, \bm{\varrho} )$ in one step, or blocks of 	$\vartheta_j^*$ from their conditional posterior.
	
	In this step the set of unique parameter values $\bm{\vartheta}$ are resampled or remixed. This speeds up convergence of the Markov chain.
	
	Sampling from  $f(\vartheta_j^* | \bm{y}, \bm{z}^*, f_z,  f_y, \bm{\varrho} )$ proceeds as described by \cite{van2011bayesian}: \newline 
    The distinct values in $\bm{\vartheta}^*$ are drawn from $H_0$. The $\vartheta_j^*$ are conditionally independent in the posterior. Denote the set of censored observations in cluster $j$ by $N_{0j} \equiv \{ i : \varrho_i = j , z_i = 0 \}$, and denote the set of uncensored observations in cluster $j$ by $N_{1j} \equiv \{ i : \varrho_i = j , z_i = 1 \}$. For any $j$, there are three possible cases:
	
	\begin{itemize}
		\item Suppose only censored observations are in cluster $j$, i.e. $N_{1j}= \emptyset$ and $N_{0j} \neq \emptyset$. Let $ \vartheta_j^* =  \{ \begin{pmatrix}		\mu_{j1}^* & \mu_{j2}^* \end{pmatrix}', \gamma_j^* , \phi_j^*  \}  $ .
		
		The new draw $\vartheta_j^*$ can be sampled as follows (using the \cite{van2011bayesian} prior $\gamma_i | \phi_i \sim \mathcal{N} (\gamma_0, \tau \phi)$ ) 
		$$\phi_j^* \sim IG(\frac{n_0}{2}, \frac{S_0}{2}) \ , \ \gamma_j^* | \phi_j \sim \mathcal{N} (\gamma_0, \tau \phi_j^*)$$
		$$ \mu_{j1}^* | \bm{y}, \bm{z}^* , f_z,  f_y, \bm{\varrho} \sim  \mathcal{N} \bigg( \frac{\sum_{i: \varrho_i = j} u_{i1} }{ \frac{1}{\omega_{11}} + n_j} \ , \ \frac{ 1 }{ \frac{1}{\omega_{11}} + n_j} \bigg)   \ , \  \mu_{j2}^* | \mu_{j1}^*  \sim \mathcal{N} \left( \frac{\omega_{12}}{\omega_{11}} \mu_{j1}^* , \omega_{22} - \frac{\omega_{12}^2}{\omega_{11}}  \right) $$
		where $n_j = \sum_{i=1}^n \mathbb{I} \{ \varrho_i = j \} $ and $\omega_{11}, \omega_{12}, \omega_{22}$ are the elements of the prior covariance matrix $\Omega$ for $\begin{pmatrix} \mu_{i1} & \mu_{i2} \end{pmatrix}' $.

If we apply the generalization of the \cite{ding2014bayesian} prior, we sample $\Omega_j$ as follows:
$$ \tilde{\Sigma}_j \sim W_2^{-1} (\nu_0, c I_2) \ , \  \Omega_j =
\begin{pmatrix}
    \frac{1}{\tilde{\sigma}_{1j} } & 0 \\
    0 & 1 \\
\end{pmatrix}   \tilde{\Sigma}_j
 \begin{pmatrix}
    \frac{1}{\tilde{\sigma}_{1j} } & 0 \\
    0 & 1 \\
\end{pmatrix}  $$

If $\nu_0 = 3$, then while the prior distribution for the correlation is a uniform distribution, the Wishart distribution defined above has an undefined expectation, and therefore it might be preferable to set $\nu_0$ to some value slightly greater than 3, for example 3.25.
\item Suppose only uncensored observations are in cluster $j$, i.e. $N_{0j}= \emptyset$ and $N_{1j} \neq \emptyset$ Then exact sampling of $\vartheta_j^*$ from the posterior is not feasible. A Gibbs sampler iteration is used to sample $\phi_j^*$, $\gamma_j^*$, and $\begin{pmatrix} \mu_{i1}^*  &\mu_{i2}^* \end{pmatrix}' $
$$\begin{pmatrix} \mu_{i1}^* & \mu_{i2}^* \end{pmatrix}' | \bm{y}, \bm{z}^*, f_z,  f_y, \bm{\varrho} , \phi_j^* , \gamma_j^*  \sim  \mathcal{N}_2 \left( [\Omega^{-1} +n_j \Sigma_j^{*-1}]^{-1}  \Sigma_j^{*-1}\sum_{i : \varrho_i=j } u_i , [\Omega^{-1}+n_j \Sigma_j^{*-1}]^{-1}   \right) $$
		$$\phi_j^* |  \bm{y}, \bm{z}^*, f_z,  f_y, \bm{\varrho} ,\begin{pmatrix} \mu_{i1}^* \\ \mu_{i2}^* \end{pmatrix} , \gamma_j^*  \sim \mathcal{IG}(\frac{n_0}{2} + \frac{n_j+1}{2}, \bar{d}) $$
		$ \text{ where } \bar{d} = \frac{S_0}{2} + \frac{{\gamma_j^*}^2}{ \tau} + \frac{1}{2}\sum_{i: \varrho_i = j} \big(u_{i2} - \mu_{j2}^* - \gamma_j^* (u_{i1} - \mu_{j1}^*)\big)^2$
		$$\gamma_j^* |  \bm{y}, \bm{z}^*, f_z,  f_y, \bm{\varrho} ,\begin{pmatrix} \mu_{i1}^* \\ \mu_{i2}^* \end{pmatrix} \ , \  \phi_j^*  \sim \mathcal{N} \Bigg( \bar{\gamma},  \frac{\phi_j^*}{\frac{1}{\tau} + \sum_{i: \varrho_i = j} (u_{i1} - \mu_{j1}^*)^2}  \Bigg) $$
        $$\text{ where } \bar{\gamma} = \frac{\sum_{i: \varrho_i = j} (u_{i1} - \mu_{j1}^*) (u_{i2} - \mu_{j2}^*) }{\frac{1}{\tau} + \sum_{i: \varrho_i = j} (u_{i1} - \mu_{j1}^*)^2}  $$
        Alternatively, we can use the fact that $\gamma_{j}^* ,  \phi_j^* | \bm{y}, \bm{z}^*, f_z,  f_y, \bm{\varrho}, \mu_{j1}^*. \mu_{j2}^*$ is normal inverse gamma distributed (see appendix \ref{phigamma_norm_ig_sec})  to sample as follows
  
Draw $ \phi_j^* \sim \Gamma^{-1} \Bigg(\frac{n_{j} + n_0 }{2}, \frac{1}{2} \Big[ S_0 + \frac{\gamma_0^2}{\tau} - \frac{\frac{\gamma_0}{\tau} + \sum_{i: \varrho_i = j} (u_{i1} - \mu_{j1}^* ) (u_{i2} - \mu_{j2}^*) }{\frac{1}{\tau} +  \sum_{i: \varrho_i = j}  (u_{i1} - \mu_{j1}^* )^2}  + \sum_{i: \varrho_i = j} (u_{i2} - \mu_{j2}^*)^2 \Big]  \Bigg) $.
        
Draw $ \gamma_j^* \sim \mathcal{N} \Bigg( \frac{\frac{\gamma_0}{\tau} + \sum_{i: \varrho_i = j} (u_{i1} - \mu_{j1}^* ) (u_{i2} - \mu_{j2}^*)  }{\frac{1}{\tau} +  \sum_{i: \varrho_i = j} (u_{i1} - \mu_{j1}^* )^2}  , \frac{\phi_j^*}{\frac{1}{\tau} +  \sum_{i: \varrho_i = j} (u_{i1} - \mu_{j1}^* )^2} \Bigg) $

If instead we apply the prior of \cite{ding2014bayesian}, first we draw $\sigma_{1j}^2 | \Omega, \dots  \sim  \{ \frac{1- \rho^2}{c} \chi_{\nu_0}^2 \}^{-1} $ , then calculate $\bm{E}_i$ for the observations for which $\varrho_i = j$,
  $ \bm{E}_i := \begin{pmatrix}
    \sigma_{1j} & 0 \\
    0 & 1
\end{pmatrix} \begin{pmatrix}
    \tilde{{z}}_i \\
    \tilde{{y}}_i \\
\end{pmatrix} =
\begin{pmatrix}
    \sigma_{1j}  \tilde{{z}}_i \\
    \tilde{{y}}_i \\
\end{pmatrix}
$
and then draw  from $\tilde{\Sigma}_j| \{\bm{E}_i\}_{i: \varrho_i = j}$.
$$\tilde{\Sigma_j } | \tilde{\bm{z}}, \tilde{\bm{y}}, \dots  \sim  \bm{W}_2^{-1} \left( n_{j} + \nu_0 , c \bm{I}_2  +   \sum_{i: \varrho_i = j}    \bm{E}_i 
\bm{E}_i^T  \right)  \ , \  \Omega_j =
\begin{pmatrix}
    \frac{1}{\tilde{\sigma}_{1j} } & 0 \\
    0 & 1 \\
\end{pmatrix}   \tilde{\Sigma}_j
 \begin{pmatrix}
    \frac{1}{\tilde{\sigma}_{1j} } & 0 \\
    0 & 1 \\
\end{pmatrix}  $$

\item Suppose $j$ contains both censored and uncensored observations.  i.e. $N_{1j} \neq \emptyset$ and $N_{0j} \neq \emptyset$. The posterior can be factored as $f(\vartheta_j^* | \bm{y}, \bm{z}^*, f_z,  f_y, \bm{\varrho} ) = f(\mu_{j1}^* | \bm{y}, \bm{z}^*, f_z,  f_y, \bm{\varrho} ) f(\mu_{j2}^*, \gamma_j^*, \phi_j^*  | \bm{y}, \bm{z}^*, f_z,  f_y, \bm{\varrho}, \mu_{j1}^*  )  $.

First sample directly from $f(\mu_{j1}^* | \bm{y}, \bm{z}^*, f_z,  f_y, \bm{\varrho} )$:
$$ \mu_{j1}^* | \bm{y}, \bm{z}^*, f_z,  f_y, \bm{\varrho} \sim \mathcal{N} \Bigg( \frac{\sum_{i: \varrho_i = j} u_{i1} }{ \frac{1}{\omega_{11}} + n_j} , \frac{ 1 }{ \frac{1}{\omega_{11}} + n_j}  \Bigg)  $$
It is not possible to sample directly from $ f(\mu_{j2}^*, \gamma_j^*, \phi_j^*  | \bm{y}, \bm{z}^*, f_z,  f_y, \bm{\varrho}, \mu_{j1}^*  ) $. A Gibbs sampler is used to sample $\mu_{j2}^*, \gamma_j^*, \phi_j^*$. The conditional posteriors are:
$$ \mu_{j2}^* | \bm{y}, \bm{z}^*, f_z,  f_y, \bm{\varrho}, \mu_{j1}^*, \gamma_j^*, \phi_j^*  \sim \mathcal{N} \left( \bar{\mu}_2 , \bar{\Omega}_2 \right)  \text{ where } \bar{\Omega}_2 = \Bigg( \frac{\omega_{11}}{\omega_{11}\omega_{22}- \omega_{12}^2 }  + \frac{n_{j1}}{\phi_j^*}  \Bigg)^{-1} $$
$$ \text{ and }  \bar{\mu}_2 = \bar{\Omega}_2 \Bigg( \frac{\omega_{11} \mu_{j1}^*}{\omega_{11}\omega_{22}- \omega_{12}^2 }  + \frac{1}{\phi_j^*} \sum_{i: \varrho_i=j , z_i = 1} \big( u_{i2} - \gamma_j^* (u_{i1} - \mu_{j1}^* )  \big)  \Bigg) $$
$$\gamma_{j}^* | \bm{y}, \bm{z}^*, f_z,  f_y, \bm{\varrho}, \mu_{j1}^*. \mu_{j2}^*,  \phi_j^*  \sim \mathcal{N} \Bigg( \bar{\gamma} , \frac{\phi_j^*}{\frac{1}{\tau}  +  \sum_{i: \varrho_i=j , z_i = 1} (u_{i1} - \mu_{j1}^* )^2 } \Bigg)  $$
$$ \text{ where }  \bar{\gamma} = \frac{ \sum_{i: \varrho_i=j , z_i = 1} (u_{i1} - \mu_{j1}^* )(u_{i2} - \mu_{j2}^* )  }{\frac{1}{\tau}  +  \sum_{i: \varrho_i=j , z_i = 1} (u_{i1} - \mu_{j1}^* )^2 }$$
$$\phi_{j}^* | \bm{y}, \bm{z}^*, f_z,  f_y, \bm{\varrho}, \mu_{j1}^*. \mu_{j2}^*,  \gamma_j^*  \sim \mathcal{IG}(\frac{n_0}{2} + \frac{n_{j1}+1}{2}, \bar{d}) $$
where $\bar{d} = \frac{S_0}{2} + \frac{{\gamma_j^*}^2}{ \tau} + \frac{1}{2}\sum_{i: \varrho_i = j, z_i = 1} \big(u_{i2} - \mu_{j2}^* - \gamma_j^* (u_{i1} - \mu_{j1}^*)\big)^2$

Alternatively, we can use the fact that $\gamma_{j}^* ,  \phi_j^* | \bm{y}, \bm{z}^*, f_z,  f_y, \bm{\varrho}, \mu_{j1}^*. \mu_{j2}^*$ is normal inverse gamma distributed (see appendix \ref{phigamma_norm_ig_sec})  to sample as follows
  
Draw $ \phi_j^* \sim \Gamma^{-1} \Bigg(\frac{n_{j1} + n_0 }{2}, \frac{1}{2} \Big[ S_0 + \frac{\gamma_0^2}{\tau} - \frac{\frac{\gamma_0}{\tau} + \sum_{i: \varrho_i = j, z_i = 1} (u_{i1} - \mu_{j1}^* ) (u_{i2} - \mu_{j2}^*) }{\frac{1}{\tau} +  \sum_{i: \varrho_i = j, z_i = 1}  (u_{i1} - \mu_{j1}^* )^2}  + \sum_{i: \varrho_i = j, z_i = 1} (u_{i2} - \mu_{j2}^*)^2 \Big]  \Bigg) $.
        
Draw $ \gamma_j^* \sim \mathcal{N} \Bigg( \frac{\frac{\gamma_0}{\tau} + \sum_{i: \varrho_i = j, z_i = 1} (u_{i1} - \mu_{j1}^* ) (u_{i2} - \mu_{j2}^*)  }{\frac{1}{\tau} +  \sum_{i: \varrho_i = j, z_i = 1} (u_{i1} - \mu_{j1}^* )^2}  , \frac{\phi_j^*}{\frac{1}{\tau} +  \sum_{i: \varrho_i = j, z_i = 1} (u_{i1} - \mu_{j1}^* )^2} \Bigg) $.

If instead we apply the (modified) \cite{ding2014bayesian} prior, first we draw $\sigma_{1j}^2 | \Omega, \dots  \sim  \{ \frac{1- \rho^2}{c} \chi_{\nu_0}^2 \}^{-1} $ then calculate $\bm{E}_i$ for the observations for which $\varrho_i = j$,
          $ \bm{E}_i := \begin{pmatrix}
            \sigma_{1j} & 0 \\
            0 & 1
        \end{pmatrix} \begin{pmatrix}
            \tilde{{z}}_i \\
            \tilde{{y}}_i \\
        \end{pmatrix} =
        \begin{pmatrix}
            \sigma_{1j}  \tilde{{z}}_i \\
            \tilde{{y}}_i \\
        \end{pmatrix}
        $
        and then draw  from $\tilde{\Sigma}_j| \{\bm{E}_i \}_{i: \varrho_i = j, z_i = 1}$.
        $$\tilde{\Sigma_j } | \tilde{\bm{z}}, \tilde{\bm{y}}, \dots  \sim  \bm{W}_2^{-1} \left( n_{j1} + \nu_0 , c \bm{I}_2  +   \sum_{i: \varrho_i = j, z_i = 1}    \bm{E}_i 
        \bm{E}_i^T  \right)  \ , \ \Omega_j =
        \begin{pmatrix}
            \frac{1}{\tilde{\sigma}_{1j} } & 0 \\
            0 & 1 \\
        \end{pmatrix}   \tilde{\Sigma}_j
         \begin{pmatrix}
            \frac{1}{\tilde{\sigma}_{1j} } & 0 \\
            0 & 1 \\
        \end{pmatrix}  $$
	\end{itemize}
	
	[Note: $ \{\vartheta_i \}_{i = 1}^n $ must be updated in each iteration of the above loop].

	%
	%
	%
	%

	\item Return to step 2 and repeat.
	
	
	
\end{enumerate}

\subsubsection{Initial values for the $\vartheta_i$ parameters}

As in \cite{escobar1995bayesian}, sample from  $f(\vartheta_i | y_i, z_i^*, f_z,  f_y )$, using the initial values of $z_i^*, f_z,  f_y$

These samples are similar to step 7 above. For each individual $i$, sample the initial $\vartheta_i^*$ values as follows

\begin{itemize}
	\item Suppose individual $i$'s outcome is censored . Let $ \vartheta_i^* =  \{ \begin{pmatrix}		\mu_{i1} & \mu_{i2} \end{pmatrix}', \gamma_i , \phi_i  \}  $

The new draw $\vartheta_i$ is sampled as follows (using the \cite{van2011bayesian} prior $\gamma_i | \phi_i \sim \mathcal{N} (\gamma_0, \tau \phi)$ ) 
$$\phi_i \sim IG(\frac{n_0}{2}, \frac{S_0}{2}) \ ,\ \gamma_i | \phi_i \sim \mathcal{N} (\gamma_0, \tau \phi_i^*) $$
$$ \mu_{i1} | y_i, z_i^* ,  f_z,  f_y   \sim  \mathcal{N} \bigg( \frac{ u_{i1} }{ \frac{1}{\omega_{11}} + 1} \ , \ \frac{ 1 }{ \frac{1}{\omega_{11}} + 1} \bigg)   \ , \  \mu_{i2} | \mu_{i1}  \sim \mathcal{N} \Bigg( \frac{\omega_{12}}{\omega_{11}} \mu_{i1} , \omega_{22} - \frac{\omega_{12}^2}{\omega_{11}}  \Bigg) $$
where  $\omega_{11}, \omega_{12}, \omega_{22}$ are the elements of the prior covariance matrix $\Omega$ for $\begin{pmatrix} \mu_{i1} \\ \mu_{i2} \end{pmatrix} $.

\item Suppose individual $i$'s outcome is uncensored. Then exact sampling of $\vartheta_i$ from the posterior is not feasible. A Gibbs sampler iteration is used to sample $\phi_i$, $\gamma_i$, and $\begin{pmatrix} \mu_{i1} & \mu_{i2} \end{pmatrix}' $
$$\begin{pmatrix} \mu_{i1} & \mu_{i2} \end{pmatrix}' | y_i, z_i^*, f_z,  f_y, \phi_i , \gamma_i  \sim  \mathcal{N}_2 \Bigg( [\Omega^{-1} + 1 \Sigma_i^{*-1}]^{-1}  \Sigma_i^{*-1}  \begin{pmatrix} u_{i1} \\ u_{i2} \end{pmatrix} , [\Omega^{-1}+  \Sigma_i^{*-1}]^{-1}   \Bigg) $$
$$\phi_i | y_i, z_i^*, f_z,  f_y,  ,\begin{pmatrix} \mu_{i1} & \mu_{i2} \end{pmatrix}' \gamma_i  \sim \mathcal{IG}(\frac{n_0}{2} + 1, \bar{d}) $$
where $\bar{d} = \frac{S_0}{2} + \frac{{\gamma_i}^2}{ \tau} + \frac{1}{2} \big(u_{i2} - \mu_{i2} - \gamma_i^* (u_{i1} - \mu_{i1}^*)\big)^2$
$$\gamma_i |  y_i, z_i^*, f_z,  f_y,  ,\begin{pmatrix} \mu_{i1} & \mu_{i2} \end{pmatrix}' , \phi_i  \sim \mathcal{N} \Bigg( \bar{\gamma},  \frac{\phi_i}{\frac{1}{\tau} +  (u_{i1} - \mu_{i1}^*)^2}  \Bigg) $$
where 
$$\bar{\gamma} = \frac{ (u_{i1} - \mu_{i1}^*) (u_{i2} - \mu_{i2}^*) }{\frac{1}{\tau} +  (u_{i1} - \mu_{i1}^*)^2}  $$

\end{itemize}

\subsubsection{Out of sample distribution of the error}

The out-of-sample errors are sampled as described by \cite{van2011bayesian}. At iteration $t \in \{ 1,...,T\} $ of the Markov chain, and given the current state $\{\vartheta_{i,t}\}_{i=1}^n$, generate an out-of-sample value $\vartheta_{\tilde{n}, t}$ according to:
$$ \vartheta_{\tilde{n}, t}  \begin{cases}
	= \vartheta_{i,t} \ \text{with probability} \ \frac{1}{\alpha + n} \ i = 1,...,n \\
	\sim H_0  \ \text{with probability} \ \frac{\alpha}{\alpha + n}
\end{cases} $$
An estimate of the posterior predictive distribution of the errors is 
$$ \hat{f}(u_1, u_2 |y, s) = \frac{1}{T} \sum_{t=1}^T f(u_1, u_2 | \vartheta_{\tilde{n},t}) $$

\section{Derivation of marginal distributions for generalization of \cite{ding2014bayesian} prior}\label{marginal_prior_app}

We begin with the following prior
$$ \tilde{\Sigma} = \begin{pmatrix}
    \tilde{\sigma}_1 & 0 \\
    0 & 1 \\
\end{pmatrix} 
\Omega  \begin{pmatrix}
    \tilde{\sigma}_1 & 0 \\
    0 & 1 \\
\end{pmatrix} 
\sim W_2^{-1} (\nu_0, c I_2) 
$$
The prior probability can be written as
$$ p(\tilde{\Sigma} )   = \frac{|c I_2|^{\frac{\nu_0}{2} }}{2^{\frac{2\nu_0}{2}} \Gamma_2(\frac{\nu_0}{2})} |\tilde{\Sigma} |^{-\frac{\nu_0 + 3}{2}} \exp \left\{ -\frac{1}{2} tr(c I \tilde{\Sigma}^{-1} )  \right\}   \propto \frac{ c^2 | I_2|^{\frac{\nu_0}{2} }}{2^{\nu_0} \Gamma_2(\frac{\nu_0}{2})} |\tilde{\Sigma} |^{-\frac{\nu_0 + 3}{2}} \exp \left\{ -\frac{1}{2} c \ tr( \tilde{\Sigma}^{-1} )  \right\}  $$
where $\Gamma_2$ is a multivariate gamma function.

Denoting the elements of  $\tilde{\Sigma}$ by $\Sigma_{1,1} = \tilde{\sigma}_1^2$, $\Sigma_{1,2} = \tilde{\sigma}_{12}$, $\Sigma_{2,2} = \tilde{\sigma}_2^2$, we can write
$$ \Omega =  \begin{pmatrix}
    1 & \omega_{12} \\
    \omega_{12} & \omega_{22} \\
\end{pmatrix} =
\begin{pmatrix}
    \frac{1}{\tilde{\sigma}_1 } & 0 \\
    0 & 1 \\
\end{pmatrix}   \tilde{\Sigma}
 \begin{pmatrix}
    \frac{1}{\tilde{\sigma}_1 } & 0 \\
    0 & 1 \\
\end{pmatrix}  = 
 \begin{pmatrix}
    1 & \frac{\tilde{\sigma}_{12}}{\tilde{\sigma}_1 }  \\
    \frac{\tilde{\sigma}_{12}}{\tilde{\sigma}_1 } & \tilde{\sigma}_2^2 \\
\end{pmatrix} $$
Therefore $\tilde{\sigma}_1^2 = \sigma_1^2$, $\tilde{\sigma}_{12} = \rho \sigma_2 \sigma_1 = \omega_{12} (\sigma_1^2)^{1/2} $, and $\tilde{\sigma}_2^2 = \sigma_2^2 = \omega_{22} $

To re-express this probability distribution in terms of $(\Omega, \sigma_1^2)$, as in \cite{ding2014bayesian}, we must consider the following Jacobian
$$ \frac{\partial  ( \tilde{\sigma}_{1}^2, \tilde{\sigma}_{12}, \tilde{\sigma}_{2}^2 )  }{ \partial ( \sigma_1^2, \omega_{12}, \omega_{22} )} =
\begin{pmatrix}
     1 & 0 & 0  \\
     \frac{1}{2} \omega_{12} (\sigma_1^2)^{-1/2} & (\sigma_1^2)^{1/2} & 0 \\
     0 & 0 & 1 \\
\end{pmatrix}
$$
The determinant of this Jacobian is $\sigma_1$. Therefore
$$ p(\sigma_1^2, \Omega )   \propto p(\tilde{\Sigma} ) \sigma_1 $$
$$\propto \frac{ c^2 |I_2|^{\frac{\nu_0}{2} }}{2^{\nu_0} \Gamma_2(\frac{\nu_0}{2})} \Bigg| \begin{pmatrix}
    {\sigma}_1 & 0 \\
    0 & 1 \\
\end{pmatrix} 
\Omega  \begin{pmatrix}
    {\sigma}_1 & 0 \\
    0 & 1 \\
\end{pmatrix}  \Bigg|^{-\frac{\nu_0 + 3}{2}} \exp \left\{ -\frac{1}{2} c \ tr \Bigg( \left( \begin{pmatrix}
    {\sigma}_1 & 0 \\
    0 & 1 \\
\end{pmatrix}  
\Omega  \begin{pmatrix}
    {\sigma}_1 & 0 \\
    0 & 1 \\
\end{pmatrix}  \right)^{-1} \Bigg)  \right\}  \sigma_1 $$
Note that 
$$\Bigg| \begin{pmatrix}
    {\sigma}_1 & 0 \\
    0 & 1 \\
\end{pmatrix} 
\Omega  \begin{pmatrix}
    {\sigma}_1 & 0 \\
    0 & 1 \\
\end{pmatrix}  \Bigg|^{-\frac{\nu_0 + 3}{2}} = \Bigg( \Bigg| \begin{pmatrix}
    {\sigma}_1 & 0 \\
    0 & 1 \\
\end{pmatrix}\Bigg|^2 
|\Omega | \Bigg)  ^{-\frac{\nu_0 + 3}{2}} = (\sigma_1^2) ^{-\frac{\nu_0 + 3}{2}} |\Omega |   ^{-\frac{\nu_0 + 3}{2}} $$
and
$$ tr \Bigg( \left( \begin{pmatrix}
    {\sigma}_1 & 0 \\
    0 & 1 \\
\end{pmatrix}  
\Omega  \begin{pmatrix}
    {\sigma}_1 & 0 \\
    0 & 1 \\
\end{pmatrix}  \right)^{-1} \Bigg) = 
 tr\Bigg( \begin{pmatrix}
    \frac{1}{{\sigma}_1^2}  & 0 \\
    0 & 1 \\
\end{pmatrix}   \Omega^{-1}  \Bigg)  =  tr\Bigg( \begin{pmatrix}
    \frac{1}{{\sigma}_1^2}  & 0 \\
    0 & 1 \\
\end{pmatrix}    \begin{pmatrix}
    \omega^{11} & \omega^{12} \\
    \omega^{12} & \omega^{22} \\
\end{pmatrix}   \Bigg) =   \frac{\omega^{11}}{\sigma_1^2} + \omega^{22} $$
where $\omega^{11},\omega^{12}, \omega^{22} $ denote the elements of $\Omega^{-1}$.
$\Omega^{-1} =  \frac{1}{\sigma_2^2 (1 - \rho^2)} \begin{pmatrix}
    \sigma_2^2 & - \rho \sigma_2 \\
    - \rho \sigma_2 & 1 \\
\end{pmatrix}  $  and  $\omega^{11} = \frac{1}{1- \rho^2}$,  $ \omega^{12} = - \frac{\rho}{\sigma_2 (1 - \rho^2)}$ and $\omega^{22} = \frac{1}{\sigma_2^2 (1 - \rho^2)}$.

The prior can therefore be expressed as \footnote{There is a small mistake in the appendix of \cite{ding2014bayesian} where $\Sigma$ should be replaced by $\Omega$.}
$$ p(\sigma_1^2, \Omega )  \propto 
\propto (\sigma_1^2) ^{-\frac{\nu_0 }{2} -1} |\Omega |   ^{-\frac{\nu_0 + 3}{2}}   \exp \left\{ -\frac{c}{2}  \left( \frac{\omega^{11}}{\sigma_1^2} + \omega^{22} \right) \right\}   $$

Now, as in \cite{ding2014bayesian} we may re-write this prior by noting that $ p(\sigma_1^2, \Omega) \propto p(\Omega) p(\sigma_1^2 | \Omega)$. First, we integrate out $\sigma_1^2$ to obtain $p(\Omega)$.
$$ p (\Omega) = \int_{0}^{\infty} p(\sigma_1^2, \Omega) d \sigma_1^2   \propto \int_{0}^{\infty} (\sigma_1^2) ^{-\frac{\nu_0 }{2} -1} |\Omega |   ^{-\frac{\nu_0 + 3}{2}}   \exp \left\{ -\frac{c}{2} \left(  \frac{\omega^{11}}{\sigma_1^2} + \omega^{22} \right)  \right\}d \sigma_1^2  $$
%
$$ \propto |\Omega |   ^{-\frac{\nu_0 + 3}{2}} \exp \left\{ -\frac{c}{2}  \omega^{22}  \right\}  \int_{0}^{\infty} \left(\frac{1}{\sigma_1^2} \right) ^{\frac{\nu_0 }{2} + 1}   \exp \left\{ -\frac{c\omega^{11} }{2}   \frac{1}{\sigma_1^2}   \right\} d \sigma_1^2  $$
Now we apply the fact that the density of an inverse gamma distribution must integrate to 1, i.e.
$$ \frac{ \left(  \frac{c\omega^{11} }{2} \right)^{\frac{\nu_0 }{2}}  }{\Gamma(\frac{\nu_0 }{2})  }  \int_{0}^{\infty} \left(\frac{1}{\sigma_1^2} \right) ^{\frac{\nu_0 }{2} + 1}   \exp \left\{ -\frac{c\omega^{11} }{2}   \frac{1}{\sigma_1^2}   \right\} d \sigma_1^2 = 1  $$
Therefore 
$$ p (\Omega) \propto \Gamma(\frac{\nu_0 }{2})  \left(  \frac{c  }{2} \right)^{-\frac{\nu_0 }{2}}   (\omega^{11})^{-\frac{\nu_0 }{2}}  |\Omega |   ^{-\frac{\nu_0 + 3}{2}} \exp \left\{ -\frac{c}{2}  \omega^{22}  \right\}     \propto    (\omega^{11})^{-\frac{\nu_0 }{2}}  |\Omega |   ^{-\frac{\nu_0 + 3}{2}} \exp \left\{ -\frac{c}{2}  \omega^{22}  \right\}    $$
Them noting that $ p(\sigma_1^2 | \Omega) = \frac{p(\sigma_1^2 , \Omega) }{ p(\Omega) } $, we obtain
$$  p(\sigma_1^2 | \Omega) \propto \frac{  (\sigma_1^2) ^{-\frac{\nu_0 }{2} -1} |\Omega |   ^{-\frac{\nu_0 + 3}{2}}   \exp \left\{ -\frac{c}{2}  \left( \frac{\omega^{11}}{\sigma_1^2} + \omega^{22} \right) \right\}  }{ (\omega^{11})^{-\frac{\nu_0 }{2}}  |\Omega |   ^{-\frac{\nu_0 + 3}{2}} \exp \left\{ -\frac{c}{2}  \omega^{22}  \right\}  }  \propto (\sigma_1^2)^{ - \frac{\nu_0}{2} - 1 } \exp \left\{ -\frac{c}{2}   \frac{\omega^{11}}{\sigma_1^2} \right\} $$
Noting that $\omega^{11} = \frac{1}{1 - \rho^2}$ we can write
$$  p(\sigma_1^2 | \Omega) 
\propto \left( \frac{1}{\sigma_1^2}\right)^{  \frac{\nu_0}{2} +
 1 } \exp \left\{ -\frac{c}{2 (1 - \rho^2)}   \frac{1}{\sigma_1^2} \right\} $$
Therefore we have a generalization of the result from \cite{ding2014bayesian}, $\sigma_1^2|\Omega \sim \Gamma^{-1} \left( \frac{\nu_0}{2}, \frac{c}{2(1-\rho^2)} \right)$ or $\sigma_1^2|\Omega \sim c \omega^{11}/ \chi_{\nu_0}^2 = \{ \frac{1- \rho^2}{c} \chi_{\nu_0}^2 \}^{-1} $

\bigskip

\bigskip

To obtain marginal prior distributions for $\sigma_2^2$ and $\rho$, it is necessary to apply a change of variables from $(\omega_{12}, \omega_{22})$ to $(\rho, \sigma_2^2)$. We begin with
$$ p (\Omega)    \propto    (\omega^{11})^{-\frac{\nu_0 }{2}}  |\Omega |   ^{-\frac{\nu_0 + 3}{2}} \exp \left\{ -\frac{c}{2}  \omega^{22}  \right\}    $$
Note that $\omega_{12} = \rho \sigma_2 = \rho (\sigma_2^2)^{1/2}$ and $\omega_{22} = \sigma_2^2$. The Jacobian for the change of variables is:
$$\frac{\partial(\omega_{12}, \omega_{22})}{\partial (\rho, \sigma_2^2 ) } = 
\begin{pmatrix}
    (\sigma_2^2)^{1/2} & \frac{1}{2} \rho (\sigma_2^2)^{-1/2} \\
    0 & 1 \\
\end{pmatrix} $$
The determinant of this Jacobian is $(\sigma_2^2)^{1/2}$. Note that $\omega^{11} = \frac{1}{1 - \rho^2}$ and $\omega^{22} = \frac{1}{\sigma_2^2 (1 - \rho^2 ) }$

Therefore
$$ p(\rho, \sigma_2^2 ) \propto p(\Omega) (\sigma_2^2)^{1/2} \propto (\omega^{11})^{-\frac{\nu_0 }{2}}  |\Omega |   ^{-\frac{\nu_0 + 3}{2}} \exp \left\{ -\frac{c}{2}  \omega^{22}  \right\} (\sigma_2^2)^{1/2} $$
$$ \propto \left( \frac{1}{1 - \rho^2} \right)^{-\frac{\nu_0}{2}}  (\sigma_2^2 (1 - \rho^2))^{- \frac{\nu_0+3}{2}}  \exp \left\{ -\frac{c}{2} \frac{1}{\sigma_2^2 (1 - \rho^2 )}   \right\} (\sigma_2^2)^{1/2} $$
%
%
$$ \propto  (\sigma_2^2 )^{- \frac{\nu_0+3}{2} + \frac{1}{2}} (1 - \rho^2)^{- \frac{3}{2} }  \exp \left\{ -\frac{c}{2} \frac{1}{\sigma_2^2 (1 - \rho^2 )}   \right\}  $$
Now we can obtain the marginal priors for $\sigma_2^2$ and $\rho$ by integration.
$$ p ( \sigma_2^2) = \int_{-1}^{1} p (\rho, \sigma_2^2) d \rho \propto (\sigma_2^2)^{ - \frac{\nu_0+3}{2} +\frac{1}{2}}  \int_{-1}^{1} (1 - \rho^2)^{\frac{3}{2}}  \exp \left\{ -\frac{c}{2} \frac{1}{\sigma_2^2 (1 - \rho^2 )}   \right\}  d \rho $$
We can separate the above integral into two regions, $\rho \in [-1,0]$ and $\rho \in [0,1]$ and apply $u$ substitution by setting $u = 1 - \rho^2$. Note that $\frac{\partial u}{\partial \rho} = - 2 \rho $ , $\rho^2 = 1- u$, $\rho = \pm \sqrt{1-u}$. In the range $\rho \in [-1,0]$, we have $u \in [0,1]$ and $\frac{\partial \rho}{\partial u} = \frac{1}{2} (1- u)^{-1/2} $. In the range $\rho \in [0,1]$, we have $u \in [1,0]$ and $\frac{\partial \rho}{\partial u} = -\frac{1}{2} (1- u)^{-1/2} $
Therefore the integral is 
$$  \int_{-1}^{1} (1 - \rho^2)^{\frac{3}{2}}  \exp \left\{ -\frac{c}{2} \frac{1}{\sigma_2^2 (1 - \rho^2 )}   \right\}  d \rho $$
$$=   \int_{0}^{1} u^{\frac{3}{2}}  \exp \left\{ -\frac{c}{2} \frac{1}{\sigma_2^2 u}   \right\} \frac{1}{2} (1- u)^{-1/2}   d u +  \int_{1}^{0} u^{\frac{3}{2}}  \exp \left\{ -\frac{c}{2} \frac{1}{\sigma_2^2 u}   \right\} (- \frac{1}{2} (1- u)^{-1/2}  ) d u $$
$$ = 2 \int_{0}^{1} u^{\frac{3}{2}} \frac{1}{2} (1- u)^{-1/2} \exp \left\{ -\frac{c}{2} \frac{1}{\sigma_2^2 u}   \right\}     d u =  \int_{0}^{1} u^{\frac{3}{2}} (1- u)^{-1/2} \exp \left\{ -\frac{c}{2} \frac{1}{\sigma_2^2 u}   \right\}     d u$$
$$ = 
\exp \left\{ -\frac{c}{2\sigma_2^2 }   \right\} \sqrt{2 \pi \sigma_2^2 \frac{1}{c} }  \quad \propto  \   \exp \left\{ -\frac{c}{2\sigma_2^2 }   \right\}(\sigma_2^2)^{1/2} $$

Overall the marginal prior is
$$ p ( \sigma_2^2)  \propto (\sigma_2^2)^{ - \frac{\nu_0+3}{2} +\frac{1}{2}} \exp \left\{ -\frac{c}{2\sigma_2^2 }   \right\}(\sigma_2^2)^{1/2}  
\propto \left(\frac{1}{\sigma_2^2}\right)^{  \left( \frac{\nu_0}{2} -\frac{1}{2}\right) + 1 } \exp \left\{ -\frac{c}{2\sigma_2^2 }   \right\} $$
$$ \sigma_2^2 \sim \Gamma^{-1}\left(  \frac{\nu_0}{2} -\frac{1}{2} , \frac{c}{2}  \right)  $$
Similarly, the marginal prior for $\rho$ can be obtained by integration:
$$ p(\rho) = \int_{0}^{\infty} p(\rho, \sigma_2^2 ) d \sigma_2^2 \propto \int_{0}^{\infty} (\sigma_2^2 )^{- \frac{\nu_0+3}{2} + \frac{1}{2}} (1 - \rho^2)^{- \frac{3}{2} }  \exp \left\{ -\frac{c}{2} \frac{1}{\sigma_2^2 (1 - \rho^2 )}   \right\} d \sigma_2^2 $$
%
%
$$ \propto (1 - \rho^2)^{- \frac{3}{2} }  \int_{0}^{\infty} \left( \frac{1}{\sigma_2^2} \right)^{\frac{\nu_0+3}{2} - \frac{1}{2}}   \exp \left\{ -\frac{c}{2 (1 - \rho^2 ) } \frac{1}{\sigma_2^2 }   \right\} d \sigma_2^2   $$
The density of an inverse gamma distribution must integrate to 1, therefore
$$ \frac{ \left( \frac{c}{2 (1 - \rho^2 ) } \right)^{\frac{\nu_0}{2} }  }{  \Gamma \left( \frac{\nu_0}{2}  \right) }   \int_{0}^{\infty} \left( \frac{1}{\sigma_2^2} \right)^{\frac{\nu_0}{2} - 1}   \exp \left\{ -\frac{c}{2 (1 - \rho^2 ) } \frac{1}{\sigma_2^2 }   \right\} d \sigma_2^2 = 1 $$
This implies that the marginal prior distribution for $\rho$ is
$$ p(\rho) \propto   (1 - \rho^2)^{- \frac{3}{2} }  \Gamma \left( \frac{\nu_0}{2}  \right)  \left( \frac{c}{2 (1 - \rho^2 ) } \right)^{- \frac{\nu_0}{2} }   \propto (1 - \rho^2)^{ \frac{\nu_0 - 3}{2} } $$
This prior is uniform when $\nu_0=3$, as noted by \cite{ding2014bayesian}.

\section{Implementation Details and Parameter Settings}\label{imp_details_app}


\subsection{Prior mean of variance of outcome}

Under the prior specification of \cite{van2011bayesian}, the mean of the prior distribution of the outcome variance $\phi + \gamma^2$ can be derived as follows:
$$ p(\gamma) = \int_0^{\infty} \frac{1}{\sqrt{2 \pi \tau}}  \frac{ \left( \frac{S_0}{2}  \right)^{\frac{n_0}{2}}  }{ \Gamma \left( \frac{n_0}{2} \right) } \left( \frac{1}{\phi} \right)^{ \frac{n_0 + 3}{2}} \exp \left( - \frac{ S_0 + \frac{1}{\tau} (\gamma - g_0)^2 }{2 \phi} \right) d\phi $$
This integral can be evaluated by u-substitution, setting $u = \phi^{-1}$. Noting that $ d \phi = - u^{-2} $ and that the limits become $u = \infty$ when $\phi = 0$, and $u = 0$ when $\phi = \infty$.
$$ p(\gamma) = \int_0^{\infty} \frac{1}{\sqrt{2 \pi \tau}}  \frac{ \left( \frac{S_0}{2}  \right)^{\frac{n_0}{2}}  }{ \Gamma \left( \frac{n_0}{2} \right) } u^{ \frac{n_0 + 3  }{2}} \exp \left( - \frac{ S_0 + \frac{1}{\tau} (\gamma - g_0)^2 }{2} u \right) u^{-2} d u $$
$$ 
= \frac{1}{\sqrt{2 \pi \tau}}  \frac{ \left( \frac{S_0}{2}  \right)^{\frac{n_0}{2}}  }{ \Gamma \left( \frac{n_0}{2} \right) } \int_0^{\infty}  u^{ \frac{n_0 -1 }{2}} \exp \left( - \frac{ S_0 + \frac{1}{\tau} (\gamma - g_0)^2 }{2} u \right) d u $$
The above integral is a (generalized) gamma integral. Therefore the marginal prior probability density for $\gamma$ is:
$$ p(\gamma) = \frac{1}{\sqrt{2 \pi \tau}}  \frac{ \left( \frac{S_0}{2}  \right)^{\frac{n_0}{2}}  }{ \Gamma \left( \frac{n_0}{2} \right) } \frac{\Gamma\left( \frac{n_0 + 1}{2} \right) }{ \left[ \frac{S_0 + \frac{1}{\tau} (\gamma - g_0)^2 }{2} \right]^{\frac{n_0+1}{2}} }  \propto \left[ \frac{S_0}{2} \left( 1 + \frac{1}{\tau S_0} (\gamma - g_0)^2  \right) \right]^{- \frac{n_0+1}{2}}  $$
$$ \propto \left[  \left( 1 + \frac{1}{\tau S_0} (\gamma - g_0)^2  \right) \right]^{- \frac{n_0+1}{2}}  \propto \left[  \left( 1 + \frac{1}{n_0} \bigg( \frac{\gamma - g_0}{\sqrt{\tau S_0/n_0}} \bigg)^2 \right) \right]^{- \frac{n_0+1}{2}}  $$
Therefore the marginal prior distribution of $\gamma$ is a location-scale t-distribution with location $g_0$, scale $\sqrt{\frac{\tau S_0}{n_0}}$, and degrees of freedom $n_0$. The mean, variance, and second moment are:
$$ E[\gamma] = g_0  \ , \ \text{Var}(\gamma) = \frac{ \tau S_0 }{n_0 - 2} \ , \ E(\gamma^2) = \frac{ \tau S_0 }{n_0 - 2} + g_0^2 $$
The mean of the inverse gamma prior for $\phi$ is $\frac{S_0}{n_0-2}$, therefore we may conclude that
$$ E[ \phi + \gamma^2 ] = (1 + \tau)\frac{S_0}{n_0-2} + g_0^2  $$
Therefore when $\tau$ is large, a considerable proportion of the variance in the outcome is attributable to sample selection. Furthermore, an uninformative prior on $\phi$ imposed by a small value of $n_0$ and/or a large value of $S_0$ implies a high variance prior for $\gamma$ and a high prior mean for $\phi + \gamma^2$. If the prior for $\phi + \gamma^2$ is data-calibrated by the sample standard deviation of the selected outcomes or the estimated outcome variance from a linear type 2 Tobit model estimated by maximum likelihood, then the prior variance of $\phi$ can be small when $\tau$ is large. 

We consider the following prior parameter values by default: $n_0=6$, $g_0 = 0$ or $g_0$ is obtained from a linear type 2 Tobit model estimated by maximum likelihood, and a data-informed value of $S_0$ is set so that $E[ \phi + \gamma^2 ] = \hat{\sigma}_{y}^2$ where $\hat{\sigma}_{y}^2$ is the sample standard deviation of the uncensored outcomes or the estimated outcome variance from a linear type 2 Tobit model estimated by maximum likelihood.
$$S_0 = (\hat{\sigma}_{y}^2 - g_0^2) \frac{n_0-2}{1+\tau} $$
A possible approach for calibration of $\tau$ is to obtain an estimate of $\widehat{\text{Var}}(\hat{\gamma}) $ from linear type 2 Tobit model estimated by maximum likelihood. If the model is parametrized int terms of $\rho$ instead of $\gamma$, then the delta method can be applied, giving $\widehat{\text{Var}}(\hat{\gamma}) = \hat{\rho} \widehat{\text{Var}}(\hat{\sigma}_y)  + \hat{\sigma}_y^2 \widehat{\text{Var}}(\hat{\rho}) + 2 \hat{\rho} \hat{\sigma}_y  \widehat{\text{Cov}}( \hat{\rho} , \hat{\sigma}_y)  $, then from the formulae for $ E[\gamma]$ and $\text{Var}(\gamma)$ above, one can set $\tau \overset{!}{=} \frac{\widehat{\text{Var}}(\hat{\gamma}) }{ \widehat{\text{Var}}(\hat{\sigma}_y)  -  \widehat{\text{Var}}(\hat{\gamma})  - g_0  }  $. However, the estimate $\widehat{\text{Var}}(\hat{\gamma})$ is potentially a very small value if there are many observations, resulting in a small value of $\tau$, whereas a larger prior variance of $\gamma$ might produce better results.

However, the true outcome equation error variance is likely to be smaller than that estimated for a linear Tobit-2 model because a non-linear model can explain a greater amount of variation in the outcome. An adjustment that scales down the data-informed value of $S_0$ would be closer in spirit to the prior hyperparameter calibration approach described by \cite{chipman2010bart} for standard BART.

An alternative approach, even more similar to the approach of \cite{chipman2010bart} would be to calibrate $S_0$ so that a particular prior quantile of $ \phi + \gamma^2 $ equals $\hat{\sigma}_{y}^2$. However, we have not obtained a convenient closed form for the prior distribution of $\phi + \gamma^2$. Nonetheless, the prior CDF of $ \phi + \gamma^2 $ can be expressed in terms of the CDF of a generalized Chi-square distribution, which does not have a closed form in general, but can be accurately and quickly approximated.

Note that $\phi + \gamma^2 = \phi + (\gamma_0 + \sqrt(\tau \phi) Z)^2$ where $Z \sim \mathcal{N}(0,1)$. 

Since $Z^2 \sim \chi^2(1)$ and $\frac{n_0}{S_0} \phi \sim \frac{1}{X}$ where $X \sim \chi^2(n_0)$, the distribution is similar to a non-central F distribution.


If $\gamma_0 = 0 $, then $\phi + \gamma^2 = \phi( 1 + \tau Z^2) = \tau S_0 \frac{\frac{1}{\tau} + Z^2}{X}$, where $X \sim \chi^2(n_0)$.

If the variable were instead expressible as $  \tau S_0 \frac{ (\frac{1}{\tau} + Z )^2}{X}$, then it would  be scaled non-central F-distributed, with degrees of freedom $n_1=1, \ n_2 = n_0$, and location $\lambda = \frac{1}{\tau}$.

The CDF of $\phi + \gamma^2$ is $F(c) = \Pr \left(\tau S_0  (\frac{1}{\tau} + Z^2) \frac{1}{X} < c \right) =  \Pr (S_0 + \tau S_0 Z^2 - cX < 0 )  $. This is the CDF of a generalized Chi-square distributed variable evaluated at 0. The parameter values are the weights $(\tau S_0, -c)$, the degrees of freedom $(1, n_0)$, the scale of normal distribution $s=0$, and offset $S_0$. This CDF does not in general have a closed form.

\bigskip

Yet another option would be to calibrate the prior on $\phi$ as in standard BART using the approach of \cite{chipman2010bart}. However, as explained above, this can potentially imply a much larger prior mean for the overall outcome variance $\phi + \gamma^2 $, particularly if $\tau$ or $g_0$ are large, resulting in bad prior calibration.

\subsection{\cite{omori2007efficient} prior calibration}

The prior of \cite{omori2007efficient} implies the following:
$$ E[\gamma^2 + \phi] = \frac{S_0}{n_0 - 2} + G_0 (1 + g_0^2)   $$
We set the following prior parameter values by default: $n_0 = 6$, $g_0 = 0$ or $g_0$ is set to the estimated covariance parameter value from a linear type 2 Tobit model estimated by maximum likelihood, and $S_0$ is calibrated so that $E[ \phi + \gamma^2 ] = \hat{\sigma}_{y}^2$ where $\hat{\sigma}_{y}^2$ is the sample standard deviation of the uncensored outcomes or the estimated outcome variance from a linear type 2 Tobit model estimated by maximum likelihood.
$$ S_0 = (n_0 - 2) ( \hat{\sigma}_{y}^2 - G_0 (1 + g_0^2) ) $$
For the Omori prior, we have not found a closed form or otherwise straightforward to compute expression for PDF or CDF of the implied prior on $\gamma^2 + \phi$. The CDF can be re-written as $\Pr (\phi + \gamma^2 <c) = \Pr(  {G_0} Z^2 + \frac{S_0}{X} < c) = \Pr(  X ( G_0 Z^2  - c ) < - S_0) $, where $X \sim \chi^2 (n_0)$. Therefore the CDF of $\gamma^2 + \phi$ can be expressed as a CDF of the product of a gamma-distributed variable and a shifted gamma distributed variable. Therefore, the results for the product of shifted generalized gamma distributed variables from \cite{rathie2013distribution} can be applied to obtain the PDF of $X ( G_0 Z^2  - c )$ . \cite{yilmaz2010outage} obtain the PDF and CDF for the more specific case of a product of (not generalized) shifted Gamma distributed variables, but the results of \cite{yilmaz2010outage} require non-negative shifts, and we require a CDF for a product with a negative shift of $-c$. The product of shifted gamma distributions is also a limiting case of the product of variance-gamma distributed variables \citep{gaunt2024cumulative}.

The PDF of $\phi^2 + \gamma^2$ can be expressed in integral form. Begin with the joint distribution of $(\gamma, \phi)$
$$ p(\gamma, \phi) = \frac{1}{\sqrt{2 \pi G_0} } \exp \left( - \frac{(\gamma - \gamma_0)^2}{2 G_0} \right) \frac{\left( \frac{S_0}{2} \right)^{\frac{n_0}{2}} }{\Gamma \left( \frac{n_0}{2} \right)} \phi^{-\left( \frac{n_0+1}{2} \right)} \exp \left( - \frac{ \left( \frac{S_0}{2} \right) }{\phi}  \right) $$
Assuming $\gamma_0= 0$ and applying the change of variable $(\gamma^2) = \gamma^2$, noting that $\gamma = (\gamma^2)^{\frac{1}{2}}$ and $\frac{\partial \gamma}{\partial (\gamma^2)} = \frac{1}{2} \gamma = \frac{1}{2} (\gamma^2)^{-\frac{1}{2}}$.
$$ p(\gamma^2, \phi) = \frac{1}{\sqrt{2 \pi G_0} } \exp \left( - \frac{ \gamma^2}{2 G_0} \right) \frac{\left( \frac{S_0}{2} \right)^{\frac{n_0}{2}} }{\Gamma \left( \frac{n_0}{2} \right)} \phi^{-\left( \frac{n_0+1}{2} \right)} \exp \left( - \frac{ \left( \frac{S_0}{2} \right) }{\phi}  \right) \frac{1}{2} (\gamma^2)^{-\frac{1}{2}} $$
Applying a second change of variable $x = \gamma^2 + \phi $, noting that $\gamma^2 = x - \phi $ and $\frac{\partial \gamma^2}{\partial x} = 1$
$$ p(x, \phi) = \frac{1}{2\sqrt{2 \pi G_0} } \exp \left( - \frac{ (x - \phi) }{2 G_0} \right) \frac{\left( \frac{S_0}{2} \right)^{\frac{n_0}{2}} }{\Gamma \left( \frac{n_0}{2} \right)} \phi^{-\left( \frac{n_0+1}{2} \right)} \exp \left( - \frac{ \left( \frac{S_0}{2} \right) }{\phi}  \right) \frac{1}{2} (x - \phi)^{-\frac{1}{2}} $$
Therefore the PDF of $x = \gamma^2 + \phi $ is
$$ p(x, \phi) = \frac{1}{2\sqrt{2 \pi G_0} }  \frac{\left( \frac{S_0}{2} \right)^{\frac{n_0}{2}} }{\Gamma \left( \frac{n_0}{2} \right)}  \exp \left( - \frac{ x  }{2 G_0} \right) \int_0^x \phi^{-\left( \frac{n_0+1}{2} \right)} \exp \left( \frac{ \phi }{2 G_0}  - \frac{ \left( \frac{S_0}{2} \right) }{\phi}  \right) \frac{1}{2} (x - \phi)^{-\frac{1}{2}} d \phi $$
The CDF is:
$$ F_x(c) =   \frac{1}{2\sqrt{2 \pi G_0} }  \frac{\left( \frac{S_0}{2} \right)^{\frac{n_0}{2}} }{\Gamma \left( \frac{n_0}{2} \right)} \int_0^c \int_0^x \exp \left( - \frac{ x  }{2 G_0} \right)  \phi^{-\left( \frac{n_0+1}{2} \right)} \exp \left( \frac{ \phi }{2 G_0}  - \frac{ \left( \frac{S_0}{2} \right) }{\phi}  \right) \frac{1}{2} (x - \phi)^{-\frac{1}{2}} d \phi dx $$
Swapping the limits of integration implies the following expressions for the probability
$$ F_x(c) =   \frac{1}{2\sqrt{2 \pi G_0} }  \frac{\left( \frac{S_0}{2} \right)^{\frac{n_0}{2}} }{\Gamma \left( \frac{n_0}{2} \right)} \int_0^c \phi^{-\left( \frac{n_0+1}{2} \right)} \exp \left( \frac{ \phi }{2 G_0}  - \frac{ \left( \frac{S_0}{2} \right) }{\phi}  \right)  \int_0^x \exp \left( - \frac{ x  }{2 G_0} \right)  \frac{1}{2} (x - \phi)^{-\frac{1}{2}} dx d \phi  $$
$$ =   \frac{1}{2}  \frac{\left( \frac{S_0}{2} \right)^{\frac{n_0}{2}} }{\Gamma \left( \frac{n_0}{2} \right)} \int_0^c \phi^{-\left( \frac{n_0+1}{2} \right)} \exp \left(  - \frac{ \left( \frac{S_0}{2} \right) }{\phi}  \right)  \Bigg[ 1  - \frac{1}{\sqrt{\pi} } \Gamma \left(\frac{1}{2}, \frac{1}{2 G_0} (c - \phi)  \right) \Bigg]  d \phi  $$
$$ =   \frac{1}{2 }  \frac{\left( \frac{S_0}{2} \right)^{\frac{n_0}{2}} }{\Gamma \left( \frac{n_0}{2} \right)} \int_0^c \phi^{-\left( \frac{n_0+1}{2} \right)} \exp \left(   - \frac{ \left( \frac{S_0}{2} \right) }{\phi}  \right)   \text{erf} \left( \sqrt{\frac{(c - \phi) }{2 G_0} } \right)   d \phi  $$

The  following expression can be converted to polar coordinates
$$ F_x(c)  =   \frac{1}{2 }  \frac{\left( \frac{S_0}{2} \right)^{\frac{n_0}{2}} }{\Gamma \left( \frac{n_0}{2} \right)} \int_0^c \phi^{-\left( \frac{n_0+1}{2} \right)} \exp \left(   - \frac{ \left( \frac{S_0}{2} \right) }{\phi}  \right)  \int_0^{ \sqrt{\frac{(c - \phi) }{2 G_0} } }  \exp(-t^2 ) d t d \phi  $$
Note that $0 \le \phi + 2 G_0 t^2 \le c$ and the area of the double integral implies $ 0 \le \theta \le \frac{\pi}{2}$. Setting $\sqrt{\phi} = r \cos\theta $ and $\sqrt{2 G_0 t^2 } = r \sin \theta $.
$$ F_x(c) =   \frac{1}{2 }  \frac{\left( \frac{S_0}{2} \right)^{\frac{n_0}{2}} }{\Gamma \left( \frac{n_0}{2} \right)} \int_0^{\frac{\pi}{2}}  \int_0^{ c }  r (r \cos \theta)^{-(n_0+1 )} \exp \left(   - \frac{ \left( \frac{S_0}{2} \right) }{r^2 \cos^2 \theta }   -r^2 \sin^2 \theta \right) d r d \theta  $$
with substitution $u=r^2$, the integral with respect to $r$ can be written as a generalized incomplete gamma function. However, this would create a complicated integral with respect to $\theta$.

\subsection{Implied prior on $\rho$ and dependence on $\tau$ for \cite{van2011bayesian} prior}

In this section, we derive the prior on $\rho$ implied by the \cite{van2011bayesian} prior on $(\phi, \gamma)$ specified by \cite{van2011bayesian}, and discuss the impact of the $\tau$ parameter. The prior introduced by \cite{van2011bayesian} was motivated by the claim that the prior introduced by \cite{omori2007efficient} places most mass of the implied prior on $\rho$ near the boundary points $\pm 1$. In fact, there are hyperparameter values for which this claim is not true, as shown in figure \ref{fig:Omorirhohist}. Nonetheless, a prior for which the scale of the $\gamma$ increases with $\phi$ is appealing.

As explained by \cite{van2005bayesian}, the value $\tau = 0.5$ was chosen because it implies a prior on $\rho$ that is close to uniform for a wide range of values of $\rho$. We show below that high values of $\tau$ result in a very non-uniform distribution that places considerable probability mass close to $\pm 1$, and therefore implies a prior similar to the \cite{omori2007efficient} prior with certain parameter values. \footnote{\cite{van2011bayesian} claims in a footnote that a value $\tau = 0$ corresponds to a prior on $\gamma$ that is independent of $\phi$. However, a value of $\tau = 0$ implies undefined prior variance of $\gamma$ and values of $\tau$ close to zero imply a prior $\gamma$ variance that is very dependent on $\phi$. }

Noting that $\rho = \frac{\gamma}{\sqrt{\gamma^2 + \phi}}$ and $\gamma = \pm \frac{\rho \sqrt{\phi}}{\sqrt{1-\rho^2}}$ and $\frac{\partial \gamma}{\partial \rho} = \phi^{1/2} (1-\rho^2)^{-3/2}$, we obtain the following

$ p(\rho, \phi) = p(\gamma, \phi) \Big|\frac{\partial(\gamma, \phi)}{\partial(\rho, \phi)} \Big| = \frac{1}{ \sqrt{2\pi  \tau} } \frac{\left( \frac{S_0}{2} \right)^{\frac{n_0}{2}} }{  \Gamma \left( \frac{n_0}{2} \right)} \left( \frac{1}{\phi}\right)^{\frac{n_0+3}{2}} \exp \left( - \frac{S_0 + \frac{1}{\tau}(\gamma - g_0)^2}{2 \phi} \right) \phi^{1/2} (1-\rho^2)^{-3/2}$ since  $ \Big|\frac{\partial(\gamma, \phi)}{\partial(\rho, \phi)} \Big| = | \frac{\partial \gamma}{\partial \rho} \frac{\partial \phi}{\partial \phi} - \frac{\partial \gamma}{\partial \phi}\frac{\partial \phi}{\partial \rho}| = |\frac{\partial \gamma}{\partial \rho} . 1 - \frac{\partial \gamma}{\partial \phi}.0| = \phi^{1/2} (1-\rho^2)^{-3/2} $

This implies that
$$ p(\rho) \propto (1-\rho^2)^{-\frac{3}{2}} \exp \left( - \frac{1}{2\tau} \frac{\rho^2}{1-\rho^2} \right)  $$
The modes can be obtained by setting the gradient of the density to zero. First applying the product rule to derive the gradient, we have $f(\rho) = (1-\rho^2)^{-\frac{3}{2}}$, $f'(\rho) = 3\rho (1-\rho^2)^{-\frac{5}{2}}$, $g(\rho) = \exp \left( - \frac{1}{2\tau} \frac{\rho^2}{1-\rho^2} \right)  $, and $g'(\rho) = \frac{\rho}{\tau (1-\rho^2)^2} \exp \left( - \frac{1}{2\tau} \frac{\rho^2}{1-\rho^2} \right)  $. Therefore,
$$ \frac{\partial p(\rho) }{\partial \rho} = f'(\rho)g(\rho) + g'(\rho)f(\rho) =$$
$$ 3\rho (1-\rho^2)^{-\frac{5}{2}} \exp \left( - \frac{1}{2\tau} \frac{\rho^2}{1-\rho^2} \right)  -  \frac{\rho}{\tau (1-\rho^2)^2} \exp \left( - \frac{1}{2\tau} \frac{\rho^2}{1-\rho^2} \right) (1-\rho^2)^{-\frac{3}{2}} =$$
$$ \frac{\left( 3(1+\rho)(1-\rho) - \frac{1}{\tau} \right)\rho}{(1 - \rho^2)^{7/2}}   \exp \left( - \frac{1}{2\tau} \frac{\rho^2}{1-\rho^2} \right)  \overset{!}{=} 0 $$
This implies (for $\rho \neq \pm 1)$) that $\left( 3(1+\rho)(1-\rho) - \frac{1}{\tau} \right)\rho = 0$ and therefore $\rho = 0$ or $ 3(1+\rho)(1-\rho) - \frac{1}{\tau} = 0$. The latter equality is equivalent to $3\tau(1-\rho^2)=1$ which can be rearranged to $\rho^2 = \frac{3\tau - 1}{3\tau}$. Therefore, in addition to the (possibly local) mode at $\rho = 0$, there are also modes at $\rho = \pm \sqrt{\frac{3\tau - 1}{3\tau}} = \pm \sqrt{ 1 - \frac{ 1}{3\tau}} $ when $\tau > 1/3$.

When $\tau = \frac{1}{3}$, the prior is unimodal at $\rho = 0$. When $\tau = \frac{1}{2}$, the prior is bimodal, with modes at $\pm \sqrt{\frac{1}{3}} $, and the probability densities at the modes are close to the probability density at $\rho = 0$. It can be verified that the distribution is close to uniform over a wide range of $\rho$ values, as claimed by \cite{van2011bayesian}. When $\tau = 5$, the prior is much more peaked at the modes $\rho = \pm \sqrt{\frac{9}{10}}$.

It is verified below that $\lim_{\rho \to \pm 1} p(\rho) = 0$:
$$ \lim_{\rho \to 1^{-}} p(\rho) = \frac{\lim_{\rho \to 1^{-}}  (1-\rho^2)^{-\frac{3}{2}}  }{\lim_{\rho \to 1^{-}}  \exp \left( - \frac{1}{2\tau} \frac{\rho^2}{1-\rho^2} \right)  }  = \frac{\infty}{\infty} $$
by L'H\^{o}pital's rule,
$$\lim_{\rho \to 1^{-}} p(\rho) = \lim_{\rho \to 1^{-}} \frac{  3\rho (1-\rho^2)^{-\frac{5}{2}} }{ \frac{\rho}{\tau (1-\rho^2)^2} \exp \left(  \frac{1}{2\tau} \frac{\rho^2}{1-\rho^2} \right) } = 0 $$
by symmetry, $\lim_{\rho \to -1^{+}} p(\rho) = 0$.

The full probability formula for $p(\rho)$, including the constant of integration, is derived as follows:
$$ p(\rho) = \int p(\rho, \phi) d \phi = \frac{1}{\sqrt{2 \pi \tau}} \frac{\left( \frac{S_0}{2} \right)^{\frac{n_0}{2}} }{  \Gamma \left( \frac{n_0}{2} \right)} \int_0^{\infty}  \left( \frac{1}{\phi}\right)^{\frac{n_0+2}{2}} \exp \left( - \frac{S_0}{2\phi} \right) d\phi  (1-\rho^2)^{-\frac{3}{2}} \exp \left( - \frac{1}{2\tau} \frac{\rho^2}{1-\rho^2} \right)     $$
$$ = \frac{1}{\sqrt{2 \pi \tau}} \frac{\left( \frac{S_0}{2} \right)^{\frac{n_0}{2}} }{  \Gamma \left( \frac{n_0}{2} \right)} \frac{  \Gamma \left( \frac{n_0}{2} \right)}{\left( \frac{S_0}{2} \right)^{\frac{n_0}{2}} }  (1-\rho^2)^{-\frac{3}{2}} \exp \left( - \frac{1}{2\tau} \frac{\rho^2}{1-\rho^2} \right) = \frac{1}{\sqrt{2 \pi \tau}} (1-\rho^2)^{-\frac{3}{2}} \exp \left( - \frac{1}{2\tau} \frac{\rho^2}{1-\rho^2} \right)$$
Therefore
$$\int_{-1}^1 p(\rho) d\rho = 1 \Rightarrow  \sqrt{2 \pi \tau} = \int_{-1}^1  (1-\rho^2)^{-\frac{3}{2}} \exp \left( - \frac{1}{2\tau} \frac{\rho^2}{1-\rho^2} \right) d \rho $$

The integral $\int_{-1}^1  (1-\rho^2)^{-\frac{3}{2}} \exp \left( - \frac{1}{2\tau} \frac{\rho^2}{1-\rho^2} \right) d \rho$ can be verified by u-substitution. If $u =  \frac{\rho^2}{1-\rho^2} $ and $\rho  = \pm \sqrt{\frac{u}{u+1}} $, then $d\rho = \pm \frac{1}{2} u^{-\frac{1}{2}} (u+1)^{-\frac{3}{2}} du$. Note that $u = \infty$ when $\rho = -1$, then $u = 0$ when $\rho = 0$, and $ u = \infty$ when $\rho = 1$. Therefore the integral with respect to $u$ can (by symmetry) be written as twice the integral from 0 to $\infty$.
$$ 2 \int_0^{\infty} (u+1)^{-\frac{3}{2}} \exp \left( - \frac{u}{\tau}  \right) \frac{2}{2} u^{-\frac{1}{2}} (u+1)^{-\frac{3}{2}}  du = \int_0^{\infty} u^{-\frac{1}{2}} \exp \left( - \frac{u}{\tau}  \right) \frac{1}{2}  du   = \frac{\Gamma\left(\frac{1}{2}\right) }{ \left( \frac{1}{2\tau} \right)^{\frac{1}{2}} }  =   \frac{\sqrt{\pi} }{ \frac{1}{\sqrt{2\tau}} } = \sqrt{2\pi \tau}  $$

The CDF of $\rho$ can be similarly derived by u-substitution, setting $u =  \frac{\rho^2}{1-\rho^2} $ and noting that $u =  \frac{c^2}{1-c^2} $ when $\rho = c$. Note also that $\rho  = - \sqrt{\frac{u}{u+1}} $ when $\rho < 0$ and $\rho  = \sqrt{\frac{u}{u+1}} $ when $\rho > 0$.
$$ F_{\rho} (c) = 
     \frac{1}{\sqrt{2 \pi \tau}} \int_{-1}^c (1-\rho^2)^{-\frac{3}{2}} \exp \left( - \frac{1}{2\tau} \frac{\rho^2}{1-\rho^2} \right) d \rho 
$$
$$ =  \begin{cases} 0.5 + \frac{1}{2}
\frac{1}{\sqrt{2 \pi \tau}}  \int_{0}^{\frac{c^2}{1- c^2}} u^{-\frac{1}{2}} \exp \left( - \frac{1}{2\tau} u \right) d u  \ \text{ if } \ c \ge 0  \\
\frac{1}{\sqrt{2 \pi \tau}} \int_{\infty}^{\frac{c^2}{1- c^2}} (-1) u^{-\frac{1}{2}} \exp \left( - \frac{1}{2\tau} u \right) d u  \ \text{ if } \ c < 0 \\
\end{cases}  = \begin{cases}
    \Phi\left( \sqrt{\frac{c^2}{\tau (1-c^2)}} \right) \ \text{ if } \ c \ge 0 \\
    1 - \Phi\left( \sqrt{\frac{c^2}{\tau (1-c^2)}} \right) \ \text{ if } \ c < 0
\end{cases}$$

\subsection{Implied prior on $\rho$ for \cite{omori2007efficient} prior}

The \cite{omori2007efficient} prior implies a less convenient density of $\rho$. 

Recall that $p(\rho, \phi) = p(\gamma, \phi) \Big| \frac{\partial \gamma}{\partial \rho} \Big|$. 
If we assume $g_0=0$, this implies:
$$ p(\rho) = \frac{\left(\frac{S_0}{2}\right)^{\frac{n_0}{2}}}{\Gamma \left( \frac{n_0}{2}\right) } \frac{1}{\sqrt{2 \pi G_0}} (1-\rho^2)^{-\frac{3}{2}} \int_0^{\infty} \phi^{- \frac{n_0+1}{2} }  \exp \left( -\frac{S_0}{2} \frac{1}{\phi} - \frac{1}{2G_0} \frac{\rho^2}{(1-\rho^2)} \phi \right) d \phi $$
Then we can apply 3.462.8 from \cite{gradshteyn2014table}: $\int_0^{\infty} x^{\nu-1} \exp\left( - \frac{\beta}{x} - \gamma x \right) dx = 2 \left( \frac{\beta}{\gamma} \right)^{\frac{\nu}{2}} K_{\nu} (2 \sqrt{\beta \gamma}) $ where $K_{\nu}$ is a modified Bessel function of the second kind.
$$ p(\rho) = \frac{\left(\frac{S_0}{2}\right)^{\frac{n_0}{2}}}{\Gamma \left( \frac{n_0}{2}\right) } \frac{1}{\sqrt{2 \pi G_0}} (1-\rho^2)^{-\frac{3}{2}} 2  \Bigg( \frac {\left( \frac{S_0}{2} \right)}{ \left( \frac{1}{2 G_0} \right) \left( \frac{\rho^2}{1 - \rho^2} \right) } \Bigg)^{- \frac{n_0 - 1}{4}}  K_{- \frac{n_0 - 1}{2}} \Bigg( 2 \left(  \frac{S_0}{2} \frac{1}{2 G_0}  \frac{\rho^2}{(1 - \rho^2)} \right)^{\frac{1}{2}}  \Bigg) $$
%
%
$$ \propto  (1-\rho^2)^{- \frac{n_0 + 5}{4} } |\rho|^{ \frac{n_0 - 1}{2}}  K_{- \frac{n_0 - 1}{2}} \Bigg(  \left(  \frac{S_0}{ G_0} \right)^{\frac{1}{2}} \left( \frac{\rho^2}{(1 - \rho^2)} \right)^{\frac{1}{2}}  \Bigg)  $$
Consider the following change of variable: $u =  \frac{\rho^2}{(1 - \rho^2)} $. This implies $\rho = \pm \sqrt{\frac{u}{u+1}}$ and $\frac{\partial \rho }{\partial u}  = \pm  \frac{1}{2} u^{-\frac{1}{2}} (u+1)^{-\frac{3}{2}}$. The variable $u$ has the following PDF:
$$ f_u(u) = f_{\rho}(\rho) \Big| \frac{\partial \rho }{\partial u} \Big| \propto  (1-\rho^2)^{- \frac{n_0 + 5}{4} } |\rho|^{ \frac{n_0 - 1}{2}}  K_{- \frac{n_0 - 1}{2}} \Bigg(  \sqrt{ \left(  \frac{S_0}{ G_0} \right) u }   \Bigg) \times \pm  \frac{1}{2} u^{-\frac{1}{2}} (u+1)^{-\frac{3}{2}} $$
$$   \propto  (u + 1)^{- \frac{n_0 + 5}{4} } (u)^{ \frac{n_0 - 1}{4}} (u + 1)^{ - \frac{n_0 - 1}{4}}     u^{-\frac{1}{2}} (u+1)^{-\frac{3}{2}} K_{- \frac{n_0 - 1}{2}} \Bigg(  \sqrt{ \left(  \frac{S_0}{ G_0} \right) u }   \Bigg) $$
$$   \propto  (u)^{ \frac{n_0 - 3}{4}} K_{- \frac{n_0 - 1}{2}} \Bigg( 2 \sqrt{ \left(  \frac{S_0}{4 G_0} \right) u }   \Bigg) $$

Therefore $u$ is $K-$distributed with parameters $\alpha = \frac{1}{2}$, $\beta = \frac{n_0}{2}$, and $\mu = \frac{n_o G_0}{S_0}$. This is a special case of a variance-gamma distribution. Alternatively, it can be shown that $x = \sqrt{u}$ is variance-gamma distributed. Note that $\frac{\partial u}{\partial x} = 2x $ and $u = x^2$. Therefore
$$ p(x) \propto x^{\frac{n_0-3}{2}} K_{\frac{1}{2} - \frac{n_0}{2}} \Bigg( \sqrt{\frac{S_0}{ G_0} } |x| \Bigg) 2x 
= x^{\frac{n_0-1}{2}} K_{\frac{n_0-1}{2}} \Bigg( \sqrt{\frac{S_0}{ G_0} } |x| \Bigg) $$
For $x>0$, this is the PDF of a variance-gamma distribution with parameters $r= n_0$, $\theta = 0 $, $\mu = 0$, and $\sigma = \sqrt{\frac{G_0}{ S_0} } $. \cite{gaunt2024cumulative} and \cite{jankov2021new} have shown that, for these parameter values, the CDF is:\footnote{There are a number of alternative expressions for this CDF. See \cite{gaunt2024cumulative, fischer2023variance, jankov2021new}. }
$$F(x) = \frac{1}{2} + \frac{x - \mu}{2\sigma} \Bigg[ K_{\frac{r-1}{2}} \Bigg( \frac{|x - \mu|}{\sigma} \Bigg)  L_{\frac{r-3}{2}} \Bigg( \frac{|x - \mu|}{\sigma} \Bigg) + L_{\frac{r-1}{2}} \Bigg( \frac{|x - \mu|}{\sigma} \Bigg)  K_{\frac{r-3}{2}} \Bigg( \frac{|x - \mu|}{\sigma} \Bigg)  \Bigg] $$
where $L_\nu(\cdot)$ is a modified Struve function of the first kind.

This allows us to obtain the probability that $\rho$ is in a certain range as follows:
$\Pr(-c < \rho < c) = \Pr(\rho^2 < c^2) = \Pr(\sqrt{ \frac{\rho^2}{(1 - \rho^2)} }  <  \sqrt{\frac{c^2}{(1 - c^2)} } ) = \Pr(x  <  \sqrt{\frac{c^2}{(1 - c^2)} } )$.

Furthermore
$$F(|x|) =  \frac{|x - \mu|}{\sigma} \Bigg[ K_{\frac{r-1}{2}} \Bigg( \frac{|x - \mu|}{\sigma} \Bigg)  L_{\frac{r-3}{2}} \Bigg( \frac{|x - \mu|}{\sigma} \Bigg) + L_{\frac{r-1}{2}} \Bigg( \frac{|x - \mu|}{\sigma} \Bigg)  K_{\frac{r-3}{2}} \Bigg( \frac{|x - \mu|}{\sigma} \Bigg)  \Bigg] $$

$$ F\left(\sqrt{ \frac{\rho^2}{(1 - \rho^2)} } \right) =  \sqrt{\frac{S_0}{G_0}}\sqrt{ \frac{\rho^2}{(1 - \rho^2)} }\Bigg[ K_{\frac{r-1}{2}} \Bigg( \Bigg|  \sqrt{\frac{S_0}{G_0}} \sqrt{ \frac{\rho^2}{(1 - \rho^2)} } \Bigg| \Bigg)  L_{\frac{r-3}{2}} \Bigg( \Bigg| \sqrt{\frac{S_0}{G_0}} \sqrt{ \frac{\rho^2}{(1 - \rho^2)} } \Bigg| \Bigg) + $$
$$ L_{\frac{r-1}{2}} \Bigg( \Bigg|  \sqrt{\frac{S_0}{G_0}}\sqrt{ \frac{\rho^2}{(1 - \rho^2)} } \Bigg| \Bigg)  K_{\frac{r-3}{2}} \Bigg( \Bigg| \sqrt{\frac{S_0}{G_0}} \sqrt{ \frac{\rho^2}{(1 - \rho^2)} } \Bigg| \Bigg)  \Bigg] $$
This implies that
$$ F_{\rho}(\rho) - F_{\rho}(-\rho) = \sqrt{\frac{S_0}{G_0}}\sqrt{ \frac{\rho^2}{(1 - \rho^2)} }\Bigg[ K_{\frac{r-1}{2}} \Bigg( \Bigg|  \sqrt{\frac{S_0}{G_0}} \sqrt{ \frac{\rho^2}{(1 - \rho^2)} } \Bigg| \Bigg)  L_{\frac{r-3}{2}} \Bigg( \Bigg| \sqrt{\frac{S_0}{G_0}} \sqrt{ \frac{\rho^2}{(1 - \rho^2)} } \Bigg| \Bigg) + $$
$$ L_{\frac{r-1}{2}} \Bigg( \Bigg|  \sqrt{\frac{S_0}{G_0}}\sqrt{ \frac{\rho^2}{(1 - \rho^2)} } \Bigg| \Bigg)  K_{\frac{r-3}{2}} \Bigg( \Bigg| \sqrt{\frac{S_0}{G_0}} \sqrt{ \frac{\rho^2}{(1 - \rho^2)} } \Bigg| \Bigg)  \Bigg] $$

By symmetry of the $\rho$ prior, $F(\rho) = \frac{1}{2}(1 - F_{\rho}(\rho) + F_{\rho}(-\rho) ) + F_{\rho}(\rho) - F_{\rho}(-\rho) = \frac{1}{2} + \frac{1}{2}(F_{\rho}(\rho) - F_{\rho}(-\rho)) $
Therefore
$$ F_{\rho}(\rho) = \frac{1}{2} +  \frac{1}{2} sign(\rho) \sqrt{\frac{S_0}{G_0}}\sqrt{ \frac{\rho^2}{(1 - \rho^2)} }\Bigg[ K_{\frac{r-1}{2}} \Bigg( \Bigg|  \sqrt{\frac{S_0}{G_0}} \sqrt{ \frac{\rho^2}{(1 - \rho^2)} } \Bigg| \Bigg)  L_{\frac{r-3}{2}} \Bigg( \Bigg| \sqrt{\frac{S_0}{G_0}} \sqrt{ \frac{\rho^2}{(1 - \rho^2)} } \Bigg| \Bigg) + $$
$$ L_{\frac{r-1}{2}} \Bigg( \Bigg|  \sqrt{\frac{S_0}{G_0}}\sqrt{ \frac{\rho^2}{(1 - \rho^2)} } \Bigg| \Bigg)  K_{\frac{r-3}{2}} \Bigg( \Bigg| \sqrt{\frac{S_0}{G_0}} \sqrt{ \frac{\rho^2}{(1 - \rho^2)} } \Bigg| \Bigg)  \Bigg] $$

Alternatively, the CDF can be derived by noting that $\frac{1}{\rho^2} = \frac{\gamma^2 + \phi}{\gamma^2} = 1 + \frac{\phi}{\gamma^2} = 1 + \frac{1}{\left( \frac{\gamma^2}{\phi}\right)} $. Since $\gamma^2 \sim G_0 \chi^2(1)$ and $\frac{1}{\phi} \sim \frac{1}{S_0} \chi^2(n_0)$, it can be observed that $\frac{\gamma^2}{\phi} \sim \frac{G_0}{S_0} \chi^2(1) \chi^2(n_0) $. This implies that $ W =\frac{S_0}{G_0} \frac{\gamma^2}{\phi} = \chi^2(1) \chi^2(n_0) $ is K-distributed with $\alpha = \frac{n_0}{2}$, $\beta = \frac{1}{2}$ and $\mu = n_0 $ \citep{wells1962distribution, joarder2013statistical}. Moreover $z = \sqrt{w}$ is variance-gamma distributed with the density
$$f(z) \propto z^{\frac{n_0 - 1}{2}} K_{\frac{n_0 - 1}{2}}(z) $$
Note that $\Pr(-c < \rho < c) = \Pr \left( \sqrt{\frac{S_0}{G_0} \frac{\rho^2}{(1 - \rho^2)}} < \sqrt{\frac{S_0}{G_0} \frac{c^2}{(1 - c^2)}} \right) = \Pr \left( |z| < \sqrt{\frac{S_0}{G_0} \frac{c^2}{(1 - c^2)}} \right)$. By symmetry about zero, we can conclude that $\Pr(\rho < c) = \frac{1}{2} \Pr(-c < \rho < c) + \frac{1}{2} $ for $ c > 0 $  and $ \Pr(\rho < c) = \frac{1}{2} + sign(c) \frac{1}{2}  F_{|z|} \left( \sqrt{\frac{S_0}{G_0} \frac{c^2}{(1 - c^2)} } \right)  $. Applying the result for the CDF of a variance-gamma distribution from \cite{gaunt2024cumulative} we obtain the same expression for $F_{\rho}(\rho)$ as that derived above. 

It is a difficult task to obtain the modal value(s) of the prior for $\rho$ and the conditions under which it is bimodal. Perhaps bounds for the mode can be obtained by differentiating the PDF for $\rho$, applying the identity $\frac{\partial K_v(x)}{\partial x} = \frac{v}{x} K_v(x) - K_{v+1}(x)$, and considering bounds (A.70) and (A.71) on $\frac{K_v(x)}{K_{v+1}(x)} $ from \cite{fischer2023variance}. However, there might not be a convenient expression for the exact mode.

\begin{figure}
\begin{subfigure}{.33\textwidth}
  \centering
  \includegraphics[width=.8\linewidth]{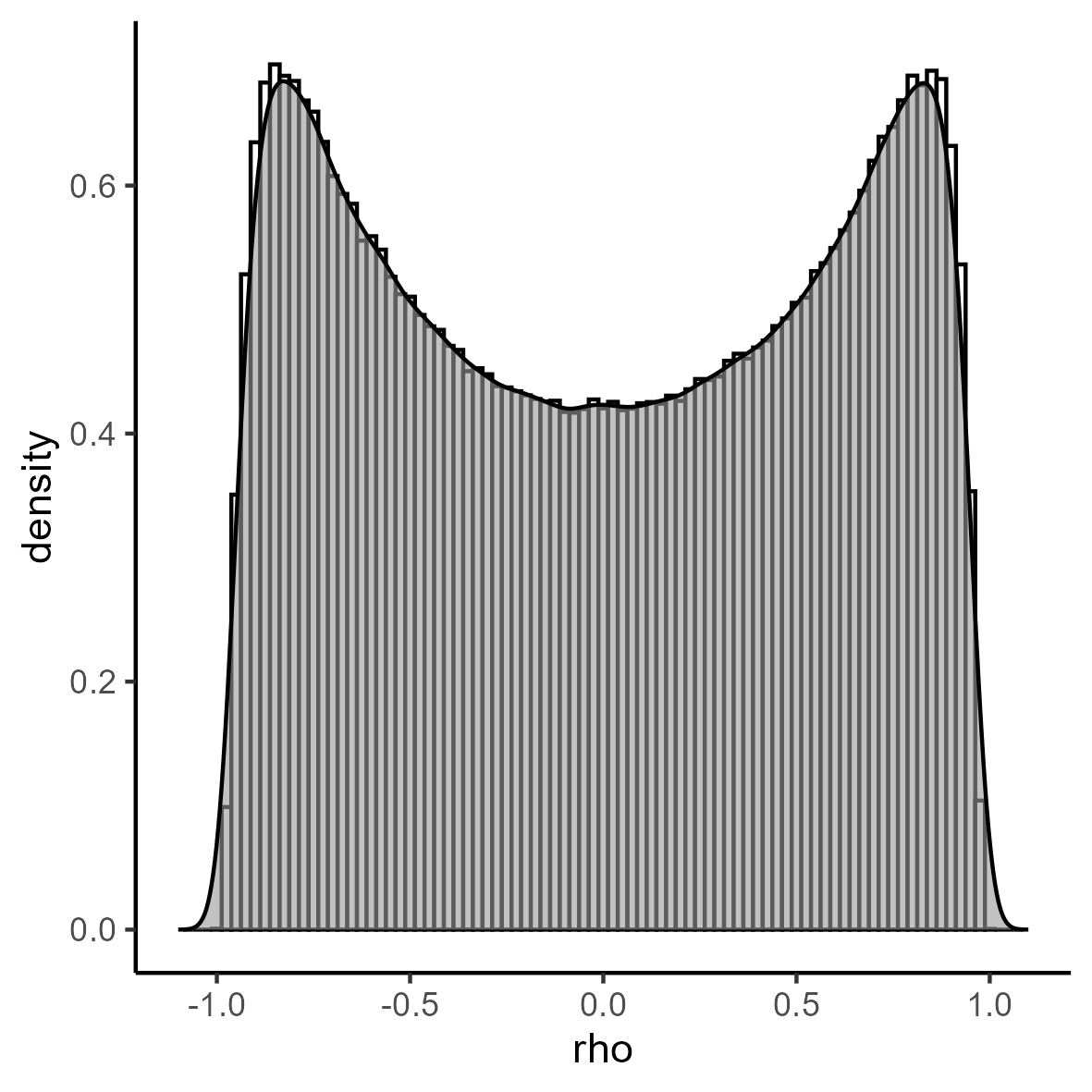}
  \caption{$n_0=6$, $G_0 = 6$, $S_0=5$.}
  \label{fig:sfig1}
\end{subfigure}%
\begin{subfigure}{.33\textwidth}
  \centering
  \includegraphics[width=.8\linewidth]{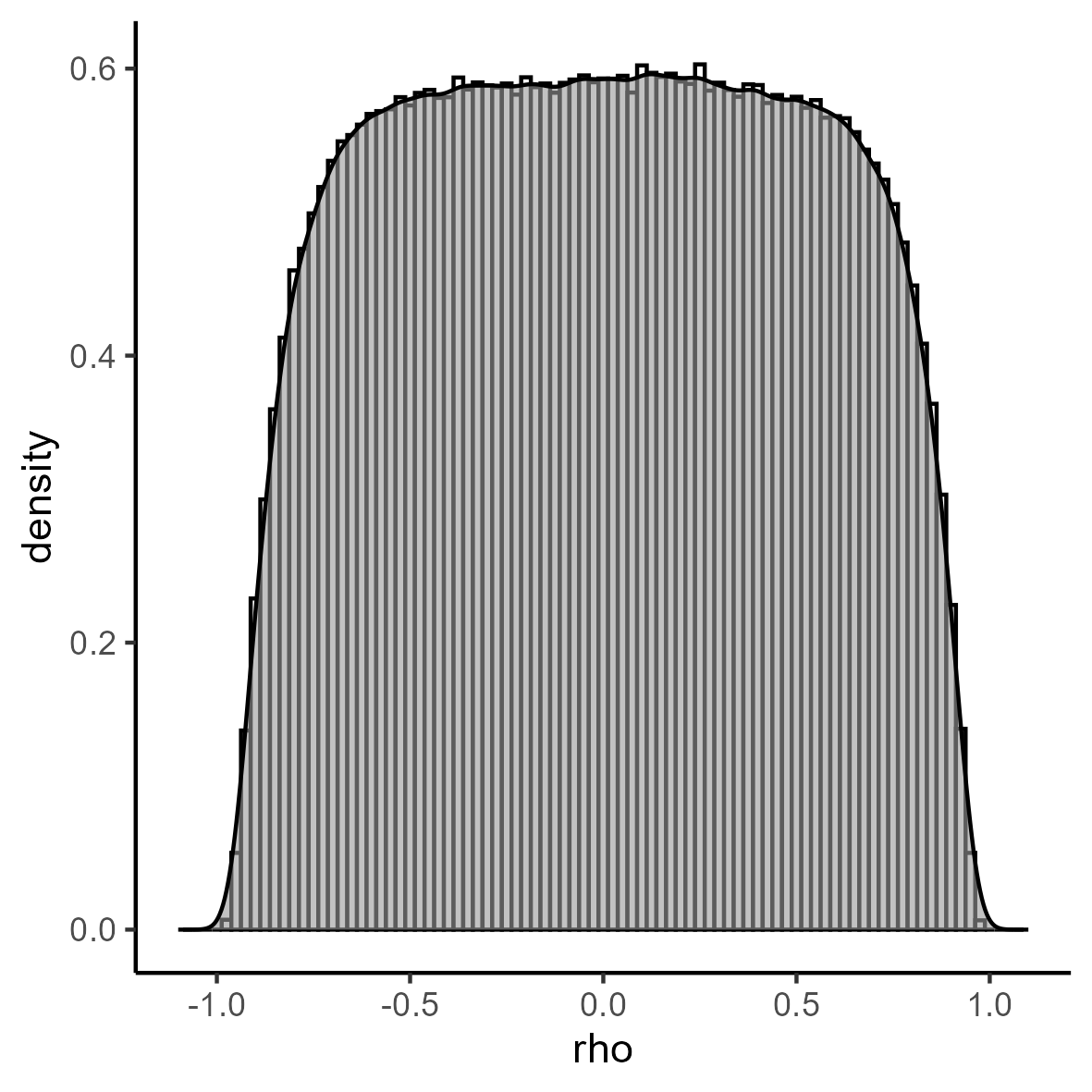}
  \caption{$n_0=6$, $G_0 = 6$, $S_0=10$.}
  \label{fig:sfig2}
\end{subfigure}
\begin{subfigure}{.33\textwidth}
  \centering
  \includegraphics[width=.8\linewidth]{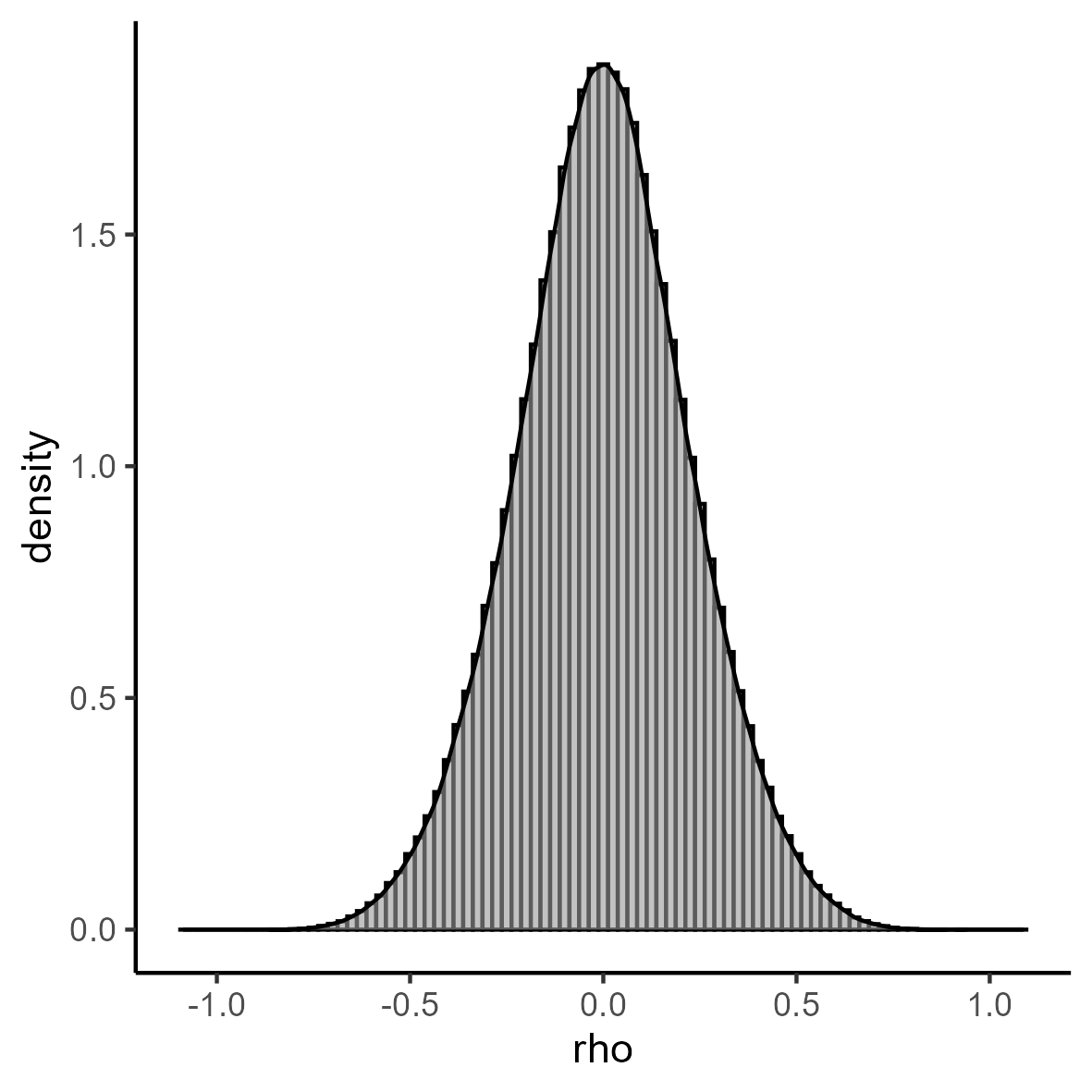}
  \caption{$n_0=6$, $G_0 = 6$, $S_0=100$.}
  \label{fig:sfig3}
\end{subfigure}

\begin{subfigure}{.33\textwidth}
  \centering
  \includegraphics[width=.8\linewidth]{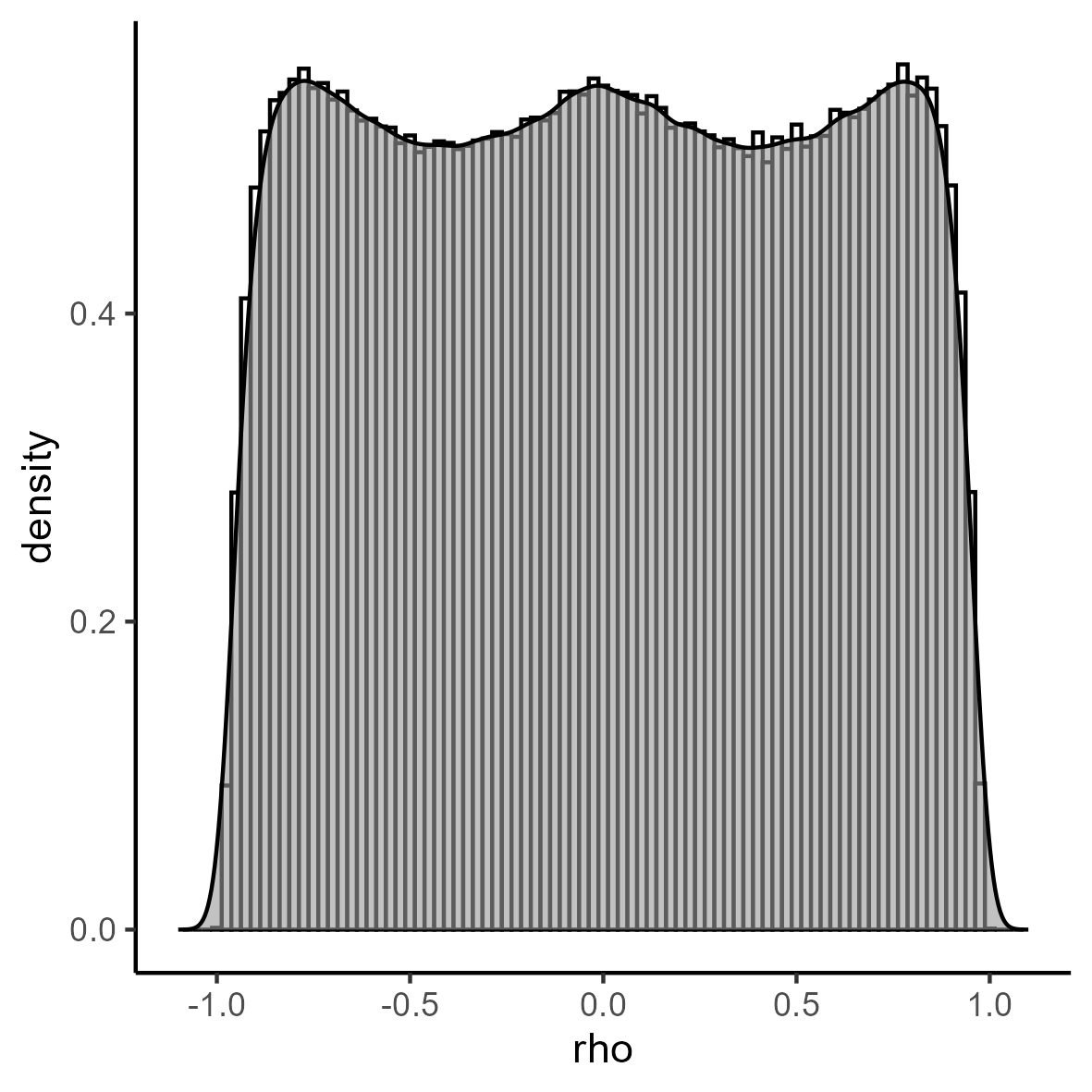}
  \caption{$n_0=3$, $G_0 = 60$, $S_0=80$.}
  \label{fig:sfig1}
\end{subfigure}%
\begin{subfigure}{.33\textwidth}
  \centering
  \includegraphics[width=.8\linewidth]{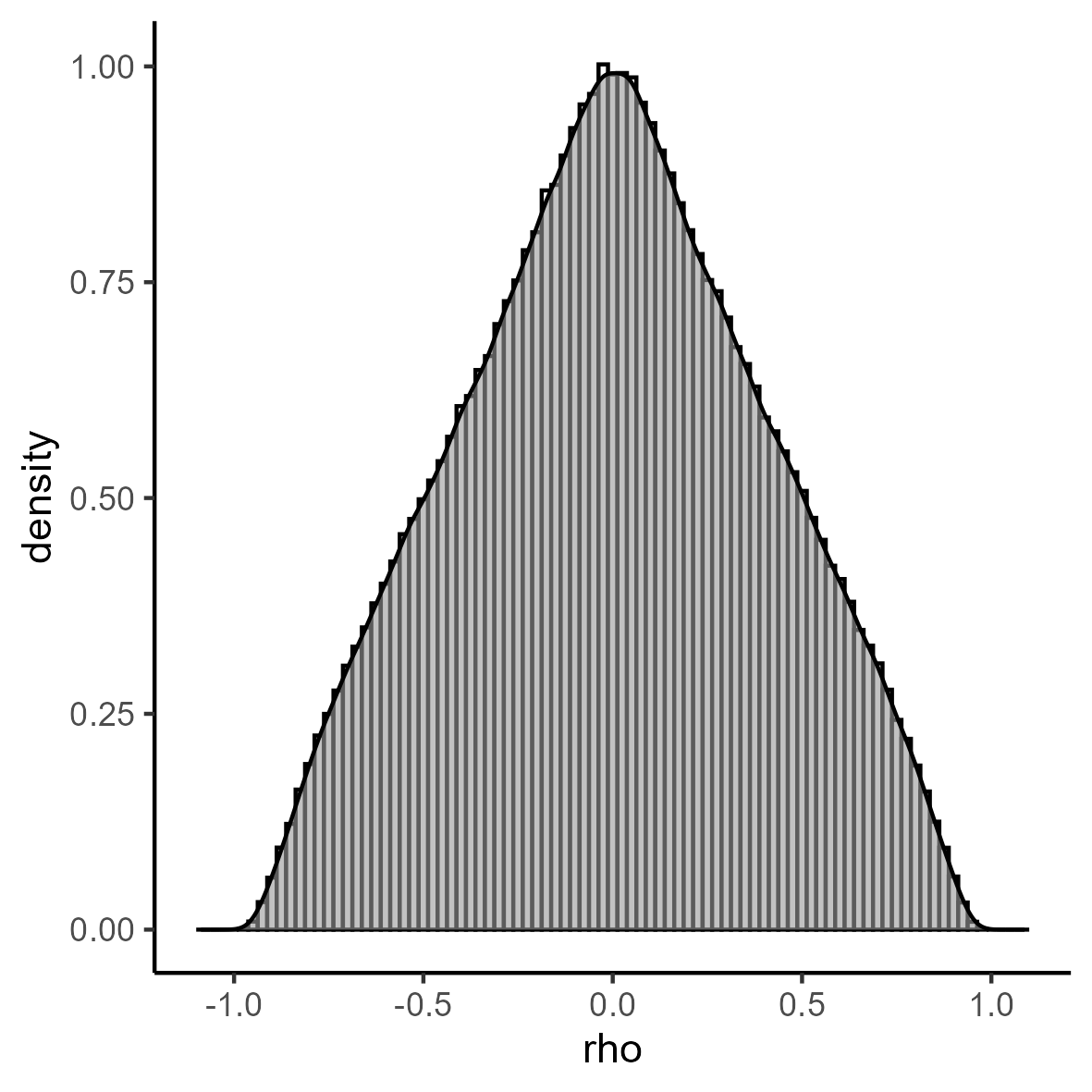}
  \caption{$n_0=3$, $G_0 = 1$, $S_0=10$.}
  \label{fig:sfig2}
\end{subfigure}
\begin{subfigure}{.33\textwidth}
  \centering
  \includegraphics[width=.8\linewidth]{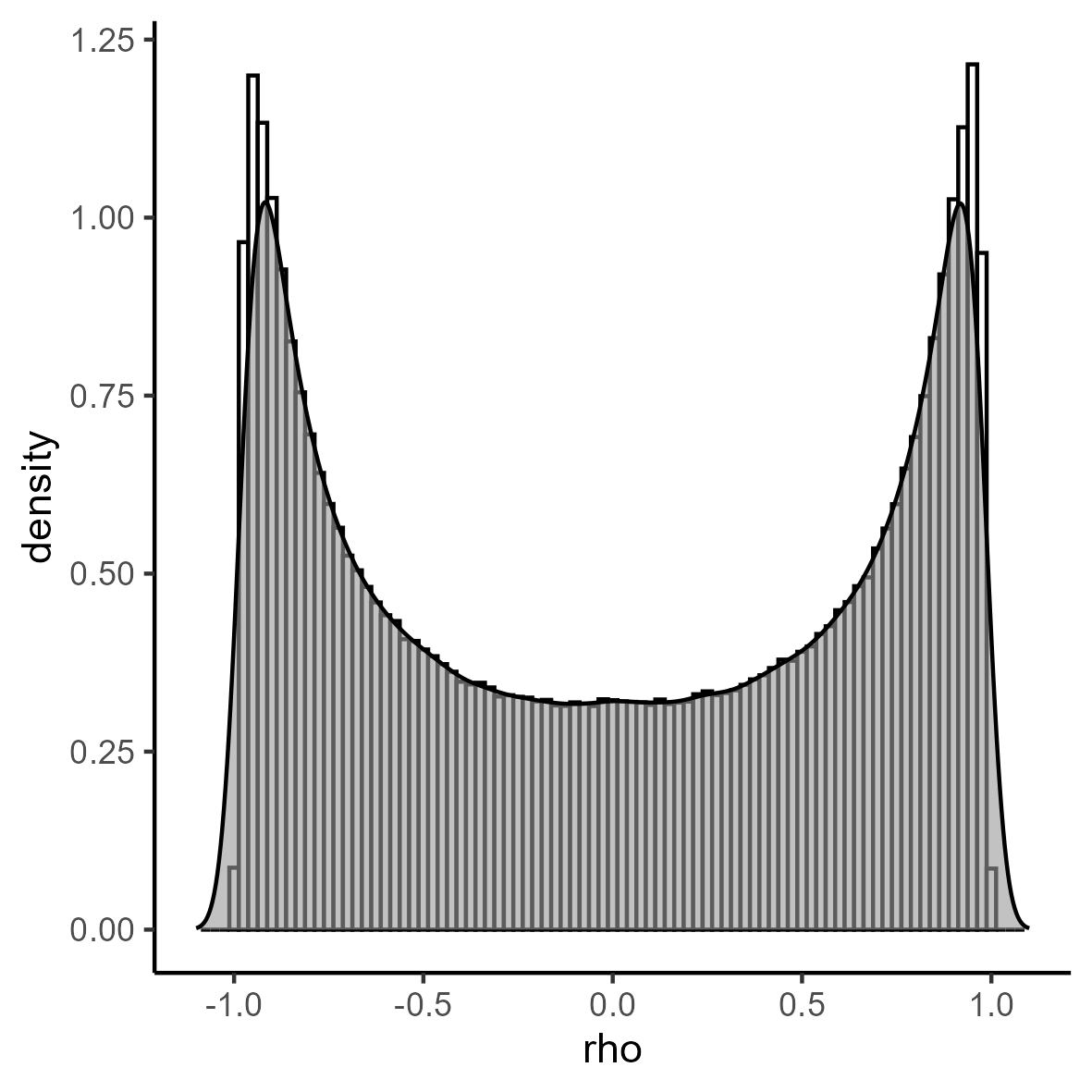}
  \caption{$n_0=3$, $G_0 = 6$, $S_0=6$.}
  \label{fig:sfig3}
\end{subfigure}

\begin{subfigure}{.33\textwidth}
  \centering
  \includegraphics[width=.8\linewidth]{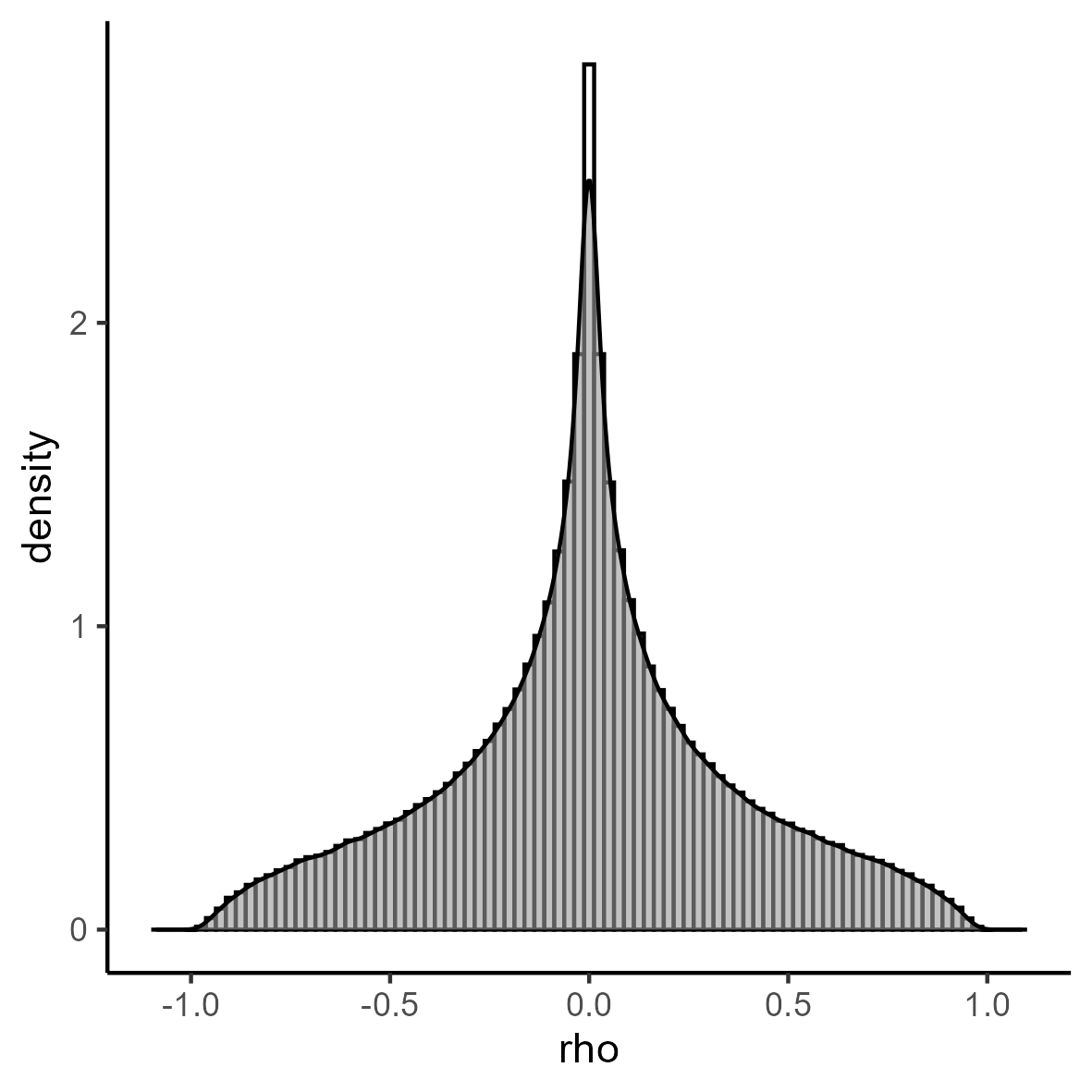}
  \caption{$n_0=1$, $G_0 = 15$, $S_0=50$.}
  \label{fig:sfig1}
\end{subfigure}%
\begin{subfigure}{.33\textwidth}
  \centering
  \includegraphics[width=.8\linewidth]{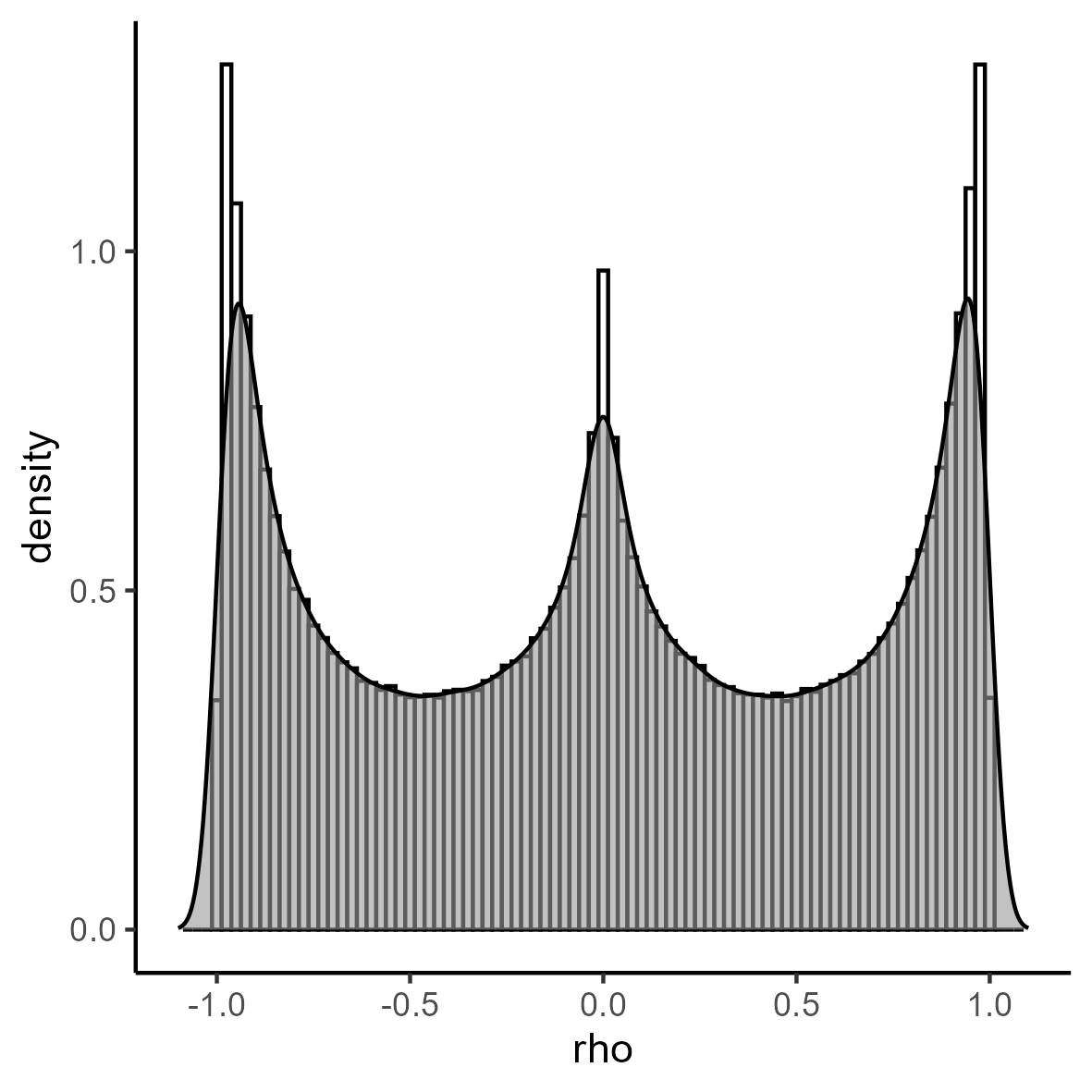}
  \caption{$n_0=1$, $G_0 = 20$, $S_0=5$.}
  \label{fig:sfig2}
\end{subfigure}
\begin{subfigure}{.33\textwidth}
  \centering
  \includegraphics[width=.8\linewidth]{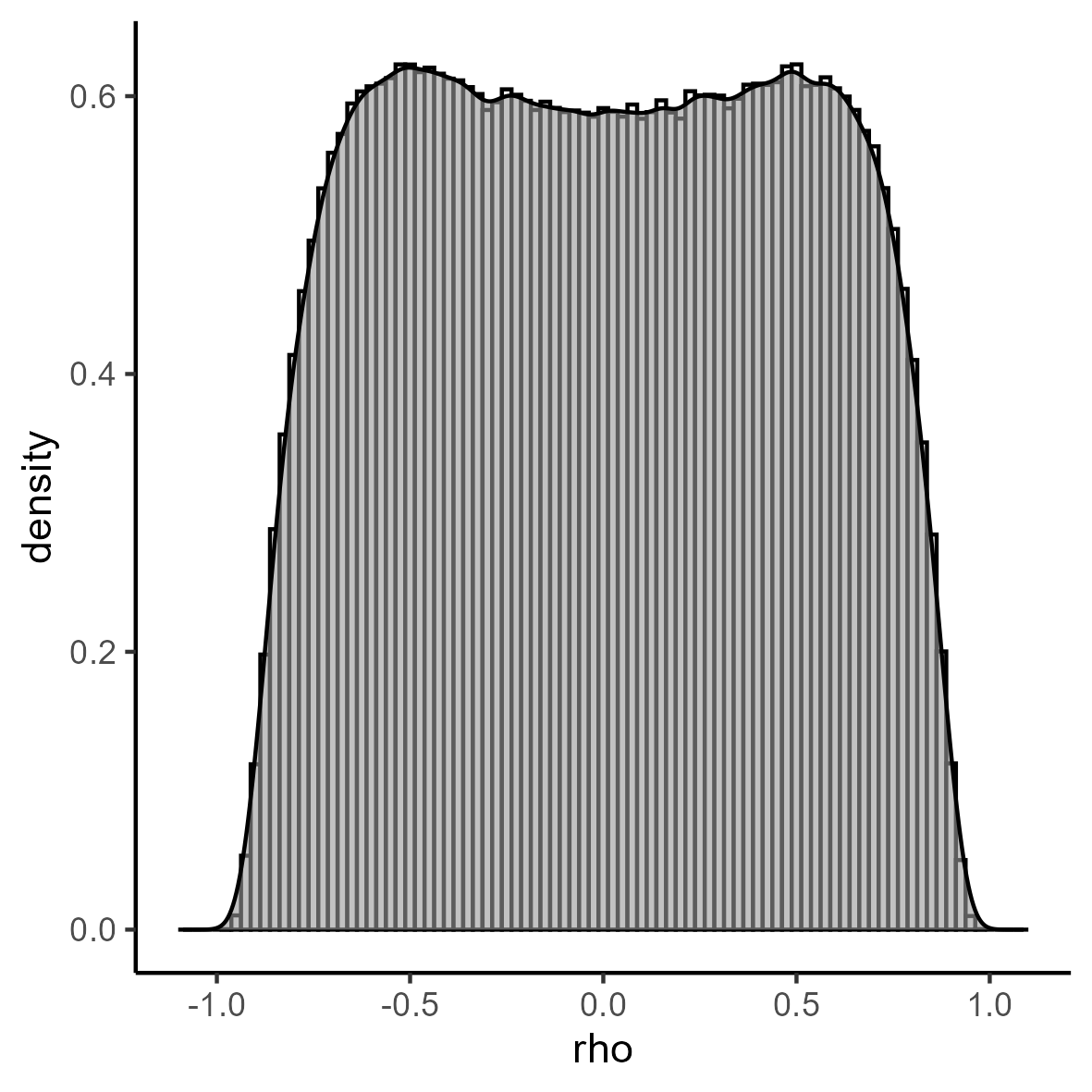}
  \caption{$n_0=10$, $G_0 = 0.5$, $S_0=20$.}
  \label{fig:sfig3}
\end{subfigure}
\caption{Histograms and empirical densities of $1,000,000$ draws of $\rho$ from \cite{omori2007efficient} prior.}
\label{fig:Omorirhohist}
\end{figure}

A wide variety of possible prior probability densities for $\rho$ implied by the \cite{omori2007efficient} prior are displayed in Figure \ref{fig:Omorirhohist}. Many possible prior distributions are notably different from the distribution concentrated near $\pm1$ claimed by \cite{van2011bayesian} (perhaps when referring only to particular parameter values).

\FloatBarrier

\newpage

\section{Full Conditional Derivations for \cite{van2011bayesian} prior}

\subsection{Full Conditional of $\phi$}

Let the prior on $\phi$ be $\mathcal{IG}(c_0,d_0)$ and let the prior $\gamma |\phi$ be $\mathcal{N}(g_0,\tau \phi) $. Note that this dependence also implies that $p(\phi|\gamma) \neq p(\phi)$ 

$$ p(\phi|\gamma)  = p(\gamma|\phi) p(\phi)/p(\gamma) $$
$$ \propto \frac{1}{\sqrt{2\pi \tau \phi}} \exp \left(-\frac{1}{2} \frac{(\gamma - g_0)^2}{\tau \phi} \right) \frac{d_0^{c_0}}{\Gamma(c_0)} \left( \frac{1}{\phi} \right)^{c_0+1} \exp\left(- \frac{d_0}{\phi}\right) $$
$$ \propto \left(\frac{1}{\phi}\right)^{c_0+\frac{1}{2}+1}  \exp \Bigg( -\frac{1}{\phi} \left( d_0 + \frac{(\gamma - g_0)^2}{2\tau} \right)  \Bigg) $$
Therefore $\phi|\gamma \sim \mathcal{IG}\left(c_0+\frac{1}{2} , d_0 + \frac{(\gamma - g_0)^2}{2\tau} \right)$ .

Consider the full conditional distribution of $\phi$. Let $\tilde{y}_i = y_i - f_y (\bm{x}_i) $ and let $\tilde{z}_i = z_i - f_z (\bm{w}_i) $.

$$p(\phi| \bm{\tilde{y}}, \bm{\tilde{z}}, \gamma) \propto  p(\bm{\tilde{y}}, \bm{\tilde{z}}| \phi, \gamma)  p(\phi|  \gamma) $$
$$ = \left( \frac{1}{\sqrt{2\pi\phi}} \right)^n \prod_{i=1}^n \exp \Bigg(  - \frac{1}{2 \frac{\phi}{\gamma^2+\phi}} \left[ \tilde{z}_i^2 - 2 \frac{\gamma}{\gamma^2+\phi} \tilde{z}_i \tilde{y}_i  + \frac{\tilde{y}_i^2}{\gamma^2+\phi} \right]  \Bigg)  \frac{(d_0 + \frac{(\gamma - g_0)^2}{2\tau} )^{c_0+\frac{1}{2}}}{\Gamma(c_0+\frac{1}{2})} \left( \frac{1}{\phi}\right)^{c_0+\frac{1}{2}+1} \times $$
$$ \exp \left( - \frac{(d_0 + \frac{(\gamma - g_0)^2}{2\tau} )}{\phi} \right)  $$
$$ = \left( \frac{1}{\sqrt{2\pi\phi}} \right)^n  \exp \Big(  - \frac{\sum_{i=1}^n \tilde{z}_i^2}{2 \frac{\phi}{\gamma^2+\phi}}    + \frac{\gamma}{\phi} \sum_{i=1}^n \tilde{z}_i \tilde{y}_i  + \frac{\sum_{i=1}^n \tilde{y}_i^2}{2\phi}  - \frac{(d_0 + \frac{(\gamma - g_0)^2}{2\tau} )}{\phi} \Big)  \frac{(d_0 + \frac{(\gamma - g_0)^2}{2\tau} )^{c_0+\frac{1}{2}}}{\Gamma(c_0+\frac{1}{2})} \left( \frac{1}{\phi}\right)^{c_0+\frac{1}{2}+1} $$
$$ = \left( \frac{1}{\sqrt{2\pi}} \right)^n  \frac{(d_0 + \frac{\gamma^2}{2\tau} )^{c_0+\frac{1}{2}}}{\Gamma(c_0+\frac{1}{2})} \left( \frac{1}{\phi}\right)^{\frac{n}{2} + c_0+\frac{1}{2}+1} \exp \Big(  - \frac{\gamma^2 \sum_{i=1}^n \tilde{z}_i^2}{2 \phi}- \frac{ \sum_{i=1}^n \tilde{z}_i^2}{2 }    + \frac{\gamma}{\phi} \sum_{i=1}^n \tilde{z}_i \tilde{y}_i  + \frac{\sum_{i=1}^n \tilde{y}_i^2}{2\phi}  - \frac{(d_0 + \frac{(\gamma - g_0)^2}{2\tau} )}{\phi} \Big)   $$
$$ \propto \left( \frac{1}{\phi}\right)^{\frac{n}{2} + c_0+\frac{1}{2}+1} \exp \Bigg(  -\frac{1}{\phi} \left( (d_0 + \frac{(\gamma - g_0)^2}{2\tau} ) + \frac{1}{2} \sum_{i=1}^n (\tilde{y}_i -\tilde{z}_i \gamma  )^2  \right) \Bigg)   $$
Therefore 
$$ \phi| \bm{\tilde{y}}, \bm{\tilde{z}}, \gamma \sim \mathcal{IG} \left( \frac{n}{2}+c_0+\frac{1}{2},  d_0 + \frac{(\gamma - g_0)^2}{2\tau}  + \frac{1}{2} \sum_{i=1}^n (\tilde{y}_i -\tilde{z}_i \gamma  )^2\right) $$
where $c_0 = \frac{n_0}{2}$ and $d_0 = \frac{S_0}{2}$.

For the prior of \cite{omori2007efficient},  $\gamma |\phi \sim \mathcal{N}(g_0, G_0) $, $p(\phi|\gamma) =p(\phi)$,  and $$ \phi| \bm{\tilde{y}}, \bm{\tilde{z}}, \gamma \sim \mathcal{IG} \left( \frac{n}{2}+c_0+\frac{1}{2},  d_0  + \frac{1}{2} \sum_{i=1}^n (\tilde{y}_i -\tilde{z}_i \gamma  )^2\right) $$

\subsection{Full Conditional of $\gamma$}
\noindent Similarly, the full conditional of $\gamma$ can be derived as follows.
$$p(\gamma| \bm{\tilde{y}}, \bm{\tilde{z}}, \phi) \propto  p(\bm{\tilde{y}}, \bm{\tilde{z}}| \phi, \gamma)  p(\gamma|  \phi) $$
$$ = \left( \frac{1}{\sqrt{2\pi\phi}} \right)^n \prod_{i=1}^n \exp \Bigg(  - \frac{1}{2 \frac{\phi}{\gamma^2+\phi}} \left[ \tilde{z}_i^2 - 2 \frac{\gamma}{\gamma^2+\phi} \tilde{z}_i \tilde{y}_i  + \frac{\tilde{y}_i^2}{\gamma^2+\phi} \right]  \Bigg)  \frac{1}{ \sqrt{2\pi \tau\phi} } \exp \left( - \frac{1}{2\tau \phi} (\gamma-\gamma_0)^2 \right)  $$
$$ \propto \exp \Bigg(  - \frac{1 }{2 \phi} \left[ (\frac{1}{\tau} + \sum_{i=1}^n \tilde{z}_i^2) \gamma^2  -  \gamma \left( \frac{2\gamma_0}{\tau} + 2\sum_{i=1}^n  \tilde{z}_i \tilde{y}_i \right)   \right]   \Bigg)    $$
Now complete the square. In general $ax^2 + bx + c = a(x+\frac{b}{2a})^2+c - \frac{b^2}{4a} $ and here we have $x=\gamma$, $a =\frac{1}{\tau} + \sum_{i=1}^n \tilde{z}_i^2$, $b = -\frac{2\gamma_0}{\tau} - 2\sum_{i=1}^n  \tilde{z}_i \tilde{y}_i $, $c=0$. Therefore
$$ (\frac{1}{\tau} + \sum_{i=1}^n \tilde{z}_i^2) \gamma^2  -  \gamma \left( \frac{2\gamma_0}{\tau} + 2\sum_{i=1}^n  \tilde{z}_i \tilde{y}_i \right) 
 = \left(\frac{1}{\tau} + \sum_{i=1}^n \tilde{z}_i^2 \right) \Bigg(\gamma - 2\frac{\frac{\gamma_0}{\tau} + \sum_{i=1}^n  \tilde{z}_i \tilde{y}_i }{2(\frac{1}{\tau} + \sum_{i=1}^n \tilde{z}_i^2) } \Bigg)^2  - \frac{4 
 (  \frac{2\gamma_0}{\tau} + 2\sum_{i=1}^n  \tilde{z}_i \tilde{y}_i )^2  }{4 (\frac{1}{\tau} + \sum_{i=1}^n \tilde{z}_i^2 )  } $$

$$p(\gamma| \bm{\tilde{y}}, \bm{\tilde{z}}, \phi) \propto exp \Bigg(  - \frac{1 }{2 \phi} \left[ \left(\frac{1}{\tau} + \sum_{i=1}^n \tilde{z}_i^2 \right) \Bigg(\gamma - \frac{\frac{\gamma_0}{\tau} + \sum_{i=1}^n  \tilde{z}_i \tilde{y}_i }{\frac{1}{\tau} + \sum_{i=1}^n \tilde{z}_i^2 } \Bigg)^2  - \frac{  (  \frac{2\gamma_0}{\tau} + 2\sum_{i=1}^n  \tilde{z}_i \tilde{y}_i )^2  }{(\frac{1}{\tau} + \sum_{i=1}^n \tilde{z}_i^2 )  }    \right]   \Bigg)    $$
%
$$ \propto exp \Bigg(  - \frac{1 }{2 \frac{\phi}{ \frac{1}{\tau} + \sum_{i=1}^n \tilde{z}_i^2  } } \left[  \Bigg(\gamma - \frac{\frac{\gamma_0}{\tau} + \sum_{i=1}^n  \tilde{z}_i \tilde{y}_i }{\frac{1}{\tau} + \sum_{i=1}^n \tilde{z}_i^2 } \Bigg)^2  \right]   \Bigg)    $$
$$ \gamma| \bm{\tilde{y}}, \bm{\tilde{z}}, \phi \sim  \mathcal{N} \Bigg( \frac{\frac{\gamma_0}{\tau} + \sum_{i=1}^n  \tilde{z}_i \tilde{y}_i }{\frac{1}{\tau} + \sum_{i=1}^n \tilde{z}_i^2 } ,   \frac{\phi}{ \frac{1}{\tau} + \sum_{i=1}^n \tilde{z}_i^2  } \Bigg)  $$
where we set $\gamma_0=0$ as the default parameter value.

For the \cite{omori2007efficient} prior,  $\gamma |\phi \sim \mathcal{N}(g_0, G_0) $, $p(\phi|\gamma) =p(\phi)$,  and
$$ \gamma| \bm{\tilde{y}}, \bm{\tilde{z}}, \phi \sim  \mathcal{N} \Bigg( \frac{\frac{\gamma_0}{ G_0 } + \frac{1}{\phi} \sum_{i=1}^n  \tilde{z}_i \tilde{y}_i }{  \frac{1}{G_0} + \frac{1}{\phi} \sum_{i=1}^n \tilde{z}_i^2  } ,   \frac{1}{ \frac{1}{G_0} + \frac{1}{\phi} \sum_{i=1}^n \tilde{z}_i^2  } \Bigg)  $$

\subsection{Joint draw of $\phi$ and $\gamma$}\label{phigamma_norm_ig_sec}
For the \cite{van2011bayesian} prior, the $\phi$ and $\gamma$ parameters can be sampled jointly. First note that the censored observations contain no information about $\phi$ and $\gamma$ and therefore do not affect the conditional distribution.
$$p(\phi, \gamma | \bm{\tilde{y}}, \bm{\tilde{z}}, \tau)  =  p( \bm{\tilde{y}}, \bm{\tilde{z}} | \phi, \gamma , \tau)p( \gamma | \phi )p(\phi) =$$
$$\propto  \Bigg\{ \prod_{i=1}^{n_{uncens}}  \frac{1}{2\pi \sqrt{\phi + \gamma^2} \sqrt{ \frac{\phi}{\gamma^2 + \phi} } } \exp \Bigg( - \frac{1}{2 \frac{\phi}{\gamma^2+\phi}   } \Big[ \tilde{z}_i^2 - 2\frac{\gamma}{ \sqrt{\gamma^2 + \phi}} \frac{\tilde{z}_i \tilde{y}_i }{\sqrt{\gamma^2 + \phi }} + \frac{\tilde{y}_i^2}{{\gamma^2 + \phi}}      \Big]  \Bigg)   \Bigg\} $$
 $$ \times \frac{1}{\sqrt{2 \pi \tau}} \frac{ (\frac{S_0}{2})^{\frac{n_0}{2}}  }{\Gamma(\frac{n_0}{2})} \left( \frac{1}{\phi} \right)^{\frac{n_0+3}{2}} \exp \left( - \frac{S_0 + \frac{1}{\tau}(\gamma- \gamma_0)^2 }{2\phi} \right)$$
$$ \propto \left( \frac{1}{\phi} \right)^{\frac{n_{uncens} + n_0+3}{2}}   \exp \Bigg( - \frac{1}{2 \phi } \Big[ S_0 + \frac{1}{\tau}(\gamma- \gamma_0)^2 + \sum_{i=1}^{n_{uncens}} \left(  (\gamma^2+\phi) \tilde{z}_i^2 - 2 \gamma  \tilde{z}_i \tilde{y}_i  + \tilde{y}_i^2  \right)    \Big]  \Bigg)  $$
$$ \propto \left( \frac{1}{\phi} \right)^{\frac{n_{uncens} + n_0+3}{2}}   \exp \Bigg( - \frac{1}{2 \phi } \Big[ 
(\frac{1}{\tau} +  \sum_{i=1}^{n_{uncens}} \tilde{z}_i^2 ) \gamma^2 - 2 (\frac{\gamma_0}{\tau} + \sum_{i=1}^{n_{uncens}} \tilde{z}_i \tilde{y}_i  )\gamma + (\sum_{i=1}^{n_{uncens}}  \tilde{y}_i^2 ) + S_0 + \frac{\gamma_0^2}{\tau} \Big]  \Bigg)  $$
The exponential function contains a quadratic in $\gamma$. We can complete the sqaure by setting $a = \frac{1}{\tau} +  \sum_{i=1}^{n_{uncens}} \tilde{z}_i^2$, $b = -2 (\frac{\gamma_0}{\tau} + \sum_{i=1}^{n_{uncens}} \tilde{z}_i \tilde{y}_i  ) $, and $c = (\sum_{i=1}^{n_{uncens}}  \tilde{y}_i^2 ) + S_0 + \frac{\gamma_0^2}{\tau}  $ . In general $a x^2 + bx + c = a(x-h)^2 + k$, where $h = -\frac{b}{2a} $ and $k = c-\frac{b^2}{4a}$. This gives
$$ p(\phi, \gamma | \bm{\tilde{y}}, \bm{\tilde{z}}, \tau)  \propto $$
$$  \left( \frac{1}{\phi} \right)^{\frac{n_{uncens} + n_0+3}{2}}   \exp \Bigg( - \frac{1}{2 \phi } \Bigg[ \left(  \frac{1}{\tau} +  \sum_{i=1}^{n_{uncens}} \tilde{z}_i^2 \right) \left( \gamma - \frac{\frac{\gamma_0}{\tau} + \sum_{i=1}^{n_{uncens}} \tilde{z}_i \tilde{y}_i  }{\frac{1}{\tau} +  \sum_{i=1}^{n_{uncens}} \tilde{z}_i^2} \right)^2 + $$
$$ (\sum_{i=1}^{n_{uncens}}  \tilde{y}_i^2 ) + S_0 + \frac{\gamma_0^2}{\tau} - \frac{\frac{\gamma_0}{\tau} + \sum_{i=1}^{n_{uncens}} \tilde{z}_i \tilde{y}_i}{\frac{1}{\tau} +  \sum_{i=1}^{n_{uncens}} \tilde{z}_i^2}
\Bigg]  \Bigg)  $$
This is proportional to the pdf of a normal-inverse-gamma distribution, therefore
$$\phi, \gamma \sim N-\Gamma^{-1} \Bigg(  \frac{\frac{\gamma_0}{\tau} + \sum_{i=1}^{n_{uncens}} \tilde{z}_i \tilde{y}_i  }{\frac{1}{\tau} +  \sum_{i=1}^{n_{uncens}} \tilde{z}_i^2} , \frac{1}{\tau} +  \sum_{i=1}^{n_{uncens}} \tilde{z}_i^2, \frac{n_{uncens} + n_0 }{2}, \frac{1}{2} \Big[ S_0 + \frac{\gamma_0^2}{\tau} - \frac{\frac{\gamma_0}{\tau} + \sum_{i=1}^{n_{uncens}} \tilde{z}_i \tilde{y}_i}{\frac{1}{\tau} +  \sum_{i=1}^{n_{uncens}} \tilde{z}_i^2}  + \sum_{i=1}^{n_{uncens}}  \tilde{y}_i^2 \Big]  \Bigg) $$
This distribution is sampled as follows:
\begin{itemize}
    \item Draw $ \phi \sim \Gamma^{-1} \Bigg(\frac{n_{uncens} + n_0}{2}, \frac{1}{2} \Big[ S_0 + \frac{\gamma_0^2}{\tau} - \frac{\frac{\gamma_0}{\tau} + \sum_{i=1}^{n_{uncens}} \tilde{z}_i \tilde{y}_i}{\frac{1}{\tau} +  \sum_{i=1}^{n_{uncens}} \tilde{z}_i^2}  + \sum_{i=1}^{n_{uncens}}  \tilde{y}_i^2 \Big]  \Bigg) $.
    \item Draw $ \gamma \sim \mathcal{N} \Bigg( \frac{\frac{\gamma_0}{\tau} + \sum_{i=1}^{n_{uncens}} \tilde{z}_i \tilde{y}_i  }{\frac{1}{\tau} +  \sum_{i=1}^{n_{uncens}} \tilde{z}_i^2}  , \frac{\phi}{\frac{1}{\tau} +  \sum_{i=1}^{n_{uncens}} \tilde{z}_i^2} \Bigg) $
\end{itemize}


\section{Derivation of full conditional sampler for \cite{ding2014bayesian} prior}
As described by \cite{ding2014bayesian}, parameter expansion can be applied. Note that
$$  \begin{pmatrix}
    \tilde{{z}}_i \\
    \tilde{{y}}_i \\
\end{pmatrix} 
\sim \mathcal{N} \left(  \bm{0}_2, 
\begin{pmatrix}
    \frac{1}{\sigma_1} & 0 \\
    0 & 1
\end{pmatrix}
\tilde{\Sigma}
\begin{pmatrix}
    \frac{1}{\sigma_1} & 0 \\
    0 & 1
\end{pmatrix}
\right)
$$
The model is re-parametrized by defining 
$$ \bm{E}_i := \begin{pmatrix}
    \sigma_1 & 0 \\
    0 & 1
\end{pmatrix} \begin{pmatrix}
    \tilde{{z}}_i \\
    \tilde{{y}}_i \\
\end{pmatrix} =
\begin{pmatrix}
    \sigma_1  \tilde{{z}}_i \\
    \tilde{{y}}_i \\
\end{pmatrix}
\sim \mathcal{N} \left(  \bm{0}_2, 
\tilde{\Sigma}
\right)
$$
First we draw $\sigma_1^2 | \Omega, \dots  \sim  \{ \frac{1- \rho^2}{c} \chi_{\nu_0}^2 \}^{-1} $ , then calculate $\bm{E}_i$, and then draw  from $\tilde{\Sigma}| \bm{E}_i$.
$$p( \tilde{\Sigma } | \bm{E}_i, \dots )  \propto p(\tilde{\Sigma}) \prod_{i=1}^{n_{uncens}} p( \bm{E}_i | \tilde{\Sigma} , \dots ) $$
$$ \propto  \frac{ c^2 | I_2|^{\frac{\nu_0}{2} }}{2^{\nu_0} \Gamma_2(\frac{\nu_0}{2})} |\tilde{\Sigma} |^{-\frac{\nu_0 + 3}{2}} \exp \left\{ -\frac{1}{2} c \ tr( \tilde{\Sigma}^{-1} )  \right\}  \prod_{i = 1}^{n_{uncens}} \Bigg( |\tilde{\Sigma}|^{-1/2} \exp \left\{ - \frac{1}{2}       \bm{E}_i^T \tilde{\Sigma}^{-1} \bm{E}_i  \right\} \Bigg)  $$
$$ \propto |\tilde{\Sigma} |^{-\frac{n_{uncens} + \nu_0 + 3}{2}}    \exp \left\{ -\frac{1}{2} c \ tr( \tilde{\Sigma}^{-1} )   - \frac{1}{2}  \sum_{i = 1}^{n_{uncens}}    \bm{E}_i^T  \tilde{\Sigma}^{-1}
\bm{E}_i \right\}   $$
$$ \propto |\tilde{\Sigma} |^{-\frac{n_{uncens} + \nu_0 + 3}{2}}    \exp \left\{ -  \frac{1}{2} \  tr( \tilde{\Sigma}^{-1} (c \bm{I}_2  +   \sum_{i = 1}^{n_{uncens}}    \bm{E}_i 
\bm{E}_i^T  ) )   \right\}   $$
Therefore
$$\tilde{\Sigma } | \tilde{\bm{z}}, \tilde{\bm{y}}, \dots  \sim  \bm{W}_2^{-1} \left( n_{uncens} + \nu_0 , c \bm{I}_2  +   \sum_{i = 1}^{n_{uncens}}    \bm{E}_i  \bm{E}_i^T  \right)  $$

\section{Alternative TOBART-2 sampler}\label{TOBART2Marg_app}
\cite{van2011bayesian, van2005bayesian} notes the high correlation between draws of outcome equation parameters and the parameter $\gamma$. A joint sample of the outcome equation coefficients and $\gamma$ addresses this issue.

Similarly, for TOBART-2, the outcome equation draws of tree structures, $T_1^y , T_2^y , \dots , T_{m_y}^y $ and terminal node parameters \\ $\bm{\mu}_{y, 1:m_y} = (\bm{\mu}_{y, 1}, \dots, \bm{\mu}_{y, m_y} )' = ({\mu}_{y, 1,1},\dots,{\mu}_{y, 1,\ell_{y, 1}}, {\mu}_{y, 2,1},\dots,{\mu}_{y, 2,\ell_{y, 2}} \dots, {\mu}_{y, m_y,1},\dots,{\mu}_{y, m_y,\ell_{y, m_y}} )' $, 
\\$f_y$ are dependent on the draw $\gamma$. 

To avoid this issue, we extend the improved BART sampler introduced by \cite{collins2023improved} and draw trees  $T_1^y , T_2^y , \dots , T_{m_y}^y $  while marginalizing out both $\bm{\mu}_{y, 1:m_y}$ and $\gamma$. This is followed by a joint full conditional draw of $(\bm{\mu}_{y, 1:m_y}, \gamma)$.

Denote the number of terminal nodes in tree $T_k^y$ by $\ell_{y, k} $, and denote the total number of terminal nodes in $T_1^y , T_2^y , \dots , T_{m_y}^y $ by  $ \ell_{y, 1:m_y} =  \sum_{k=1}^{m_{y}} \ell_{y, k}$. Let the  $n_1 \times \sum_{k=1}^{m_{y}} \ell_{y, k}  $ matrix of terminal node indicator variables for selected observations be denoted by $\bm{B}_y $. Let $\bm{y}$ be the vector of outcomes for selected observations, let   $\tilde{\bm{z}} = \bm{z} - \bm{f}_z(\bm{W})  $ be a vector of selection equation residuals for selected observations, and define the combined matrix $\tilde{\bm{B}}_y = [\bm{B}_y \ \tilde{\bm{z}} ] $. The prior variance of the terminal nodes in the sum of trees $\bm{f}_y$ is denoted by $\sigma_{0,y}^2$.

The trees $T_1^y , T_2^y , \dots , T_{m_y}^y $ are drawn from the (marginalized) conditional $T_j^y | \bm{y}, \bm{X}, \bm{W}, T_{-j}^y, \tilde{\bm{z}} , \phi  $. The tree proposal and prior is the same as for standard BART, and the likelihood for the Metropolis-Hastings accept-reject step is:
$$ p(\bm{y}| \tilde{\bm{B}}_y  \bm{X}, \bm{W}, \phi)  \propto  \left( \frac{1}{\sigma_{0,y}^2} \right)^{ \frac{\ell_{y, 1:m_y}}{2} } \Bigg| \frac{1}{\phi} \tilde{\bm{B}}_y^T \tilde{\bm{B}}_y + \begin{pmatrix}
    \frac{1}{\sigma_{0,y}^2}  I_{\ell_{y, 1:m_y}  } & \bm{0}_{\ell_{y, 1:m_y} }  \\
    \bm{0}_{\ell_{y, 1:m_y}}^T & \frac{1}{\tau \phi} \\
\end{pmatrix}   \Bigg|^{-1/2} \times $$
$$ \exp \Bigg( \frac{1}{2\phi} \bm{y}^T \tilde{\bm{B}}_y \left(  \frac{1}{\phi} 
 \tilde{\bm{B}}_y^T \tilde{\bm{B}}_y +   
\begin{pmatrix}
    \frac{1}{\sigma_{0,y}^2}  I_{\ell_{y, 1:m_y}  } & \bm{0}_{\ell_{y, 1:m_y} }  \\
    \bm{0}_{\ell_{y, 1:m_y}}^T & \frac{1}{\tau \phi} \\
\end{pmatrix}
\right)^{-1}   \frac{1}{\phi} \tilde{\bm{B}}_y^T \bm{y}  \Bigg)  $$
The joint full conditional draw of $(\bm{\mu}_{y, 1:m_y}, \gamma)$ is:
$$ \begin{bmatrix}
    \bm{\mu}_{y, 1:m_y} \\
    \gamma \\
\end{bmatrix}  |  T_1^y , T_2^y , \dots , T_{m_y}^y , \bm{y}, \bm{X}, \bm{W}, \tilde{\bm{z}} , \phi $$
%
$$ \sim  \mathcal{MVN} \Bigg( \left(  \frac{1}{\phi} \tilde{\bm{B}}_y^T \tilde{\bm{B}}_y +   
\begin{pmatrix}
    \frac{1}{\sigma_{0,y}^2}  I_{\ell_{y, 1:m_y}  } & \bm{0}_{\ell_{y, 1:m_y} }  \\
    \bm{0}_{\ell_{y, 1:m_y}}^T & \frac{1}{\tau \phi} \\
\end{pmatrix}
\right)^{-1}   \frac{1}{\phi}\tilde{\bm{B}}_y^T \bm{y}  ,   \left(  \frac{1}{\phi} \tilde{\bm{B}}_y^T \tilde{\bm{B}}_y +   
\begin{pmatrix}
    \frac{1}{\sigma_{0,y}^2}  I_{\ell_{y, 1:m_y}  } & \bm{0}_{\ell_{y, 1:m_y} }  \\
    \bm{0}_{\ell_{y, 1:m_y}}^T & \frac{1}{\tau \phi} \\
\end{pmatrix}
\right)^{-1}   \Bigg)  $$
The draw of $f_z$ can be standard, or involve first defining
$$ \breve{z}_{i} = \begin{cases}
			z_i^*   & \text{if $i$ censored,} \\
			z_i^* -  ( y_i^* -  \sum_{k=1}^{m_y} g_k^y (\bm{x}_i)  )\frac{\gamma}{\gamma^2 + \phi}  & \text{if $i$ uncensored}  \end{cases} $$
Define $  \mathbf{\breve{z}}  = (\breve{z}_{1}, \breve{z}_{2}, \dots, \breve{z}_{n})'  $ . The variance of $ \breve{z}_{i}$ is 1 for censored/unselected observations and $ \frac{\phi}{\gamma^2 + \phi} $ for uncensored/selected observations. Trees $T_1^z,T_2^z, \dots, T_{m_z}^z$ can be drawn by marginalizing out $\bm{\mu}_{z, 1:m_z} = (\bm{\mu}_{z, 1}, \dots, \bm{\mu}_{z, m_z} )' = ({\mu}_{z, 1,1},\dots,{\mu}_{z, 1,\ell_{z, 1}}, {\mu}_{z, 2,1},\dots,{\mu}_{z, 2,\ell_{z, 2}} \dots, {\mu}_{z, m_z,1},\dots,{\mu}_{z, m_z,\ell_{z, m_z}} )' $ conditional on $\gamma, \phi$ and $ \tilde{\bm{y}} =  \bm{y}^* -  \bm{f}_y ( \bm{X}) $. This gives a weighted regression extension of the result from \cite{collins2023improved} with variance 1, and weights equal to $1$ for censored observations and $ \frac{\gamma^2 + \phi}{\phi} $ for uncensored observations.

The results below are standard results for Bayesian weighted linear regression:

\medskip

Let $V$ denote a diagonal matrix with diagonal elements equal to 1 for censored observations and $ \frac{\phi}{\gamma^2 + \phi} $ for uncensored observations. The prior variance of the terminal node parameters in the sum-oftrees $\bm{f}_z$ is denoted by $\sigma_{0,z}^2$.

\medskip

The likelihood for the MH accept-reject steps in the draws of  $T_1^z,T_2^z, \dots, T_{m_z}^z$ marginalizing out $\bm{\mu}_{z, 1:m_z}$, but conditional on $f_y, \phi, \gamma$ is:
$$ p(\bm{z} | B_z, f_y, \phi, \gamma) = \frac{1}{(2 \pi)^{n/2}} |V|^{-1/2} \exp \Bigg( - \frac{1}{2} \Big[ \mathbf{\breve{z}}^T V^{-1} \mathbf{\breve{z}}  -\mathbf{\breve{z}}^T V^{-1} \bm{B}_z \left( 
 \bm{B}_z^T  V^{-1} \bm{B}_z + \frac{1}{\sigma_{0,z}^2} I_{\ell_{m,1:m_z}}  \right)^{-1} \bm{B}_z^T  
 V^{-1}  \mathbf{\breve{z}}   \Big]   \Bigg) $$
 $$  \times  \frac{1}{\sigma_{0,z}^{\ell_{m,1:m_z}}}\Big| \bm{B}_z^T  V^{-1} \bm{B}_z + \frac{1}{\sigma_{0,z}^2} I_{\ell_{m,1:m_z}} \Big|^{-1/2}  $$
$$ \propto  \frac{1}{\sigma_{0,z}^{\ell_{m,1:m_z}}} \Big| \bm{B}_z^T  V^{-1} \bm{B}_z + \frac{1}{\sigma_{0,z}^2} I_{\ell_{m,1:m_z}} \Big|^{-1/2}   \exp \Bigg( \frac{1}{2} \mathbf{\breve{z}}^T V^{-1} \bm{B}_z \left( 
 \bm{B}_z^T  V^{-1} \bm{B}_z + \frac{1}{\sigma_{0,z}^2} I_{\ell_{m,1:m_z}}  \right)^{-1} \bm{B}_z^T  
 V^{-1}  \mathbf{\breve{z}}      \Bigg) $$

If we order the observations of the matrix $\bm{B}_{z}$ so that the first $n_1$ observations are uncensored, and the last $n_0$ observations are censored, so that   $\bm{B}_{z} = \begin{bmatrix}
    \bm{B}_{z, u}\\
    \bm{B}_{z, c} \\
\end{bmatrix}$ , and similarly order $\mathbf{\breve{z}}$,  then the likelihood can be re-written as
$$ p(\bm{z} | B_z, f_y, \phi, \gamma) \propto  \frac{1}{\sigma_{0,z}^{\ell_{m,1:m_z}}} \Big| \frac{\gamma^2 + \phi }{\phi}  \bm{B}_{z, u}^T   \bm{B}_{z, u} + \bm{B}_{z, c}^T   \bm{B}_{z, c} + \frac{1}{\sigma_{0,z}^2} I_{\ell_{m,1:m_z}} \Big|^{-1/2}   \times $$
$$ \exp \Bigg( \frac{1}{2} \mathbf{\breve{z}}^T 
\begin{bmatrix}
     \frac{\gamma^2 + \phi }{\phi}  \bm{B}_{z, u}\\
       \bm{B}_{z, c} \\
\end{bmatrix}
\left( 
  \frac{\gamma^2 + \phi }{\phi}  \bm{B}_{z, u}^T   \bm{B}_{z, u} + \bm{B}_{z, c}^T   \bm{B}_{z, c} +  \frac{1}{\sigma_{0,z}^2} I_{\ell_{m,1:m_z}}     \right)^{-1} \begin{bmatrix}
     \frac{\gamma^2 + \phi }{\phi}  \bm{B}_{z, u}^T  &     \bm{B}_{z, c}^T 
\end{bmatrix}  \mathbf{\breve{z}}  \Bigg)  $$
The draw of $  \bm{\mu}_{z, 1:m_z} $ is:
 $$   \bm{\mu}_{z, 1:m_z} |  B_z, f_y, \phi, \gamma \sim   \mathcal{N}_{\ell_{m,1:m_z}} \Bigg(  \bm{B}_z^T \left( \bm{B}_z \bm{B}_z^T + \frac{1}{\sigma_{0,z}^2} V \right)^{-1} \mathbf{\breve{z}} ,  \sigma_{0,z}^2  \left( I_{\ell_{m,1:m_z}} - \bm{B}_z^T \left( \bm{B}_z \bm{B}_z^T + \frac{1}{\sigma_{0,z}^2} V \right)^{-1} \bm{B}_z  \right)  \Bigg) $$
which can also be written as:
 $$   \bm{\mu}_{z, 1:m_z} |  B_z, f_y, \phi, \gamma \sim   \mathcal{N}_{\ell_{m,1:m_z}} \Bigg(   \left( 
 \bm{B}_z^T  V^{-1} \bm{B}_z + \frac{1}{\sigma_{0,z}^2} I_{\ell_{m,1:m_z}}  \right)^{-1} \bm{B}_z^T  
 V^{-1}  \mathbf{\breve{z}}  ,     \left( 
 \bm{B}_z^T  V^{-1} \bm{B}_z + \frac{1}{\sigma_{0,z}^2} I_{\ell_{m,1:m_z}}  \right)^{-1}    \Bigg) $$
or
 $$   \bm{\mu}_{z, 1:m_z} |  B_z, f_y, \phi, \gamma \sim   \mathcal{N}_{\ell_{m,1:m_z}} \Bigg(   \left( 
  \frac{\gamma^2 + \phi }{\phi}  \bm{B}_{z, u}^T   \bm{B}_{z, u} + \bm{B}_{z, c}^T   \bm{B}_{z, c} + \frac{1}{\sigma_{0,z}^2} I_{\ell_{m,1:m_z}}  \right)^{-1} \begin{bmatrix}
     \frac{\gamma^2 + \phi }{\phi}  \bm{B}_{z, u}^T  &     \bm{B}_{z, c}^T 
\end{bmatrix}  \mathbf{\breve{z}}  ,  $$
$$ \left( 
\frac{\gamma^2 + \phi }{\phi}  \bm{B}_{z, u}^T   \bm{B}_{z, u} + \bm{B}_{z, c}^T   \bm{B}_{z, c} + \frac{1}{\sigma_{0,z}^2} I_{\ell_{m,1:m_z}}  \right)^{-1}    \Bigg) $$

\subsection{Derivation of $f_z$ sampler }

The joint distribution of $ \mathbf{\breve{z}} , \bm{\mu}_{z, 1:m_z} $ is:

$$\begin{bmatrix}
     \mathbf{\breve{z}} \\ 
     \bm{\mu}_{z, 1:m_z}
\end{bmatrix}|  B_z, f_y, \phi, \gamma \sim
\mathcal{N}_{n + \ell_{m,1:m_z}}\Bigg( 
\begin{bmatrix}
     \bm{0}_n \\ 
     \bm{0}_{\ell_{m,1:m_z}}
\end{bmatrix},
\begin{bmatrix}
    \sigma_{0,z}^2 \bm{B}_z \bm{B}_z^T + V & \sigma_{0,z}^2 \bm{B}_z  \\ 
    \sigma_{0,z}^2 \bm{B}_z^T &  \sigma_{0,z}^2 \bm{I}_{\ell_{m,1:m_z}} \\
\end{bmatrix}
\Bigg)
$$
where we use the fact that $Var(\mathbf{\breve{z}}) = Var( \bm{B}_z \bm{\mu}_{z, 1:m_z} + \varepsilon_z  )  =  \bm{B}_z Var(  \bm{\mu}_{z, 1:m_z}) \bm{B}_z^T + Var(\varepsilon_z  )  $ and $Cov(\bm{\mu}_{z, 1:m_z}, \mathbf{\breve{z}})= Cov(\bm{\mu}_{z, 1:m_z}, \bm{B}_z \bm{\mu}_{z, 1:m_z} + \varepsilon_z ) =  \bm{B}_z Var(  \bm{\mu}_{z, 1:m_z})$

Then, from a standard result for the conditional distribution of a multivariate normally distributed variable:
 $$   \bm{\mu}_{z, 1:m_z} | \mathbf{\breve{z}}, B_z, f_y, \phi, \gamma \sim   \mathcal{N}_{\ell_{m,1:m_z}} \Bigg(  \sigma_{0,z}^2 \bm{B}_z^T \left( \sigma_{0,z}^2 \bm{B}_z \bm{B}_z^T +  V \right)^{-1} \mathbf{\breve{z}} ,  \sigma_{0,z}^2   I_{\ell_{m,1:m_z}} -  \sigma_{0,z}^2\bm{B}_z^T \left(  \sigma_{0,z}^2 \bm{B}_z \bm{B}_z^T +  V \right)^{-1} \sigma_{0,z}^2 \bm{B}_z    \Bigg) $$
 which can be rearranged to 
 $$   \bm{\mu}_{z, 1:m_z} | \mathbf{\breve{z}},  B_z, f_y, \phi, \gamma \sim   \mathcal{N}_{\ell_{m,1:m_z}} \Bigg(   \Big( \bm{B}_z^T \bm{B}_z + \frac{1}{\sigma_{0,z}^2} V \Big)^{-1} \bm{B}_z^T \mathbf{\breve{z}} ,     \left( \bm{B}_z^T  V^{-1} \bm{B}_z + \frac{1}{\sigma_{0,z}^2} I_{\ell_{m,1:m_z}}  \right)^{ -1}  \Bigg) $$

\bigskip

The marginal likelihood $p(\bm{z} | B_z, f_y, \phi, \gamma) $ can be derived as follows:
$$ p(\bm{z} | B_z, f_y, \phi, \gamma) = \int \mathcal{N} (\mathbf{\breve{z}} | B_z  \bm{\mu}_{z, 1:m_z}  , V ) \mathcal{N}(\bm{\mu}_{z, 1:m_z} | \bm{0}, \sigma_{0,z}^2   I_{\ell_{m,1:m_z}}    )  d  \bm{\mu}_{z, 1:m_z}  $$
$$   = \int  (2 \pi)^{-n/2}   |V|^{-1/2} \exp \Bigg( -\frac{1}{2} (\mathbf{\breve{z}} -  B_z  \bm{\mu}_{z, 1:m_z} )^T V^{-1} (\mathbf{\breve{z}} -  B_z  \bm{\mu}_{z, 1:m_z} )   \Bigg) \times $$
$$ (2 \pi)^{-\ell_{m,1:m_z}/2}  \frac{1}{\sigma_{0,z}^{\ell_{m,1:m_z}}} \exp \Bigg( -\frac{1}{2 \sigma_{0,z}^2} \bm{\mu}_{z, 1:m_z}^T   \bm{\mu}_{z, 1:m_z}  \Bigg)   d  \bm{\mu}_{z, 1:m_z}   $$
$$   = (2 \pi)^{-n/2}   |V|^{-1/2} (2 \pi)^{-\ell_{m,1:m_z}/2}  \frac{1}{\sigma_{0,z}^{\ell_{m,1:m_z}}}  \times $$
$$  \int   \exp \Bigg( -\frac{1}{2} (\mathbf{\breve{z}} -  B_z  \bm{\mu}_{z, 1:m_z} )^T V^{-1} (\mathbf{\breve{z}} -  B_z  \bm{\mu}_{z, 1:m_z} ) -\frac{1}{2 \sigma_{0,z}^2} \bm{\mu}_{z, 1:m_z}^T   \bm{\mu}_{z, 1:m_z}  \Bigg)   d  \bm{\mu}_{z, 1:m_z}   $$
Then, rearranging the argument of the exponential function to make it a quadratic function of $\bm{\mu}_{z, 1:m_z}$ we note
$$ -\frac{1}{2} (\mathbf{\breve{z}} -  B_z  \bm{\mu}_{z, 1:m_z} )^T V^{-1} (\mathbf{\breve{z}} -  B_z  \bm{\mu}_{z, 1:m_z} ) -\frac{1}{2 \sigma_{0,z}^2} \bm{\mu}_{z, 1:m_z}^T   \bm{\mu}_{z, 1:m_z} = $$ 
$$ -\frac{1}{2} \Bigg( \bm{\mu}_{z, 1:m_z}^T  \left( \bm{B}_z^T  V^{-1} \bm{B}_z + \frac{1}{\sigma_{0,z}^2} I_{\ell_{m,1:m_z}}  \right) \bm{\mu}_{z, 1:m_z} - 2 \mathbf{\breve{z}}^T   V^{-1} \bm{B}_z  \bm{\mu}_{z, 1:m_z}  \Bigg) - \frac{1}{2}   \mathbf{\breve{z}}^T   V^{-1}  \mathbf{\breve{z}} $$
Then by completing the (multivariate) square we have
$$ = -\frac{1}{2} \Bigg[  \Bigg( \bm{\mu}_{z, 1:m_z} -  \left( \bm{B}_z^T  V^{-1} \bm{B}_z + \frac{1}{\sigma_{0,z}^2} I_{\ell_{m,1:m_z}}  \right)^{-1} \bm{B}_z^T  
 V^{-1}  \mathbf{\breve{z}}  \Bigg) ^T    
\left( \bm{B}_z^T  V^{-1} \bm{B}_z + \frac{1}{\sigma_{0,z}^2} I_{\ell_{m,1:m_z}}  \right) \times $$
$$ \Bigg( \bm{\mu}_{z, 1:m_z} -  \left( \bm{B}_z^T  V^{-1} \bm{B}_z + \frac{1}{\sigma_{0,z}^2} I_{\ell_{m,1:m_z}}  \right)^{-1} \bm{B}_z^T  
 V^{-1}  \mathbf{\breve{z}}  \Bigg)
 -  \mathbf{\breve{z}}^T   V^{-1}    \bm{B}_z   \left( \bm{B}_z^T  V^{-1} \bm{B}_z + \frac{1}{\sigma_{0,z}^2} I_{\ell_{m,1:m_z}}  \right)^{-1}  \bm{B}_z^T  
 V^{-1}  \mathbf{\breve{z}}  \Bigg]  $$
$$ - \frac{1}{2}   \mathbf{\breve{z}}^T   V^{-1}  \mathbf{\breve{z}}  $$
Therefore
$$ p(\bm{z} | B_z, f_y, \phi, \gamma) =  (2 \pi)^{-n/2}   |V|^{-1/2} (2 \pi)^{-\ell_{m,1:m_z}/2}  \frac{1}{\sigma_{0,z}^{\ell_{m,1:m_z}}}  \times $$
$$  \exp \Bigg(  - \frac{1}{2} \left[  \mathbf{\breve{z}}^T   V^{-1}  \mathbf{\breve{z}} - \mathbf{\breve{z}}^T   V^{-1}    \bm{B}_z   \left( \bm{B}_z^T  V^{-1} \bm{B}_z + \frac{1}{\sigma_{0,z}^2} I_{\ell_{m,1:m_z}}  \right)^{-1}  \bm{B}_z^T  
 V^{-1}  \mathbf{\breve{z}} \right] \Bigg) \times $$
$$  \int   \exp \Bigg( -\frac{1}{2} \Bigg( \bm{\mu}_{z, 1:m_z} -  \left( \bm{B}_z^T  V^{-1} \bm{B}_z + \frac{1}{\sigma_{0,z}^2} I_{\ell_{m,1:m_z}}  \right)^{-1} \bm{B}_z^T  
 V^{-1}  \mathbf{\breve{z}}  \Bigg) ^T    
\left( \bm{B}_z^T  V^{-1} \bm{B}_z + \frac{1}{\sigma_{0,z}^2} I_{\ell_{m,1:m_z}}  \right) \times $$
$$ \Bigg( \bm{\mu}_{z, 1:m_z} -  \left( \bm{B}_z^T  V^{-1} \bm{B}_z + \frac{1}{\sigma_{0,z}^2} I_{\ell_{m,1:m_z}}  \right)^{-1} \bm{B}_z^T  
 V^{-1}  \mathbf{\breve{z}}  \Bigg) \Bigg)   d  \bm{\mu}_{z, 1:m_z} $$
The integral in the above expression must evaluate to the reciprocal of the normalizing constant of a multivariate normal density with covariance matrix $\left( \bm{B}_z^T  V^{-1} \bm{B}_z + \frac{1}{\sigma_{0,z}^2} I_{\ell_{m,1:m_z}}  \right)^{-1}$, and is therefore $(2 \pi)^{\ell_{m,1:m_z}/2} \Big| \bm{B}_z^T  V^{-1} \bm{B}_z + \frac{1}{\sigma_{0,z}^2} I_{\ell_{m,1:m_z}} \Big|^{-1/2}$ . Therefore, overall we have:
$$ p(\bm{z} | B_z, f_y, \phi, \gamma) = \frac{1}{(2 \pi)^{n/2}} |V|^{-1/2} \exp \Bigg( - \frac{1}{2} \Big[ \mathbf{\breve{z}}^T V^{-1} \mathbf{\breve{z}}  -\mathbf{\breve{z}}^T V^{-1} \bm{B}_z \left( 
 \bm{B}_z^T  V^{-1} \bm{B}_z + \frac{1}{\sigma_{0,z}^2} I_{\ell_{m,1:m_z}}  \right)^{-1} \bm{B}_z^T  
 V^{-1}  \mathbf{\breve{z}}   \Big]   \Bigg) $$
 $$  \times  \frac{1}{\sigma_{0,z}^{\ell_{m,1:m_z}}}\Big| \bm{B}_z^T  V^{-1} \bm{B}_z + \frac{1}{\sigma_{0,z}^2} I_{\ell_{m,1:m_z}} \Big|^{-1/2}  $$

\subsection{Derivation of $f_y$ sampler }

$$ p(\bm{y}| \tilde{\bm{B}}_y  \bm{X}, \bm{W}, \phi)  \propto $$
$$ \int_{
    [\bm{\mu}_{y, 1:m_y}^T 
    \gamma]^T   \in \mathbb{R}^{\ell_{y, 1:m_y} + 1} } \mathcal{N} \left( \bm{y} | \tilde{\bm{B}}_{y}  \begin{bmatrix}
    \bm{\mu}_{y, 1:m_y} \\
    \gamma \\
\end{bmatrix} , \phi I_{n_1} \right)  \mathcal{N} \left(  \begin{bmatrix}
    \bm{\mu}_{y, 1:m_y} \\
    \gamma \\
\end{bmatrix} \Big|  \bm{0}_{\ell_{y, 1:m_y} + 1},  \begin{pmatrix}
    \sigma_{0,y}^2  I_{\ell_{y, 1:m_y}  } & \bm{0}_{\ell_{y, 1:m_y} }  \\
    \bm{0}_{\ell_{y, 1:m_y}}^T & \tau \phi \\
\end{pmatrix} \right) d\bm{\mu}_{y, 1:m_y} d \gamma     $$
$$ = \int (2 \pi)^{-n_1/2} | \phi I|^{-1/2} \exp \Bigg( -\frac{1}{2}  \left(  \bm{y} - \tilde{\bm{B}}_{y}  \begin{bmatrix}
    \bm{\mu}_{y, 1:m_y} \\
    \gamma \\
\end{bmatrix}\right)^T  (\phi I)^{-1}   \left(  \bm{y} - \tilde{\bm{B}}_{y}  \begin{bmatrix}
    \bm{\mu}_{y, 1:m_y} \\
    \gamma \\
\end{bmatrix}\right)  \Bigg) $$
$$(2 \pi)^{-\frac{\ell_{y, 1:m_y} + 1}{2}} \begin{vmatrix}
    \sigma_{0,y}^2  I_{\ell_{y, 1:m_y}  } & \bm{0}_{\ell_{y, 1:m_y} }  \\
    \bm{0}_{\ell_{y, 1:m_y}}^T & \tau \phi \\
\end{vmatrix}^{-1/2}  \exp \Bigg( - \frac{1}{2} [\bm{\mu}_{y, 1:m_y}^T 
    \gamma]  
     \begin{pmatrix}
    \sigma_{0,y}^2  I_{\ell_{y, 1:m_y}  } & \bm{0}_{\ell_{y, 1:m_y} }  \\
    \bm{0}_{\ell_{y, 1:m_y}}^T & \tau \phi \\
\end{pmatrix}^{-1}
    \begin{bmatrix}
    \bm{\mu}_{y, 1:m_y} \\
    \gamma \\
\end{bmatrix}  \Bigg)   d\bm{\mu}_{y, 1:m_y} d \gamma   $$
$$ = (2 \pi)^{-n_1/2} | \phi I|^{-1/2}  \begin{vmatrix}
    \sigma_{0,y}^2  I_{\ell_{y, 1:m_y}  } & \bm{0}_{\ell_{y, 1:m_y} }  \\
    \bm{0}_{\ell_{y, 1:m_y}}^T & \tau \phi \\
\end{vmatrix}^{-1/2}  (2 \pi)^{-\frac{\ell_{y, 1:m_y} + 1}{2}} $$
$$\int  \exp \Bigg(   -\frac{1}{2} \Bigg( \left(  \bm{y} - \tilde{\bm{B}}_{y}  \begin{bmatrix}
    \bm{\mu}_{y, 1:m_y} \\
    \gamma \\
\end{bmatrix}\right)^T  (\phi I)^{-1}   \left(  \bm{y} - \tilde{\bm{B}}_{y}  
\begin{bmatrix}
    \bm{\mu}_{y, 1:m_y} \\
    \gamma \\
\end{bmatrix}\right)  $$
$$ +[\bm{\mu}_{y, 1:m_y}^T 
    \gamma]  
\begin{pmatrix}
    \sigma_{0,y}^2  I_{\ell_{y, 1:m_y}  } & \bm{0}_{\ell_{y, 1:m_y} }  \\
    \bm{0}_{\ell_{y, 1:m_y}}^T & \tau \phi \\
\end{pmatrix}^{-1}
\begin{bmatrix}
    \bm{\mu}_{y, 1:m_y} \\
    \gamma \\
\end{bmatrix} \Bigg) \Bigg)   d\bm{\mu}_{y, 1:m_y} d \gamma   $$
$$ = (2 \pi)^{ - \frac{n_1 + \ell_{y, 1:m_y} + 1}{2} }  \left( \frac{1}{\phi} \right)^{\frac{n_1}{2}} \left(\frac{1}{\sigma_{0,y}^2}  \right)^{\frac{\ell_{y, 1:m_y}}{2}} \left(\frac{1}{\tau \phi}  \right)^{\frac{1}{2}}  $$
$$\int  \exp \Bigg(  -\frac{1}{2} \Bigg\{ [\bm{\mu}_{y, 1:m_y}^T 
    \gamma] 
    \Bigg( \tilde{\bm{B}}_{y}^T (\phi I)^{-1} \tilde{\bm{B}}_{y} +
     \begin{pmatrix}
    \sigma_{0,y}^2  I_{\ell_{y, 1:m_y}  } & \bm{0}_{\ell_{y, 1:m_y} }  \\
    \bm{0}_{\ell_{y, 1:m_y}}^T & \tau \phi \\
\end{pmatrix}^{-1} \Bigg)
    \begin{bmatrix}
    \bm{\mu}_{y, 1:m_y} \\
    \gamma \\
\end{bmatrix}   - $$
$$ 2 \bm{y}^T (\phi I)^{-1}  \tilde{\bm{B}}_{y} \begin{bmatrix}
    \bm{\mu}_{y, 1:m_y} \\
    \gamma \\
\end{bmatrix} \Bigg\} - \frac{1}{2} \bm{y}^T (\phi I)^{-1} \bm{y}     \Bigg)  d\bm{\mu}_{y, 1:m_y} d \gamma   $$
Now complete the (multivariate) square
$$ = (2 \pi)^{ - \frac{n_1 + \ell_{y, 1:m_y} + 1}{2} }  \left( \frac{1}{\phi} \right)^{\frac{n_1}{2}} \left(\frac{1}{\sigma_{0,y}^2}  \right)^{\frac{\ell_{y, 1:m_y}}{2}} \left(\frac{1}{\tau \phi}  \right)^{\frac{1}{2}}    $$
$$\int  \exp \Bigg( - \frac{1}{2} \Bigg\{ \Bigg( \begin{bmatrix}
    \bm{\mu}_{y, 1:m_y} \\
    \gamma \\
\end{bmatrix}   -
    \Bigg( \tilde{\bm{B}}_{y}^T (\phi I)^{-1} \tilde{\bm{B}}_{y} +
     \begin{pmatrix}
    \sigma_{0,y}^2  I_{\ell_{y, 1:m_y}  } & \bm{0}_{\ell_{y, 1:m_y} }  \\
    \bm{0}_{\ell_{y, 1:m_y}}^T & \tau \phi \\
\end{pmatrix}^{-1} \Bigg)^{-1}  \tilde{\bm{B}}_{y} (\phi I)^{-1} \bm{y} \Bigg)^T \times $$
$$\Bigg( \tilde{\bm{B}}_{y}^T (\phi I)^{-1} \tilde{\bm{B}}_{y} +
     \begin{pmatrix}
    \sigma_{0,y}^2  I_{\ell_{y, 1:m_y}  } & \bm{0}_{\ell_{y, 1:m_y} }  \\
    \bm{0}_{\ell_{y, 1:m_y}}^T & \tau \phi \\
\end{pmatrix}^{-1} \Bigg)   \times $$
$$\Bigg( \begin{bmatrix}
    \bm{\mu}_{y, 1:m_y} \\
    \gamma \\
\end{bmatrix}   -
    \Bigg( \tilde{\bm{B}}_{y}^T (\phi I)^{-1} \tilde{\bm{B}}_{y} +
     \begin{pmatrix}
    \sigma_{0,y}^2  I_{\ell_{y, 1:m_y}  } & \bm{0}_{\ell_{y, 1:m_y} }  \\
    \bm{0}_{\ell_{y, 1:m_y}}^T & \tau \phi \\
\end{pmatrix}^{-1} \Bigg)^{-1}  \tilde{\bm{B}}_{y} (\phi I)^{-1} \bm{y} \Bigg)  $$
$$  -\bm{y}^T (\phi I)^{-1}  \tilde{\bm{B}}_{y} \Bigg( \tilde{\bm{B}}_{y}^T (\phi I)^{-1} \tilde{\bm{B}}_{y} +
     \begin{pmatrix}
    \sigma_{0,y}^2  I_{\ell_{y, 1:m_y}  } & \bm{0}_{\ell_{y, 1:m_y} }  \\
    \bm{0}_{\ell_{y, 1:m_y}}^T & \tau \phi \\
\end{pmatrix}^{-1} \Bigg)^{-1} \tilde{\bm{B}}_{y}^T (\phi I)^{-1} \bm{y} \Bigg\}
    - \frac{1}{2} \bm{y}^T (\phi I)^{-1} \bm{y}     \Bigg)  d\bm{\mu}_{y, 1:m_y} d \gamma   $$
Taking a constant out of the integral gives
$$ = (2 \pi)^{ - \frac{n_1 + \ell_{y, 1:m_y} + 1}{2} }  \left( \frac{1}{\phi} \right)^{\frac{n_1}{2}} \left(\frac{1}{\sigma_{0,y}^2}  \right)^{\frac{\ell_{y, 1:m_y}}{2}} \left(\frac{1}{\tau \phi}  \right)^{\frac{1}{2}}    $$
$$ \exp \Bigg(   \frac{1}{2} \bm{y}^T (\phi I)^{-1}  \tilde{\bm{B}}_{y} \Bigg( \tilde{\bm{B}}_{y}^T (\phi I)^{-1} \tilde{\bm{B}}_{y} +
     \begin{pmatrix}
    \sigma_{0,y}^2  I_{\ell_{y, 1:m_y}  } & \bm{0}_{\ell_{y, 1:m_y} }  \\
    \bm{0}_{\ell_{y, 1:m_y}}^T & \tau \phi \\
\end{pmatrix}^{-1} \Bigg)^{-1} \tilde{\bm{B}}_{y}^T (\phi I)^{-1} \bm{y}
    - \frac{1}{2} \bm{y}^T (\phi I)^{-1} \bm{y} \Bigg)  $$
$$\int  \exp \Bigg( - \frac{1}{2} \Bigg( \begin{bmatrix}
    \bm{\mu}_{y, 1:m_y} \\
    \gamma \\
\end{bmatrix}   -
    \Bigg( \tilde{\bm{B}}_{y}^T (\phi I)^{-1} \tilde{\bm{B}}_{y} +
     \begin{pmatrix}
    \sigma_{0,y}^2  I_{\ell_{y, 1:m_y}  } & \bm{0}_{\ell_{y, 1:m_y} }  \\
    \bm{0}_{\ell_{y, 1:m_y}}^T & \tau \phi \\
\end{pmatrix}^{-1} \Bigg)^{-1}  \tilde{\bm{B}}_{y} (\phi I)^{-1} \bm{y} \Bigg)^T \times $$
$$\Bigg( \tilde{\bm{B}}_{y}^T (\phi I)^{-1} \tilde{\bm{B}}_{y} +
     \begin{pmatrix}
    \sigma_{0,y}^2  I_{\ell_{y, 1:m_y}  } & \bm{0}_{\ell_{y, 1:m_y} }  \\
    \bm{0}_{\ell_{y, 1:m_y}}^T & \tau \phi \\
\end{pmatrix}^{-1} \Bigg)   \times $$
$$\Bigg( \begin{bmatrix}
    \bm{\mu}_{y, 1:m_y} \\
    \gamma \\
\end{bmatrix}   -
    \Bigg( \tilde{\bm{B}}_{y}^T (\phi I)^{-1} \tilde{\bm{B}}_{y} +
     \begin{pmatrix}
    \sigma_{0,y}^2  I_{\ell_{y, 1:m_y}  } & \bm{0}_{\ell_{y, 1:m_y} }  \\
    \bm{0}_{\ell_{y, 1:m_y}}^T & \tau \phi \\
\end{pmatrix}^{-1} \Bigg)^{-1}  \tilde{\bm{B}}_{y} (\phi I)^{-1} \bm{y} \Bigg)        \Bigg)  d\bm{\mu}_{y, 1:m_y} d \gamma   $$
The integral must evaluate to the reciprocal of the normalizing constant of a multivariate normal density with covariance matrix
$ \Bigg( \tilde{\bm{B}}_{y}^T (\phi I)^{-1} \tilde{\bm{B}}_{y} +
     \begin{pmatrix}
    \sigma_{0,y}^2  I_{\ell_{y, 1:m_y}  } & \bm{0}_{\ell_{y, 1:m_y} }  \\
    \bm{0}_{\ell_{y, 1:m_y}}^T & \tau \phi \\
\end{pmatrix}^{-1} \Bigg)^{-1}  $ 
and is therefore  $ (2 \pi)^{  \frac{ \ell_{y, 1:m_y} + 1}{2} }   \begin{vmatrix}
     \tilde{\bm{B}}_{y}^T (\phi I)^{-1} \tilde{\bm{B}}_{y} +
     \begin{pmatrix}
    \sigma_{0,y}^2  I_{\ell_{y, 1:m_y}  } & \bm{0}_{\ell_{y, 1:m_y} }  \\
    \bm{0}_{\ell_{y, 1:m_y}}^T & \tau \phi \\
\end{pmatrix}^{-1} 
\end{vmatrix}^{-\frac{1}{2}}  $ giving an overall probability of 
$$ p(\bm{y}| \tilde{\bm{B}}_y  \bm{X}, \bm{W}, \phi)  \propto (2 \pi)^{ -\frac{n_1}{2} }  \left( \frac{1}{\phi} \right)^{\frac{n_1}{2}} \left(\frac{1}{\sigma_{0,y}^2}  \right)^{\frac{\ell_{y, 1:m_y}}{2}} \left(\frac{1}{\tau \phi}  \right)^{\frac{1}{2}}   \times $$
$$ \exp \Bigg(   \frac{1}{2} \bm{y}^T (\phi I)^{-1}  \tilde{\bm{B}}_{y} \Bigg( \tilde{\bm{B}}_{y}^T (\phi I)^{-1} \tilde{\bm{B}}_{y} +
     \begin{pmatrix}
    \sigma_{0,y}^2  I_{\ell_{y, 1:m_y}  } & \bm{0}_{\ell_{y, 1:m_y} }  \\
    \bm{0}_{\ell_{y, 1:m_y}}^T & \tau \phi \\
\end{pmatrix}^{-1} \Bigg)^{-1} \tilde{\bm{B}}_{y}^T (\phi I)^{-1} \bm{y}
    - \frac{1}{2} \bm{y}^T (\phi I)^{-1} \bm{y} \Bigg)  $$
$$ \begin{vmatrix}
     \tilde{\bm{B}}_{y}^T (\phi I)^{-1} \tilde{\bm{B}}_{y} +
     \begin{pmatrix}
    \sigma_{0,y}^2  I_{\ell_{y, 1:m_y}  } & \bm{0}_{\ell_{y, 1:m_y} }  \\
    \bm{0}_{\ell_{y, 1:m_y}}^T & \tau \phi \\
\end{pmatrix}^{-1} 
\end{vmatrix}^{-\frac{1}{2}}  $$

\section{Further Simulation Results}\label{further_sims}

\subsubsection{  \cite{iqbal2023bayesian} Simulations}

As in \cite{iqbal2023bayesian} we generate data as follows: \\
$$Y_i^* = 0.5 + \bm{x}_i \bm{\beta}  + \eta_i \ , \  \bm{\beta}  = (0.25,0.5,1,0,\dots,0) \in \mathcal{R}^p$$ 
$$ Z_{i}^* =  \alpha_0 + \bm{x}_i \bm{\alpha} + \xi_i  \ , \ \bm{\alpha}  = (0.5,1,1.5,0,\dots,0)/\sqrt{2} \in \mathcal{R}^p$$ 
The intercept $\alpha_0$ is set to a value that implies a proportion of 0.3 of the training data outcomes are unselected. The errors $\eta_i$ and $\xi_i$ are bivariate standard normal with correlation $\rho =0.5 $. There is no excluded instrument.

\subsubsection{  \cite{iqbal2023bayesian} Full Simulation Study Results}

\begin{table}[ht]
\centering
\begin{tabular}{p{0.5cm}p{0.3cm}p{0.7cm}p{0.9cm}p{0.9cm}p{0.9cm}p{0.9cm}p{0.9cm}p{0.9cm}p{0.9cm}p{1.1cm}p{1.1cm}p{1.1cm}p{1.1cm}}
  \hline
n & p & corr & Tobit VH orig & Tobit Ding & Tobit Omori & BART & Soft BART & Sparse Bayes Tobit & Heck ML & TO BART 2 Ding & TO BART 2 Ding not sparse & TO BART 2 marg & Soft TOBART 2 VH orig \\ 
  \hline
500 & 10 & 0.50 & 0.005 & 0.005 & 0.005 & 0.004 & 0.004 & 0.002 & 0.004 & 0.016 & 0.012 & 0.007 & 0.004 \\ 
  1000 & 10 & 0.50 & 0.003 & 0.003 & 0.003 & 0.003 & 0.002 & 0.001 & 0.002 & 0.009 & 0.010 & 0.005 & 0.003 \\ 
  500 & 25 & 0.50 & 0.011 & 0.011 & 0.011 & 0.005 & 0.004 & 0.002 & 0.011 & 0.019 & 0.018 & 0.008 & 0.004 \\ 
  1000 & 25 & 0.50 & 0.005 & 0.005 & 0.005 & 0.003 & 0.002 & 0.001 & 0.004 & 0.008 & 0.013 & 0.007 & 0.002 \\ 
  500 & 50 & 0.50 & 0.032 & 0.021 & 0.032 & 0.005 & 0.005 & 0.002 & 0.024 & 0.016 & 0.023 & 0.010 & 0.011 \\ 
  1000 & 50 & 0.50 & 0.009 & 0.010 & 0.010 & 0.003 & 0.002 & 0.001 & 0.009 & 0.008 & 0.018 & 0.006 & 0.002 \\ 
   \hline
\end{tabular}
\caption{Selection Probability RMSE, \cite{iqbal2023bayesian} simulations}
\end{table}


\begin{table}[ht]
\centering
\begin{tabular}{p{0.5cm}p{0.3cm}p{0.7cm}p{0.9cm}p{0.9cm}p{0.9cm}p{0.9cm}p{0.9cm}p{0.9cm}p{0.9cm}p{1.1cm}p{1.1cm}p{1.1cm}p{1.1cm}}
  \hline
n & p & corr & Tobit VH orig & Tobit Ding & Tobit Omori & BART & Soft BART & Sparse Bayes Tobit & Heck ML & TO BART 2 Ding & TO BART 2 Ding not sparse & TO BART 2 marg & Soft TOBART 2 VH orig \\ 
  \hline
500 & 10 & 0.50 & 0.337 & 0.337 & 0.351 & 0.634 & 0.657 & 0.163 & 0.219 & 1.181 & 0.671 & 0.805 & 0.731 \\ 
  1000 & 10 & 0.50 & 0.308 & 0.291 & 0.309 & 0.544 & 0.607 & 0.112 & 0.154 & 0.988 & 0.617 & 0.827 & 0.648 \\ 
  500 & 25 & 0.50 & 0.439 & 0.403 & 0.484 & 0.710 & 0.666 & 0.216 & 0.325 & 1.278 & 0.786 & 0.966 & 0.763 \\ 
  1000 & 25 & 0.50 & 0.361 & 0.341 & 0.372 & 0.569 & 0.630 & 0.094 & 0.203 & 0.929 & 0.675 & 0.834 & 0.613 \\ 
  500 & 50 & 0.50 & 0.687 & 0.518 & 0.689 & 0.604 & 0.646 & 0.196 & 0.509 & 1.356 & 0.772 & 0.970 & 0.695 \\ 
  1000 & 50 & 0.50 & 0.393 & 0.392 & 0.401 & 0.612 & 0.664 & 0.100 & 0.306 & 1.015 & 0.775 & 0.996 & 0.742 \\ 
   \hline
\end{tabular}
\caption{$f_y(\bm{x})$ prediction RMSE}
\end{table}

\begin{table}[ht]
\centering
\begin{tabular}{p{0.5cm}p{0.3cm}p{0.7cm}p{0.9cm}p{0.9cm}p{0.9cm}p{0.9cm}p{0.9cm}p{0.9cm}p{0.9cm}p{1.1cm}p{1.1cm}p{1.1cm}p{1.1cm}}
  \hline
n & p & corr & Tobit VH orig & Tobit Ding & Tobit Omori & BART & Soft BART & Sparse Bayes Tobit & Heck ML & TO BART 2 Ding & TO BART 2 Ding not sparse & TO BART 2 marg & Soft TOBART 2 VH orig \\ 
  \hline
500 & 10 & 0.50 & 0.944 & 0.940 & 0.947 & 0.901 & 0.894 & 0.950 & 0.973 & 0.861 & 0.912 & 0.900 & 0.914 \\ 
  1000 & 10 & 0.50 & 0.955 & 0.951 & 0.955 & 0.916 & 0.907 & 0.949 & 0.970 & 0.869 & 0.920 & 0.892 & 0.917 \\ 
  500 & 25 & 0.50 & 0.951 & 0.947 & 0.953 & 0.887 & 0.896 & 0.949 & 0.977 & 0.852 & 0.899 & 0.871 & 0.910 \\ 
  1000 & 25 & 0.50 & 0.955 & 0.950 & 0.957 & 0.914 & 0.905 & 0.954 & 0.980 & 0.879 & 0.914 & 0.886 & 0.926 \\ 
  500 & 50 & 0.50 & 0.897 & 0.944 & 0.898 & 0.903 & 0.902 & 0.947 &  & 0.850 & 0.902 & 0.870 & 0.913 \\ 
  1000 & 50 & 0.50 & 0.946 & 0.942 & 0.948 & 0.896 & 0.889 & 0.944 & 0.975 & 0.857 & 0.892 & 0.850 & 0.903 \\ 
   \hline
\end{tabular}
\caption{Latent outcome prediction interval coverage}

\end{table}

\begin{table}[ht]
\centering
\begin{tabular}{p{0.5cm}p{0.3cm}p{0.7cm}p{0.9cm}p{0.9cm}p{0.9cm}p{0.9cm}p{0.9cm}p{0.9cm}p{0.9cm}p{1.1cm}p{1.1cm}p{1.1cm}p{1.1cm}}
  \hline
n & p & corr & Tobit VH orig & Tobit Ding & Tobit Omori & BART & Soft BART & Sparse Bayes Tobit & Heck ML & TO BART 2 Ding & TO BART 2 Ding not sparse & TO BART 2 marg & Soft TOBART 2 VH orig \\ 
  \hline
500 & 10 & 0.50 & 4.129 & 4.024 & 4.205 & 3.899 & 3.877 & 3.988 & 4.634 & 4.548 & 4.097 & 4.207 & 4.226 \\ 
  1000 & 10 & 0.50 & 4.198 & 4.110 & 4.199 & 3.940 & 3.905 & 3.972 & 4.469 & 4.216 & 4.105 & 4.125 & 4.187 \\ 
  500 & 25 & 0.50 & 4.267 & 4.155 & 4.417 & 3.865 & 3.871 & 3.948 & 4.853 & 4.709 & 4.134 & 4.168 & 4.260 \\ 
  1000 & 25 & 0.50 & 4.279 & 4.178 & 4.316 & 3.950 & 3.937 & 4.019 & 4.728 & 4.187 & 4.159 & 4.124 & 4.197 \\ 
  500 & 50 & 0.50 & 3.995 & 4.342 & 3.996 & 3.849 & 3.886 & 3.999 &  & 4.906 & 4.146 & 4.171 & 4.258 \\ 
  1000 & 50 & 0.50 & 4.200 & 4.155 & 4.243 & 3.854 & 3.855 & 3.933 & 4.811 & 4.167 & 4.088 & 4.046 & 4.091 \\ 
   \hline
\end{tabular}
\caption{Latent outcome prediction interval length}

\end{table}

\begin{table}[ht]
\centering
\begin{tabular}{p{0.5cm}p{0.3cm}p{0.7cm}p{0.9cm}p{0.9cm}p{0.9cm}p{0.9cm}p{0.9cm}p{0.9cm}p{0.9cm}p{1.1cm}p{1.1cm}p{1.1cm}p{1.1cm}}
  \hline
n & p & corr & Tobit VH orig & Tobit Ding & Tobit Omori & BART & Soft BART & Sparse Bayes Tobit & Heck ML & TO BART 2 Ding & TO BART 2 Ding not sparse & TO BART 2 marg & Soft TOBART 2 VH orig \\ 
  \hline
500 & 10 & 0.50 & 0.046 & 0.025 & 0.061 & 0.250 & 0.250 & 0.027 & 0.034 & 1.216 & 0.411 & 0.887 & 0.669 \\ 
  1000 & 10 & 0.50 & 0.031 & 0.008 & 0.032 & 0.250 & 0.250 & 0.017 & 0.017 & 1.051 & 0.344 & 0.902 & 0.582 \\ 
  500 & 25 & 0.50 & 0.069 & 0.038 & 0.097 & 0.250 & 0.250 & 0.099 & 0.151 & 1.348 & 0.732 & 1.482 & 0.783 \\ 
  1000 & 25 & 0.50 & 0.036 & 0.009 & 0.044 & 0.250 & 0.250 & 0.012 & 0.015 & 0.899 & 0.424 & 0.883 & 0.439 \\ 
  500 & 50 & 0.50 & 0.179 & 0.043 & 0.179 & 0.250 & 0.250 & 0.021 & 0.145 & 1.556 & 0.606 & 1.515 & 0.741 \\ 
  1000 & 50 & 0.50 & 0.021 & 0.011 & 0.032 & 0.250 & 0.250 & 0.012 & 0.017 & 1.009 & 0.616 & 1.320 & 0.601 \\ 
   \hline
\end{tabular}
\caption{Correlation estimate RMSE}

\end{table}

\begin{table}[ht]
\centering
\begin{tabular}{p{0.5cm}p{0.3cm}p{0.7cm}p{0.9cm}p{0.9cm}p{0.9cm}p{0.9cm}p{0.9cm}p{0.9cm}p{0.9cm}p{1.1cm}p{1.1cm}p{1.1cm}p{1.1cm}}
  \hline
n & p & corr & Tobit VH orig & Tobit Ding & Tobit Omori & BART & Soft BART & Sparse Bayes Tobit & Heck ML & TO BART 2 Ding & TO BART 2 Ding not sparse & TO BART 2 marg & Soft TOBART 2 VH orig \\ 
  \hline
500 & 10 & 0.50 & 0.675 & 0.483 & 0.699 & 0.000 & 0.000 & 0.513 & 0.594 & -0.582 & -0.063 & -0.266 & -0.106 \\ 
  1000 & 10 & 0.50 & 0.655 & 0.528 & 0.657 & 0.000 & 0.000 & 0.436 & 0.461 & -0.514 & -0.067 & -0.409 & -0.189 \\ 
  500 & 25 & 0.50 & 0.657 & 0.487 & 0.721 & 0.000 & 0.000 & 0.335 & 0.396 & -0.633 & -0.333 & -0.661 & -0.258 \\ 
  1000 & 25 & 0.50 & 0.670 & 0.527 & 0.680 & 0.000 & 0.000 & 0.450 & 0.492 & -0.440 & -0.132 & -0.305 & -0.037 \\ 
  500 & 50 & 0.50 & 0.844 & 0.626 & 0.848 & 0.000 & 0.000 & 0.501 & 0.731 & -0.730 & -0.247 & -0.554 & 0.121 \\ 
  1000 & 50 & 0.50 & 0.620 & 0.528 & 0.637 & 0.000 & 0.000 & 0.442 & 0.496 & -0.495 & -0.256 & -0.642 & -0.240 \\ 
   \hline
\end{tabular}
\caption{Mean Correlation Estimate}

\end{table}

\FloatBarrier

\subsubsection{  \cite{brewer2024addressing} Simulations}

We simulate the data generating processes described by \cite{brewer2024addressing}, with less observations and variables to ensure computational feasibility. We also consider non-normally distributed errors. 
Covariates $X_1,...,X_{10}$ are generated from a multivariate normal distribution with $\text{Cov}(X_{ki},X_{ji}) = 0.3^{|k-j|}$. The excluded instrument, $W_{exc,i}$, that appears in the selection equation, but not the outcome equation is generated by $W_{exc,i} = \sum_{j=1}^{10}  0.05 X_{ji} + e_i \ , \ e_{w,i} \sim \mathcal{N}(0,0.75) $. 
%
%

We consider the following DGPs (we do not inlude DGP 4 from \cite{brewer2024addressing}):

\begin{enumerate}
    \item Outcome: $Y_i^* = \sum_{j=1}^{10} \frac{0.4}{j^2} X_{ji}  + \eta_i$ \\
    Selection: $ Z_{i}^* =  1.25 + \sum_{j=1}^{10} \frac{0.1}{(10.5 - j)^2} X_{jo} + W_{exc,i} + \xi_i  $. The errors $\eta_i$ and $\xi_i$ are bivariate standard normal with correlation $\rho \in \{ 0,0.45,0.9 \} $.
    \item Outcome: $Y_i^* = -0.25  1.25 \sin \left( \frac{\pi}{4} +  0.75 \pi \sum_{j=1}^{10} \frac{0.4}{j^2} X_{ji} \right) + \eta_i$ \\
    Selection: $ Z_{i}^* =  1.25 + \sum_{j=1}^{10} \frac{0.1}{(10.5 - j)^2} X_{jo} + W_{exc,i} + \xi_i  $. The errors $\eta_i$ and $\xi_i$ are bivariate standard normal with correlation $\rho \in \{ 0,0.45,0.9 \} $.
    \item Outcome: $Y_i^* = -0.25  1.25 \sin \left( \frac{\pi}{4} +  0.75 \pi \sum_{j=1}^{10} \frac{0.4}{j^2} X_{ji} \right) + \eta_i$ \\
    Selection: $ Z_{i}^* =  1.25 + \sum_{j=1}^{10} \frac{0.1}{(10.5 - j)^2} X_{jo} + 0.001 W_{exc,i} + \xi_i  $. The errors $\eta_i$ and $\xi_i$ are bivariate standard normal with correlation $\rho \in \{ 0,0.45,0.9 \} $.
    
    \item[5.] Outcome: $Y_i^* = \sum_{j=1}^{10} \frac{0.4}{j^2} X_{ji}  + \eta_i$ \\
    Selection: $ Z_{i}^* =  1.25 + \sum_{j=1}^{10} \frac{0.1}{(10.5 - j)^2} X_{jo} + W_{exc,i} + \xi_i  $. The errors are generated by $\eta_i \sim \rho V_i + \sqrt{1 - \rho^2} [0.8 \mathcal{N}(-0.5,0.2) + 0.2 \mathcal{N}(2,0.4) ]$ and $\xi_i \sim 0.8 \mathcal{N}(-0.5,0.2) + 0.2 \mathcal{N}(2,0.4) $, with correlation $\rho \in \{ 0,0.45,0.9 \} $.
    \end{enumerate}

We also consider the above DGPs with the following non-normal distributions for the errors.
\begin{itemize}
    \item $\eta_i \sim \rho V_i + \rho t_{\nu = 5} $ and $\xi_i \sim t_{\nu = 5}  $, where $t_{nu = 5}$ is a standard t-distribution with $5$ degrees of freedom. The correlation is $\rho = \frac{1}{sqrt{2}} $.
    \item The errors are generated from a mixture of standard bivariate normals $ 0.3 \mathcal{MVN}_2 ( (0,-2.1)', \Sigma) + 0.7 \mathcal{MVN}_2 ( (0,0.9)', \Sigma) $  where $\Sigma_{12} = 0.85$ , and the implied correlation is $0.5$.
    \end{itemize}

The numbers of training and test observations are set to $2500$ and $500$ respective;y, and the number of repetitions for each simulation scenario is set to $10$.

\subsubsection{\cite{brewer2024addressing} Full Simulation Study Results   }

\begin{table}[ht]
\centering
\begin{tabular}{p{0.6cm}p{0.6cm}|p{0.9cm}p{0.9cm}p{0.9cm}p{0.9cm}p{0.9cm}p{0.9cm}p{0.9cm}p{1.1cm}p{1.1cm}p{1.1cm}p{1.1cm}}
  \hline
DGP & corr & Tobit VH orig & Tobit Ding & Tobit Omori & BART & Soft BART & Sparse Tobit & TO BART 2 VH orig & TO BART 2 Ding & TO BART 2 marg & Soft TO BART 2 VH orig \\ 
  \hline
1 & 0.00 & 2.154 & 2.151 & 2.153 & 1.000 & 0.727 & 0.249 & 1.409 & 1.454 & 1.808 & 0.735 \\ 
  2 & 0.00 & 2.143 & 2.144 & 2.155 & 1.000 & 0.727 & 0.252 & 1.387 & 1.423 & 1.775 & 0.716 \\ 
  3 & 0.00 & 2.041 & 2.041 & 2.037 & 1.000 & 0.780 & 0.357 & 1.365 & 1.456 & 1.587 & 0.758 \\ 
  5 & 0.00 & 0.965 & 0.963 & 0.962 & 1.000 & 0.943 & 0.377 & 1.244 & 1.266 & 1.156 & 0.942 \\ 
  1 & 0.45 & 2.115 & 2.139 & 2.122 & 1.000 & 0.724 & 0.196 & 1.556 & 1.565 & 1.533 & 0.686 \\ 
  2 & 0.45 & 2.128 & 2.143 & 2.134 & 1.000 & 0.724 & 0.213 & 1.539 & 1.589 & 1.667 & 0.687 \\ 
  3 & 0.45 & 1.832 & 1.834 & 1.829 & 1.000 & 0.685 & 0.311 & 1.166 & 1.229 & 1.369 & 0.673 \\ 
  5 & 0.45 & 0.754 & 0.972 & 0.767 & 1.000 & 0.979 & 0.203 & 1.064 & 1.232 & 0.894 & 0.714 \\ 
  1 & 0.90 & 2.158 & 2.258 & 2.155 & 1.000 & 0.625 & 0.232 & 1.370 & 1.422 & 1.684 & 0.566 \\ 
  2 & 0.90 & 2.200 & 2.274 & 2.200 & 1.000 & 0.625 & 0.248 & 1.380 & 1.464 & 1.680 & 0.603 \\ 
  3 & 0.90 & 2.120 & 2.191 & 2.131 & 1.000 & 0.777 & 0.278 & 1.232 & 1.374 & 1.411 & 0.701 \\ 
  5 & 0.90 & 0.673 & 0.871 & 0.672 & 1.000 & 0.913 & 0.115 & 0.782 & 0.983 & 0.602 & 0.274 \\ 
   \hline
\end{tabular}
\caption{Selection Probability RMSE, relative to BART}

\end{table}

\begin{table}[ht]
\centering
\begin{tabular}{p{0.6cm}p{0.6cm}|p{0.9cm}p{0.9cm}p{0.9cm}p{0.9cm}p{0.9cm}p{0.9cm}p{0.9cm}p{1.1cm}p{1.1cm}p{1.1cm}p{1.1cm}}
  \hline
DGP & corr & Tobit VH orig & Tobit Ding & Tobit Omori & BART & Soft BART & Sparse Tobit & TO BART 2 VH orig & TO BART 2 Ding & TO BART 2 marg & Soft TO BART 2 VH orig \\ 
  \hline
1 & 0.00 & 0.677 & 0.666 & 0.676 & 1.000 & 0.554 & 0.365 & 0.854 & 0.865 & 0.850 & 0.576 \\ 
  2 & 0.00 & 2.350 & 2.349 & 2.350 & 1.000 & 0.746 & 2.556 & 0.997 & 0.998 & 0.986 & 0.760 \\ 
  3 & 0.00 & 0.837 & 0.684 & 0.860 & 1.000 & 0.573 & 0.409 & 0.870 & 0.875 & 0.911 & 0.600 \\ 
  5 & 0.00 & 0.709 & 0.705 & 0.709 & 1.000 & 0.587 & 0.371 & 0.903 & 0.945 & 0.870 & 0.595 \\ 
  1 & 0.45 & 0.396 & 0.440 & 0.395 & 1.000 & 0.792 & 0.212 & 0.496 & 0.557 & 0.487 & 0.361 \\ 
  2 & 0.45 & 1.543 & 1.551 & 1.543 & 1.000 & 0.832 & 1.679 & 0.645 & 0.680 & 0.643 & 0.505 \\ 
  3 & 0.45 & 0.492 & 0.758 & 0.511 & 1.000 & 0.796 & 0.263 & 0.551 & 0.662 & 0.570 & 0.417 \\ 
  5 & 0.45 & 0.349 & 0.310 & 0.346 & 1.000 & 0.811 & 0.247 & 0.374 & 0.387 & 0.411 & 0.284 \\ 
  1 & 0.90 & 0.240 & 0.272 & 0.239 & 1.000 & 0.897 & 0.123 & 0.257 & 0.306 & 0.262 & 0.190 \\ 
  2 & 0.90 & 0.989 & 1.001 & 0.989 & 1.000 & 0.913 & 1.075 & 0.389 & 0.414 & 0.401 & 0.310 \\ 
  3 & 0.90 & 0.298 & 0.749 & 0.284 & 1.000 & 0.903 & 0.148 & 0.299 & 0.817 & 0.314 & 0.224 \\ 
  5 & 0.90 & 0.187 & 0.174 & 0.187 & 1.000 & 0.928 & 0.122 & 0.195 & 0.198 & 0.223 & 0.167 \\ 
   \hline
\end{tabular}
\caption{$f_y(\bm{x})$ prediction RMSE, relative to BART}
\end{table}

\begin{table}[ht]
\centering
\begin{tabular}{p{0.6cm}p{0.6cm}|p{0.9cm}p{0.9cm}p{0.9cm}p{0.9cm}p{0.9cm}p{0.9cm}p{0.9cm}p{1.1cm}p{1.1cm}p{1.1cm}p{1.1cm}}
  \hline
DGP & corr & Tobit VH orig & Tobit Ding & Tobit Omori & BART & Soft BART & Sparse Tobit & TO BART 2 VH orig & TO BART 2 Ding & TO BART 2 marg & Soft TO BART 2 VH orig \\ 
  \hline
1 & 0.00 & 0.992 & 0.992 & 0.992 & 1.000 & 0.991 & 0.989 & 0.995 & 0.995 & 0.997 & 0.991 \\ 
  2 & 0.00 & 1.097 & 1.097 & 1.097 & 1.000 & 0.992 & 1.118 & 1.000 & 1.001 & 1.003 & 0.993 \\ 
  3 & 0.00 & 0.997 & 0.994 & 0.997 & 1.000 & 0.994 & 0.991 & 0.998 & 0.998 & 1.001 & 0.994 \\ 
  5 & 0.00 & 0.996 & 0.996 & 0.996 & 1.000 & 0.993 & 0.990 & 0.996 & 0.997 & 0.999 & 0.993 \\ 
  1 & 0.45 & 0.973 & 0.973 & 0.973 & 1.000 & 0.987 & 0.966 & 0.974 & 0.976 & 0.976 & 0.971 \\ 
  2 & 0.45 & 1.070 & 1.071 & 1.070 & 1.000 & 0.986 & 1.087 & 0.978 & 0.978 & 0.982 & 0.968 \\ 
  3 & 0.45 & 0.983 & 0.990 & 0.984 & 1.000 & 0.992 & 0.976 & 0.983 & 0.987 & 0.987 & 0.980 \\ 
  5 & 0.45 & 0.944 & 0.943 & 0.944 & 1.000 & 0.979 & 0.941 & 0.947 & 0.949 & 0.945 & 0.942 \\ 
  1 & 0.90 & 0.914 & 0.915 & 0.914 & 1.000 & 0.982 & 0.909 & 0.915 & 0.917 & 0.916 & 0.911 \\ 
  2 & 0.90 & 1.001 & 1.003 & 1.001 & 1.000 & 0.986 & 1.018 & 0.919 & 0.921 & 0.921 & 0.914 \\ 
  3 & 0.90 & 0.942 & 0.971 & 0.941 & 1.000 & 0.987 & 0.936 & 0.940 & 0.978 & 0.940 & 0.937 \\ 
  5 & 0.90 & 0.835 & 0.834 & 0.835 & 1.000 & 0.979 & 0.831 & 0.837 & 0.837 & 0.838 & 0.834 \\ 
   \hline
\end{tabular}
\caption{True latent outcome prediction RMSE, relative to BART}
\end{table}

\begin{table}[ht]
\centering
\begin{tabular}{p{0.6cm}p{0.6cm}|p{0.9cm}p{0.9cm}p{0.9cm}p{0.9cm}p{0.9cm}p{0.9cm}p{0.9cm}p{1.1cm}p{1.1cm}p{1.1cm}p{1.1cm}}
  \hline
DGP & corr & Tobit VH orig & Tobit Ding & Tobit Omori & BART & Soft BART & Sparse Tobit & TO BART 2 VH orig & TO BART 2 Ding & TO BART 2 marg & Soft TO BART 2 VH orig \\ 
  \hline
1 & 0.00 & 0.871 & 0.868 & 0.870 & 0.985 & 0.982 & 0.956 & 0.953 & 0.949 & 0.972 & 0.978 \\ 
  2 & 0.00 & 0.211 & 0.208 & 0.213 & 0.973 & 0.934 & 0.157 & 0.933 & 0.930 & 0.943 & 0.934 \\ 
  3 & 0.00 & 0.889 & 0.874 & 0.888 & 0.984 & 0.972 & 0.945 & 0.960 & 0.957 & 0.981 & 0.974 \\ 
  5 & 0.00 & 0.862 & 0.862 & 0.861 & 0.983 & 0.976 & 0.950 & 0.958 & 0.938 & 0.977 & 0.977 \\ 
  1 & 0.45 & 0.853 & 0.781 & 0.854 & 0.885 & 0.684 & 0.946 & 0.954 & 0.914 & 0.972 & 0.966 \\ 
  2 & 0.45 & 0.212 & 0.226 & 0.211 & 0.884 & 0.737 & 0.153 & 0.928 & 0.908 & 0.950 & 0.931 \\ 
  3 & 0.45 & 0.874 & 0.559 & 0.874 & 0.889 & 0.718 & 0.933 & 0.962 & 0.909 & 0.978 & 0.953 \\ 
  5 & 0.45 & 0.808 & 0.855 & 0.815 & 0.853 & 0.715 & 0.845 & 0.957 & 0.945 & 0.977 & 0.961 \\ 
  1 & 0.90 & 0.809 & 0.728 & 0.811 & 0.647 & 0.511 & 0.895 & 0.947 & 0.909 & 0.985 & 0.952 \\ 
  2 & 0.90 & 0.197 & 0.225 & 0.196 & 0.683 & 0.550 & 0.125 & 0.897 & 0.896 & 0.932 & 0.910 \\ 
  3 & 0.90 & 0.821 & 0.312 & 0.816 & 0.678 & 0.472 & 0.892 & 0.961 & 0.680 & 0.981 & 0.955 \\ 
  5 & 0.90 & 0.794 & 0.829 & 0.795 & 0.712 & 0.631 & 0.763 & 0.936 & 0.938 & 0.942 & 0.882 \\ 
   \hline
\end{tabular}
\caption{$f_y(\bm{x})$ 95\% prediction interval mean coverage}
\end{table}

\begin{table}[ht]
\centering
\begin{tabular}{p{0.6cm}p{0.6cm}|p{0.9cm}p{0.9cm}p{0.9cm}p{0.9cm}p{0.9cm}p{0.9cm}p{0.9cm}p{1.1cm}p{1.1cm}p{1.1cm}p{1.1cm}}
  \hline
DGP & corr & Tobit VH orig & Tobit Ding & Tobit Omori & BART & Soft BART & Sparse Tobit & TO BART 2 VH orig & TO BART 2 Ding & TO BART 2 marg & Soft TO BART 2 VH orig \\ 
  \hline
1 & 0.00 & 0.943 & 0.942 & 0.943 & 0.941 & 0.941 & 0.941 & 0.943 & 0.943 & 0.940 & 0.941 \\ 
  2 & 0.00 & 0.944 & 0.944 & 0.945 & 0.940 & 0.941 & 0.943 & 0.940 & 0.939 & 0.937 & 0.942 \\ 
  3 & 0.00 & 0.946 & 0.943 & 0.946 & 0.943 & 0.942 & 0.941 & 0.945 & 0.945 & 0.942 & 0.942 \\ 
  5 & 0.00 & 0.926 & 0.926 & 0.925 & 0.928 & 0.927 & 0.926 & 0.928 & 0.928 & 0.925 & 0.927 \\ 
  1 & 0.45 & 0.953 & 0.948 & 0.954 & 0.944 & 0.945 & 0.953 & 0.953 & 0.949 & 0.953 & 0.954 \\ 
  2 & 0.45 & 0.956 & 0.954 & 0.956 & 0.940 & 0.946 & 0.958 & 0.953 & 0.950 & 0.953 & 0.956 \\ 
  3 & 0.45 & 0.954 & 0.946 & 0.953 & 0.943 & 0.945 & 0.953 & 0.952 & 0.946 & 0.951 & 0.954 \\ 
  5 & 0.45 & 0.948 & 0.950 & 0.947 & 0.948 & 0.952 & 0.947 & 0.949 & 0.948 & 0.947 & 0.946 \\ 
  1 & 0.90 & 0.950 & 0.932 & 0.949 & 0.892 & 0.894 & 0.949 & 0.947 & 0.934 & 0.949 & 0.949 \\ 
  2 & 0.90 & 0.948 & 0.936 & 0.948 & 0.892 & 0.898 & 0.948 & 0.946 & 0.931 & 0.947 & 0.947 \\ 
  3 & 0.90 & 0.950 & 0.907 & 0.950 & 0.898 & 0.899 & 0.949 & 0.948 & 0.907 & 0.950 & 0.949 \\ 
  5 & 0.90 & 0.957 & 0.955 & 0.958 & 0.891 & 0.900 & 0.957 & 0.954 & 0.952 & 0.956 & 0.957 \\ 
   \hline
\end{tabular}
\caption{True latent outcome $95\%$ prediction interval mean coverage}
\end{table}

\begin{table}[ht]
\centering
\begin{tabular}{p{0.6cm}p{0.6cm}|p{0.9cm}p{0.9cm}p{0.9cm}p{0.9cm}p{0.9cm}p{0.9cm}p{0.9cm}p{1.1cm}p{1.1cm}p{1.1cm}p{1.1cm}}
  \hline
DGP & corr & Tobit VH orig & Tobit Ding & Tobit Omori & BART & Soft BART & Sparse Tobit & TO BART 2 VH orig & TO BART 2 Ding & TO BART 2 marg & Soft TO BART 2 VH orig \\ 
  \hline
1 & 0.00 & 3.956 & 3.955 & 3.956 & 3.964 & 3.948 & 3.941 & 3.971 & 3.970 & 3.934 & 3.948 \\ 
  2 & 0.00 & 4.392 & 4.389 & 4.392 & 3.993 & 3.981 & 4.470 & 4.008 & 4.007 & 3.961 & 3.983 \\ 
  3 & 0.00 & 4.004 & 3.953 & 4.009 & 3.958 & 3.945 & 3.946 & 3.970 & 3.965 & 3.970 & 3.957 \\ 
  5 & 0.00 & 4.393 & 4.395 & 4.392 & 4.410 & 4.391 & 4.377 & 4.406 & 4.407 & 4.370 & 4.385 \\ 
  1 & 0.45 & 3.931 & 3.870 & 3.931 & 3.851 & 3.834 & 3.916 & 3.931 & 3.881 & 3.922 & 3.920 \\ 
  2 & 0.45 & 4.406 & 4.353 & 4.407 & 3.894 & 3.881 & 4.481 & 3.980 & 3.929 & 3.957 & 3.967 \\ 
  3 & 0.45 & 3.960 & 3.831 & 3.963 & 3.834 & 3.823 & 3.917 & 3.934 & 3.860 & 3.927 & 3.927 \\ 
  5 & 0.45 & 4.540 & 4.452 & 4.536 & 4.408 & 4.388 & 4.556 & 4.498 & 4.463 & 4.531 & 4.516 \\ 
  1 & 0.90 & 3.945 & 3.691 & 3.946 & 3.539 & 3.518 & 3.925 & 3.920 & 3.690 & 3.959 & 3.951 \\ 
  2 & 0.90 & 4.352 & 4.156 & 4.353 & 3.578 & 3.557 & 4.417 & 3.952 & 3.734 & 3.978 & 3.984 \\ 
  3 & 0.90 & 3.952 & 3.512 & 3.950 & 3.505 & 3.491 & 3.911 & 3.921 & 3.528 & 3.942 & 3.943 \\ 
  5 & 0.90 & 4.723 & 4.600 & 4.722 & 4.321 & 4.299 & 4.726 & 4.689 & 4.600 & 4.746 & 4.747 \\ 
   \hline
\end{tabular}
\caption{True latent outcome $95\%$ prediction interval mean length}
\end{table}

\begin{table}[ht]
\centering
\begin{tabular}{p{0.6cm}p{0.6cm}|p{1.1cm}p{1.1cm}p{1.1cm}p{1.1cm}p{1.1cm}p{1.1cm}p{1.1cm}p{1.1cm}p{1.1cm}p{1.1cm}p{1.1cm}}
  \hline
DGP & corr & Tobit VH orig & Tobit Ding & Tobit Omori & BART & Soft BART & Sparse Tobit & TO BART 2 VH orig & TO BART 2 Ding & TO BART 2 marg & Soft TO BART 2 VH orig \\ 
  \hline
1 & 0.00 & -0.001 & 0.000 & -0.001 & 0.000 & 0.000 & -0.018 & -0.040 & -0.024 & -0.004 & -0.016 \\ 
  2 & 0.00 & -0.004 & -0.001 & -0.003 & 0.000 & 0.000 & -0.035 & -0.032 & -0.019 & -0.009 & -0.023 \\ 
  3 & 0.00 & -0.021 & -0.006 & -0.060 & 0.000 & 0.000 & -0.037 & -0.048 & -0.021 & -0.038 & -0.038 \\ 
  5 & 0.00 & 0.008 & 0.005 & 0.012 & 0.000 & 0.000 & -0.001 & -0.024 & -0.016 & 0.005 & -0.007 \\ 
  1 & 0.45 & 0.418 & 0.242 & 0.418 & 0.000 & 0.000 & 0.416 & 0.410 & 0.240 & 0.454 & 0.427 \\ 
  2 & 0.45 & 0.373 & 0.219 & 0.376 & 0.000 & 0.000 & 0.350 & 0.413 & 0.244 & 0.461 & 0.427 \\ 
  3 & 0.45 & 0.451 & 0.045 & 0.423 & 0.000 & 0.000 & 0.426 & 0.420 & 0.159 & 0.434 & 0.442 \\ 
  5 & 0.45 & 0.678 & 0.462 & 0.670 & 0.000 & 0.000 & 0.716 & 0.568 & 0.447 & 0.710 & 0.650 \\ 
  1 & 0.90 & 0.894 & 0.703 & 0.894 & 0.000 & 0.000 & 0.891 & 0.894 & 0.692 & 0.933 & 0.911 \\ 
  2 & 0.90 & 0.751 & 0.542 & 0.751 & 0.000 & 0.000 & 0.709 & 0.884 & 0.682 & 0.923 & 0.900 \\ 
  3 & 0.90 & 0.855 & 0.125 & 0.890 & 0.000 & 0.000 & 0.879 & 0.878 & 0.137 & 0.917 & 0.899 \\ 
  5 & 0.90 & 0.933 & 0.872 & 0.933 & 0.000 & 0.000 & 0.948 & 0.929 & 0.876 & 0.984 & 0.967 \\ 
   \hline
\end{tabular}
\caption{Mean correlation estimate for selection and outcome equation errors.}
\end{table}

\FloatBarrier


\subsection{\cite{brewer2024addressing} Simulation Study with non-normal errors}

Table \ref{Brewer_res_table_np} below includes simulation study results for DGP 1 and DGP 2 from the simulation study of \cite{brewer2024addressing}, with t-distributed and non-normally distributed errors. The purpose of these additional simulations is to investigate the effectiveness of TOBART-2-NP at modeling non-normal error terms.

It can be observed that, for these DGPs, the Dirichlet process mixture of normal distributions in the TOBART-2-NP DGP does not result in great improvements in predictions. The RMSE of the estimated mean of the latent outcome does not improve. There are some improvements in the coverage of credible intervals for the modified \cite{ding2014bayesian} prior and small potential improvements in the modeling of the correlation in the error terms for the \cite{van2011bayesian} prior. 

Possible explanations for this result include poor mixing of the Markov chain and a requirement for a greater number of observations. Perhaps the Dirichlet Process Mixture approach provides more notable improvements for error distributions that depart even further from the Gaussian distribtuion.

\begin{table}[ht]
\centering
\begin{tabular}{p{0.6cm}p{0.6cm}|p{1.1cm}p{1.1cm}p{1.1cm}p{1.1cm}p{1.1cm}p{1.1cm}p{1.1cm}p{1.1cm}}
  \hline
DGP & corr & Tobit VH  & Tobit Ding & BART & Sparse Tobit & TO BART 2 VH  & TO BART 2 Ding  & TO BART 2 VH NP & TO BART 2 Ding NP \\ 
  \hline
  \hline
  \hline
  \multicolumn{10}{c}{$f_y(\bm{x})$ prediction RMSE, relative to BART}\\
  \hline
  \hline
  \multicolumn{10}{c}{t distribution}\\
1 & 0.71 & 0.372 & 0.533 & 1.000 & 0.203 & 0.443 & 0.607 & 0.562 & 0.663 \\ 
  2 & 0.71 & 1.313 & 1.358 & 1.000 & 1.420 & 0.572 & 0.730 & 0.622 & 0.734 \\ 
    \multicolumn{10}{c}{Mixture of normals}\\
1 & 0.50 & 0.609 & 0.705 & 1.000 & 0.395 & 0.637 & 0.791 & 0.734 & 0.873 \\ 
  2 & 0.50 & 1.916 & 1.934 & 1.000 & 2.060 & 0.807 & 0.904 & 0.817 & 0.908 \\ 
   \hline
    \hline
  \hline
    \multicolumn{10}{c}{$f_y(\bm{x})$ 95\% prediction interval mean coverage}\\
   \hline  
    \hline
    \multicolumn{10}{c}{t distribution}\\
1 & 0.71 & 0.870 & 0.638 & 0.820 & 0.930 & 0.970 & 0.857 & 1.000 & 0.968 \\ 
  2 & 0.71 & 0.275 & 0.340 & 0.841 & 0.175 & 0.950 & 0.879 & 0.995 & 0.954 \\ 
    \multicolumn{10}{c}{Mixture of normals}\\
1 & 0.50 & 0.813 & 0.682 & 0.895 & 0.854 & 0.956 & 0.880 & 0.999 & 0.947 \\ 
  2 & 0.50 & 0.231 & 0.264 & 0.904 & 0.165 & 0.926 & 0.886 & 0.991 & 0.924 \\ 
     \hline
  \hline
\multicolumn{10}{c}{Selection Probability MSE, relative to BART}\\
  \hline
   \hline
       \multicolumn{10}{c}{t distribution}\\
1 & 0.71 & 1.903 & 1.845 & 1.000 & 0.541 & 1.303 & 1.248 & 1.340 & 1.369 \\ 
  2 & 0.71 & 1.869 & 1.846 & 1.000 & 0.512 & 1.308 & 1.249 & 1.382 & 1.348 \\
    \multicolumn{10}{c}{Mixture of normals}\\
    1 & 0.50 & 1.034 & 1.028 & 1.000 & 0.828 & 1.001 & 1.052 & 1.066 & 1.100 \\ 
  2 & 0.50 & 1.029 & 1.030 & 1.000 & 0.829 & 1.005 & 1.047 & 1.075 & 1.141 \\ 
     \hline
  \hline
         \multicolumn{10}{c}{Mean correlation estimate for selection and outcome equation errors}\\
           \hline
  \hline
             \multicolumn{10}{c}{t distribution}\\
1 & 0.71 & 0.666 & 0.365 & 0.000 & 0.667 & 0.664 & 0.364 & 0.724 & 0.478 \\ 
  2 & 0.71 & 0.608 & 0.331 & 0.000 & 0.591 & 0.667 & 0.362 & 0.751 & 0.517 \\ 
    \multicolumn{10}{c}{Mixture of normals}\\
1 & 0.50 & 0.498 & 0.136 & 0.000 & 0.491 & 0.460 & 0.137 & 0.456 & 0.154 \\ 
  2 & 0.50 & 0.412 & 0.124 & 0.000 & 0.377 & 0.475 & 0.143 & 0.514 & 0.184 \\ 
  \hline
  \hline
\end{tabular}
\caption{Results for \cite{brewer2024addressing} simulation study with non-normal errors. RMSE and MSE results are relative to BART.} \label{Brewer_res_table_np}
\end{table}

\FloatBarrier

\section{Cross-validation of models applied to health-insurance data}\label{data_app}

\noindent \textbf{Accuracy of Predictions in 5-fold Cross-validation}

\begin{table}[H]
\centering
\begin{tabular}{l|p{1cm}p{1cm}p{1cm}p{1cm}}
  \hline
 Method & Hit Rate & Brier & AUC & MSE cond. \\ 
   \hline
  
  Tobit-2Step & 0.769 & 0.168 & 0.656 & 1.367 \\ 
  Tobit-ML & 0.768 & 0.168 & 0.656 & 1.368 \\ 
  Logistic Regression & 0.769 & 0.168 & 0.657 &  \\ 
  RF & 0.771 & 0.165 & 0.69 & 1.365 \\ 
  BART & 0.771 & 0.166 & 0.687 & 1.358 \\ 
  Soft BART & 0.769 & 0.166 & 0.685 & 1.364 \\ 
  Bayesian Splines & 0.767 & 0.179 & 0.627 & 1.618 \\ 
  TOBART 2 & 0.769 & 0.166 & 0.689 & 1.361 \\ 
  Soft TOBART 2 & 0.769 & 0.166 & 0.687 & 1.361 \\ 
   \hline
\end{tabular}
\caption{Predictive Accuracy of models applied to RAND Health Insurance Experiment data. The second column gives the hit rate for binary censoring predictions, the third and fourth columns give the Brier score and Area under the Receiver Operating Characteristics Curve for predicted probabilities of censoring. The fifth column gives the mean squared error for predictions of the observed outcomes. All measures are evaluated in validations samples in 5-fold cross-fold validation.}
	\label{rand_cv_accuracy_table}
\end{table}


\begin{table}[H]
\centering
\begin{tabular}{l|p{2.5cm}p{2.5cm}}
  \hline
Method & Observation Interval Coverage & Observation Interval Length\\ 
  \hline
BART & 0.95 & 5.49 \\ 
Soft BART & 0.947 & 5.511 \\ 
TOBART 2 & 0.95 & 5.49 \\ 
Soft TOBART 2 & 0.951 & 5.498 \\ 
   \hline
\end{tabular}
\caption{Coverage and length of observed outcome 95\% prediction intervals produced by models applied to RAND Health Insurance Experiment data. Intervals are evaluated in validations samples in 5-fold cross-fold validation.}
	\label{rand_cv_cov_table}
\end{table}

\section{Further Applications}\label{further_application_app}

\subsection{\cite{mroz1987sensitivity} Female Labour Supply}

We apply TOBART-2 to the data studied by \cite{mroz1987sensitivity}. We are interested in the effect of a range of decisions on the female labour force participation decision (selection equation) and offered log wages (outcome equation). If wages are equal to zero (and therefore the log of wages is undefined), then the offered wage is unobserved, i.e. the observation is not selected.

The selection equation includes age, the square of age, family income, number of kids, and education level. The outcome equation includes education level, years of work experience, the square of years of work experience, and a binary variable equal to 1 if the respondent lives in an urban area.

Although when considering observed MSE conditional on selection, we have no reason to expect TOBART-2 to outperform machine learning methods naively trained on the selected observations, it can be observed in Table 	\label{mroz_cv_cov_table} that TOBART-2 slightly outperforms BART in terms of MSE and the Brier scores for predictive selection probabilities are very similar.

The estimated average effects of having kids on the decision to work and observed wages (conditional on selection) are given in Table \ref{mroz_kids_ate}, and with 95\% intervals in Table \ref{mroz_kids_ate_ints}. The variable kids does not enter the outcome equation by assumption, and therefore the estimated effect for the latent outcome is not reported. 

The estimated average effects of living in a city, which enters both the selection and outcome equations, are displayed in Tables \ref{mroz_city_ate} and \ref{mroz_city_ate_int}. It can be observed that the estimated effects from the TOBART-2 model are generally smaller in magnitude than the linear model and are indistinguishable from zero.

The expected values of offered wage by years of education are given in Table \ref{mroz_educ_latent_exp_wage}, and 95\% credible intervals are displayed in Table \ref{mroz_educ_latent_exp_wage_int}. It can be observed that  TOBART-2 estimates that a large effect occurs from 11 to 12 years of education, and the effects are negligible  from 12 to 14 years of education. This nonlinearity is not captured by the linear model.

\begin{table}[ht]
\centering
\begin{tabular}{l|p{1cm}p{1cm}p{1cm}p{1cm}}
  \hline
Method & Hit Rate & Brier & AUC & MSE cond. \\ 
  \hline
Tobit-2Step & 0.596 & 0.233 & 0.621 & 0.812 \\ 
Tobit-ML & 0.596 & 0.234 & 0.632 & 0.816 \\ 
Logistic Regression & 0.598 & 0.234 & 0.621 &  \\ 
RF & 0.583 & 0.251 & 0.584 & 0.689 \\ 
BART & 0.590 & 0.236 & 0.610 & 0.666 \\ 
Bayesian Splines & 0.550 & 0.274 & 0.614 & 1.044 \\ 
TOBART-2 & 0.608 & 0.238 & 0.608 & 0.655 \\ 
   \hline
\end{tabular}
\caption{Predictive Accuracy of models applied to female wages data \citep{mroz1987sensitivity}. The second column gives the hit rate for binary censoring predictions, the third and fourth columns give the Brier score and Area under the Receiver Operating Characteristics Curve for predicted probabilities of censoring. The fifth column gives the mean squared error for predictions of the observed outcomes. All measures are evaluated in validations samples in 5-fold cross-fold validation.}
	\label{mroz_cv_cov_table}
\end{table}

\begin{table}[ht]
\centering
\begin{tabular}{l|p{1.15cm}p{1.15cm}p{1.15cm}p{1.15cm}p{1.5cm}p{1.5cm}}
  \hline
  \hline
  & Tobit 2 step & Tobit ML & BART & Bayes Splines & TOBART 2 & TOBART 2 NP \\ 
    \hline
Prob. Censored & -0.162 & -0.106 & -0.168 & 0.358 & -0.113 & -0.125 \\ 
Observed Outcome & -0.232 & -0.180 & 0 &  & -0.128 & -0.155 \\ 
   \hline
\end{tabular}
\caption{Average effect of having kids on the probability of working and the log of wages conditional on working. The data is from \cite{mroz1987sensitivity}.}
\label{mroz_kids_ate}
\end{table}

\begin{table}[ht]
\centering
\begin{tabular}{l|p{1.15cm}p{1.15cm}p{2.75cm}p{3cm}}
  \hline
  \hline
 & Tobit 2 step & Tobit ML  & TOBART 2 & TOBART 2 NP \\ 
   \hline
 Observed Outcome & -0.232 & -0.180  & -0.128 & -0.155\\ 
Interval &  &   & (-0.217,-0.038) & (-0.269,-0.0406) \\ 
   \hline
\end{tabular}
\caption{Average effect of having kids on the log of wages conditional on working, and 95\% credible intervals for the estimated effects. The data is from \cite{mroz1987sensitivity}.}
\label{mroz_kids_ate_ints}
\end{table}

\begin{table}[ht]
\centering
\begin{tabular}{l|p{1.15cm}p{1.15cm}p{1.15cm}p{1.15cm}p{1.5cm}p{1.5cm}}
  \hline
  \hline
 & Tobit 2 step & Tobit ML & BART & Bayes Splines & TOBART 2 & TOBART 2 NP \\ 
   \hline
Latent Outcome & 0.052 & 0.056 &  & 0.043 & 0.004 & -0.056 \\ 
Observed Outcome & 0.030 & 0.031 & 0.027 &  & 0.004 & -0.077 \\ 
   \hline
\end{tabular}
\caption{Average effect of living in a city on the probability of working, the log of offered wage, and the log of wages conditional on working. The data is from \cite{mroz1987sensitivity}.}
\label{mroz_city_ate}
\end{table}

\begin{table}[ht]
\centering
\begin{tabular}{l|p{1.15cm}p{1.15cm}p{2.8cm}p{2.8cm}p{2.8cm}}
  \hline
  \hline
 & Tobit 2-step & Tobit ML & BART & TOBART 2 & TOBART 2 NP \\
   \hline
Latent Outcome & 0.052 & 0.056 &  & 0.004 & -0.056 \\ 
Interval &  &  &  & (-0.132,0.140) & (-0.167,0.057) \\ 
Observed Outcome & 0.030 & 0.031 & 0.027 & 0.004 & -0.077 \\ 
Interval &  &  & (-0.116,0.189) & (-0.132,0.140) & (-0.240,0.091) \\ 
   \hline
\end{tabular}
\caption{Average effect of living in a city on the probability of working, the log of offered wage, and the log of wages conditional on working, with 95\% credible intervals. The data is from \cite{mroz1987sensitivity}.}
\label{mroz_city_ate_int}
\end{table}

\begin{table}[ht]
\centering
\begin{tabular}{lp{1.15cm}p{1.15cm}p{1.15cm}p{1.5cm}p{1.5cm}}
  \hline
  \hline
 & Tobit 2-step & Tobit ML & Bayes Splines & TOBART 2 & TOBART 2 NP \\ 
   \hline
11 years education & 1.245 & 1.493 & 1.809 & 1.420 & 1.288 \\ 
12 years education & 1.329 & 1.557 & 1.889 & 1.562 & 1.433 \\ 
13 years education & 1.413 & 1.622 & 1.949 & 1.567 & 1.506 \\ 
14 years education & 1.497 & 1.687 & 2.017 & 1.562 & 1.573 \\ 
   \hline
\end{tabular}
\caption{Expectation of (latent) offered wage by education level. The data is from \cite{mroz1987sensitivity}.}
\label{mroz_educ_latent_exp_wage}
\end{table}

\begin{table}[ht]
\centering
\begin{tabular}{l|p{3cm}p{3cm}}
  \hline
  \hline
  & TOBART 2 & TOBART 2 NP \\ 
    \hline
11 years education & (1.151, 1.694) & (1.064, 1.512) \\ 
12 years education & (1.425, 1.696) & (1.300, 1.570) \\ 
13 years education & (1.348, 1.771) & (1.334, 1.699) \\ 
14 years education & (1.337, 1.775) & (1.382, 1.772) \\ 
   \hline
\end{tabular}
\caption{Expectation of (latent) offered wage by education level with 95\% credible intervals. The data is from \cite{mroz1987sensitivity}.}
\label{mroz_educ_latent_exp_wage_int}
\end{table}

\FloatBarrier

\subsection{\cite{schochet2001national} Job Corps training}

The Job Corps experiment involved random allocation of disadvantaged youths to academic and vocational training \citep{schochet2001national}. We apply sample selection models to estimate the effect of training allocation and training participation on the log of weekly wages four years after allocation to the training program. The set of covariates contains survey responses related to demographics, education, employment history, and health. We include the same set of covariates in the selection and outcome equations. i.e there is no exclusion restriction.

\begin{table}[ht]
\centering
\begin{tabular}{l|p{1cm}p{1cm}p{1cm}p{1cm}}
  \hline
Method & Hit Rate & Brier & AUC & MSE cond. \\ 
  \hline
Tobit-2Step & 0.828 & 0.139 & 0.629 & 183.474 \\ 
Tobit-ML & 0.828 & 0.140 & 0.623 & 181.073 \\ 
Logistic Regression & 0.828 & 0.139 & 0.629 &  \\ 
RF & 0.828 & 0.140 & 0.614 & 180.677 \\ 
BART & 0.828 & 0.138 & 0.639 & 179.664 \\ 
TOBART-2 & 0.828 & 0.138 & 0.637 & 179.841 \\ 
   \hline
\end{tabular}
\caption{Predictive Accuracy of models applied to Job Corps Experiment data \citep{schochet2001national}. The second column gives the hit rate for binary censoring predictions, the third and fourth columns give the Brier score and Area under the Receiver Operating Characteristics Curve for predicted probabilities of censoring. The fifth column gives the mean squared error for predictions of the observed outcomes. All measures are evaluated in validations samples in 5-fold cross-fold validation.}
	\label{JC_alloc_cv_cov_table}
\end{table}

\subsubsection{Effect of Training Allocation}

It can be observed in Tables \ref{JC_alloc_ATE_tab} and \ref{JC_alloc_ATE_int_tab} that the effect of training program allocation on offered wages is estimated by TOBART-2 to be larger than it is estimated to be by a linear model. The effect on the observed outcome estimated by TOBART-2 is smaller than the effect estimated by standard BART applied to selected observations, and also much smaller than the TOBART-2 effect on the latent outcome (offered wage). This suggests that there is a large negative sample selection bias. However, the 95\% credible intervals for both effects, reported in Table \ref{JC_alloc_ATE_int_tab} are very wide.

\begin{table}[ht]
\centering
\begin{tabular}{rlllll}
  \hline
& Tobit 2-step & Tobit ML & BART & TOBART 2 \\ 
  \hline
Prob. Censored & 0.027 & 0.022 & 0.026 & 0.032 \\ 
Latent Outcome & 16.946 & 19.159 &  & 22.884 \\ 
Observed Outcome & 19.188 & 16.319 & 16.864 & 12.900 \\ 
   \hline
\end{tabular}
\caption{Average effect of allocation to training program on probability of working (or survey response), log of offered weekly earnings 4 years later, and log of observed weekly earnings 4 years later (conditional on selection/response).}
\label{JC_alloc_ATE_tab}
\end{table}

\begin{table}[ht]
\centering
\begin{tabular}{llllll}
  \hline
  \hline
  & Tobit 2-step & Tobit ML & BART & TOBART-2 \\ 
    \hline
Latent Outcome & 16.946 & 19.159 &  & 22.884 \\ 
Interval &  &  &  & (14.8403, 31.572) \\ 
Observed Outcome & 19.188 & 16.319 & 16.864 & 12.900 \\ 
Interval &  &  & (10.238, 21.721) & (6.965, 19.299) \\ 
   \hline
\end{tabular}
\caption{Average effect of allocation to training program on probability of working (or survey response), log of offered weekly earnings 4 years later, and log of observed weekly earnings 4 years later (conditional on selection/response), with 95\% credible intervals.}
\label{JC_alloc_ATE_int_tab}
\end{table}

\FloatBarrier

\subsubsection{Effect of Training Participation}

The results for the effect of observed participation in the training program, which is possibly confounded if the training participation decision is confounded by unobservables, are displayed below. The average effects on the observed outcome (i.e. conditional on selection) are similar for BART and TOBART-2 are similar. This is unsurprising because, as explained elsewhere in this paper, BART trained on selected outcomes should accurately predict outcomes conditional on selection. The estimated effect on the latent outcome (i.e. the offered wage) is much larger than the effect estimate for the linear model. However, these effects should be viewed cautiously considering the possibility of unobserved confounding of treatment and also poor identification of the sample selection decision due to the lack of excluded variable in the sample selection equation.


\begin{table}[ht]
\centering
\begin{tabular}{rlllll}
  \hline
& Tobit 2-step & Tobit ML & BART & TOBART-2 \\ 
  \hline
Prob. Censored & 0.055 & 0.055 & 0.055 & 0.049 \\ 
Latent Outcome & 17.335 & 18.913 &  & 35.441 \\ 
Observed Outcome & 28.884 & 28.864 & 20.093 & 20.115 \\  
   \hline
\end{tabular}
\caption{Average effect of participation in training program on probability of working (or survey response), log of offered weekly earnings 4 years later, and log of observed weekly earnings 4 years later (conditional on selection/response).}
\label{JC_partic_ATE_tab}
\end{table}

\begin{table}[ht]
\centering
\begin{tabular}{llllll}
  \hline
  \hline
  & Tobit 2-step & Tobit ML & BART & TOBART-2 \\ 
    \hline
Latent Outcome & 17.335 & 18.913 &  & 35.441 \\ 
Interval & &  &  & (26.530, 45.111) \\ 
Observed Outcome & 28.884 & 28.864 & 20.093 & 20.115 \\ 
Interval &  &  & (14.360, 62.973) & (13.691, 26.990) \\ 
   \hline
\end{tabular}
\caption{Average effect of participation in training program on probability of working (or survey response), log of offered weekly earnings 4 years later, and log of observed weekly earnings 4 years later (conditional on selection/response), with 95\% credible intervals.}
\label{JC_partic_ATE_int_tab}
\end{table}


\end{document}